\documentstyle[11pt]{article}
\begin{document}
\hfuzz=7.0pt
\chardef\ii="10
\def\ih{{\i\over\hbar}}
\def\CD{{\cal D}}
\def\CL{{\cal L}}
\def\CP{{\cal P}}
\def\half{{1\over2}}
\def\bhalf{\hbox{$\half$}}
\def\viert{{1\over4}}
\def\bviert{\hbox{$\viert$}}
\def\Ai{\hbox{Ai}}
\def\SU{\hbox{SU}}
\def\SO{\hbox{SO}}
\def\OO{\hbox{O}}
\def\diag{\hbox{diag}}
\def\det{\hbox{det}}
\def\cn{\hbox{cn}}
\def\dn{\hbox{dn}}
\def\sn{\hbox{sn}}
\def\ps{\hbox{ps}}
\def\Ps{\hbox{Ps}}
\def\bbbone{{\mathchoice {\rm 1\mskip-4mu l} {\rm 1\mskip-4mu l}
{\rm 1\mskip-4.5mu l} {\rm 1\mskip-5mu l}}}
\def\hi{{\hbar\over\i}}
\def\hbarm{{\hbar^2\over2M}}
\def\ezwei{{E^{(2)}}}
\def\edrei{{E^{(3)}}}
\def\bbbr{{\rm I\!R}} 
\def\bbbn{{\rm I\!N}} 
\def\bbbc{{\mathchoice {\setbox0=\hbox{$\displaystyle\rm C$}\hbox{\hbox
to0pt{\kern0.4\wd0\vrule height0.9\ht0\hss}\box0}}
{\setbox0=\hbox{$\textstyle\rm C$}\hbox{\hbox
to0pt{\kern0.4\wd0\vrule height0.9\ht0\hss}\box0}}
{\setbox0=\hbox{$\scriptstyle\rm C$}\hbox{\hbox
to0pt{\kern0.4\wd0\vrule height0.9\ht0\hss}\box0}}
{\setbox0=\hbox{$\scriptscriptstyle\rm C$}\hbox{\hbox
to0pt{\kern0.4\wd0\vrule height0.9\ht0\hss}\box0}}}}
\def\bbbz{{\mathchoice {\hbox{$\sf\textstyle Z\kern-0.4em Z$}}
{\hbox{$\sf\textstyle Z\kern-0.4em Z$}}
{\hbox{$\sf\scriptstyle Z\kern-0.3em Z$}}
{\hbox{$\sf\scriptscriptstyle Z\kern-0.2em Z$}}}}
\def\e{{\mathchoice {\hbox{$\rm\textstyle e$}}
{\hbox{$\rm\textstyle e$}}
{\hbox{$\rm\scriptstyle e$}}
{\hbox{$\rm\scriptscriptstyle e$}}}}
\def\i{{\mathchoice {\hbox{$\rm\textstyle i$}}
{\hbox{$\rm\textstyle i$}}
{\hbox{$\rm\scriptstyle i$}}
{\hbox{$\rm\scriptscriptstyle i$}}}}
\large
\begin{titlepage}
\centerline{\normalsize DESY 94 - 018 \hfill ISSN 0418 - 9833}
\centerline{\hfill hep-th/9402121}
\centerline{\normalsize February 1994\hfill}
\vskip.3in

\centerline{
\Large PATH INTEGRAL DISCUSSION FOR}
\vskip.1in
\centerline{\Large SMORODINSKY-WINTERNITZ POTENTIALS:}
\vskip.1in
\centerline{\Large I.\ TWO- AND THREE DIMENSIONAL EUCLIDEAN SPACE}
\vskip.5in
\begin{center}
{\Large C.\ Grosche$^*$}
\vskip.3in
{\normalsize\em II.\ Institut f\"ur Theoretische Physik}
\vskip.05in
{\normalsize\em Universit\"at Hamburg, Luruper Chaussee 149}
\vskip.05in
{\normalsize\em 22761 Hamburg, Germany}
\end{center}
\vskip.5in
\begin{center}
{\Large G.\ S.\ Pogosyan$^{**}$ and A.\ N.\ Sissakian$^{**}$}
\vskip.3in
{\normalsize\em Laboratory of Theoretical Physics}
\vskip.05in
{\normalsize\em Joint Institute for Nuclear Research (Dubna)}
\vskip.05in
{\normalsize\em 141980 Dubna, Moscow Region, Russia}
\end{center}
\normalsize
\vfill
\begin{abstract}
\noindent
Path integral formulations for the Smorodinsky-Winternitz potentials in
two- and three-dimen\-sional Euclidean space are presented. We mention
all coordinate systems which separate the Smorodinsky-Winternitz
potentials and state the  corresponding path integral formulations.
Whereas in many coordinate systems an explicit path integral
formulation is not possible, we list in all soluble cases the path
integral evaluations explicitly in terms of the propagators and the
spectral expansions into the wave-functions.
\end{abstract}

\bigskip\bigskip\noindent
\centerline{\vrule height0.25pt depth0.25pt width4cm\hfill}
\noindent
{\footnotesize
$^*$ Supported by Deutsche Forschungsgemeinschaft under contract number
     GR 1031/2--1.
\newline
$^{**}$ Supported by Heisenberg-Landau program.}
\end{titlepage}

\begin{titlepage}
\begin{center}
 \ \ \
\end{center}
\end{titlepage}


\normalsize\noindent
{\large\bf 1.~Introduction.}
\vglue0.4truecm\noindent
In the study of the Kepler problem and the harmonic oscillator it turns
out that they possess properties making them of special interest, for
instance, all finite classical trajectories are closed and all energy
eigen-values are multiply degenerated. These properties point beyond
their specific shape and give rise to ask about their ``hidden'' or
``dynamical'' group structure, namely whether it is possible to distort
them in a systematic way, and at the same time keep these properties.
Furthermore, do they belong to a class of potential problems which have
similar properties?

For instance, as has been known for a long time, at the basis of the
specific properties of the hydrogen atom lies its $\OO(4)$ symmetry, or
to be more precise, the dynamical group $\OO(4)$ for the discrete
spectrum and the Lorentz group $\OO(3,1)$ for the continuous spectrum.
All such systems have the particular property that all the
energy-levels of the system are organized in representations of the
non-invariance group which contain representations of the dynamical
subgroup realized in terms of the wave-functions of these energy-levels
\cite{FMSUW}. In the case of the hydrogen it enabled Pauli \cite{PAULI}
and Fock \cite{FOCK} to solve the quantum mechanical Kepler problem
without explicitly solving the Schr\"odinger equation. Actually, the
algebra of the dynamical symmetry of the hydrogen atom turns out to be
a centerless twisted Kac-Moody algebra \cite{DSD}.

Smorodinsky, Winternitz et al.\ started a systematic search to find
and classify potential problems in two and three dimensions which can be
seen as non-central generalizations of the Coulomb-, the harmonic
oscillator and radial barrier potentials. The classification scheme
starts from the consideration in which way integrals of motion (in
classical mechanics), respectively additional operators corresponding to
these integrals of motion which commute with the Hamiltonian, are
related to the separability of the potential problem in more than one
coordinate system. In two dimensions \cite{FMSUW} it turns out that
there are four potentials of the sought type which all have three
constants of motion (including energy), i.e.\ there are two more
operators commuting with the Hamiltonian. Smorodinsky, Winternitz et
al.\ extended their investigations in a classical paper \cite{MSVW} to
three dimensions by listing all potentials which separate in more than
one coordinate system. However, in three dimensions one must distinguish
two cases, namely minimally super-integrable systems, where four,
and maximally super-integrable systems, where five functionally
independent integrals of motion exist.

Actually in $D$ dimensions a system is called minimally super-integrable
if it has $2D-2$ functionally independent integrals of motions, and it
is called maximally super-integrable if it has $2D-1$ functionally
independent integrals of motion \cite{BDK, HIET}. A discussion of the
two-dimensional case is due to e.g.\ Winternitz et al.\ \cite{WLS},
and a detailed study of the three-dimensional case can be found in
\cite{EVA, KIWIb}. Moreover, the Smorodinsky-Winternitz potentials
allow a generalization to $D$ dimensions and are then also maximally
super-integrable \cite{EVAb}.

In the three-dimensional Euclidean space there are eleven coordinate
systems which separate the free Hamiltonian \cite{KAL, MOON, MF}. Each
potential must now be analysed according to its separability in these
coordinate systems. Because each potential can be rotated
and translated with respect to itself, one has to look for the
equivalence classes among them. As it turns out there are eight
minimally super-integrable (including a ring-shaped oscillator and the
Hartmann potential) and five maximally super-integrable (including
harmonic oscillator and Coulomb) potentials. Evans \cite {EVA}
presented a list of these potentials including the corresponding
integrals of motion and (almost) all separating coordinate systems. The
group structure of the two-dimensional potentials was given in
\cite{FMSUW}, for the Hartmann potential by Gerry \cite{GERRY} and
Granovsky et al.\ \cite{GZLa}, for the ring-shaped oscillator by Quesne
\cite{QUES} and Zhedanov \cite{ZHE}, and the harmonic oscillator
including $1/x^2$-terms (a Smorodinsky-Winternitz potential, see below)
was analysed by Evans \cite{EVA}.

\newpage
The purpose of this paper is to give a path integral discussion of all
Smorodinsky-Winternitz potentials in two- and three-dimensional
Euclidean space. This includes the statement of the propagator (the
time-evolution kernel), respectively, if only available, the
corresponding Green function, and the spectral expansion into
wave-functions and energy-levels. In the discussion of the
two-dimensional Smorodinsky-Winternitz potentials the evaluation will
be somewhat more elaborated in order to demonstrate the techniques. Of
particular importance will be the basic path integral solutions of the
radial harmonic oscillator by Peak and Inomata \cite{PI} (compare also
Duru \cite{DURd} and Goovaerts \cite{GOOb}), and of the P\"oschl-Teller
potential by B\"ohm and Junker \cite{BJb} which is based on the
$\SU(2)$ path integral (compare also Duru \cite{DURb}, and Inomata and
Wilson \cite{INOWI}). All these path integral solutions are now
well-known in the literature and will not be repeated here.

One may ask, if it is possible to analyse the path integral of a
potential problem in terms of its dynamical symmetry group \cite{INO,
IKG}. In order to look at such a path integral formulation we consider
the Lagrangian $\CL(\vec x,{\dot{\vec x}})-V(\vec x)$ ($\vec x\in\bbbr^
{p+q} $) as formulated, say, in a not-necessarily positive definite
space with signature
\begin{equation}
       (g_{ab})=\diag\big(\underbrace{+1,\dots,+1}_{p\ times},
         \underbrace{-1,\dots,-1}_{q\ times}\big)\enspace.
\end{equation}
One introduces, say, polar coordinates $x_\nu=re_\nu(\theta_1,\dots,
\theta_{p+q-1})$ $(\nu=1,\dots,p+q)$, where the $\vec e$ are unit
vectors in some suitable chosen (timelike, spacelike or lightlike) set
\cite{BJb}. One then expresses the Lagrangian in terms of these polar
coordinates and seeks for an expansion of the quantity $\e^{z(\vec
e_1\cdot\vec e_2)}$ expressed in terms of group elements $g_1\,,g_2$.
If this is possible one can re-express the path integration of the
coordinates $\vec x$ into a path integration over group elements $g$
yielding \cite{BJb, DOWa, DOWb, MATE, PICK}
\begin{eqnarray}
  \int\limits_{\vec x(t')=\vec x'}^{\vec x(t'')=\vec x''}\CD\vec x(t)
  \exp\left[\ih\int_{t'}^{t''}\CL(\vec x,{\dot{\vec x}})dt\right]
       & \mapsto&
  \int dE_\lambda d_\lambda\sum_{mn}
  {\hat f}_{mn}^\lambda D_{mn}^\lambda({g'}^{-1}g'')
               \\
       & = &
  \int dE_\lambda d_\lambda\sum_{mn}{\hat f}_{mn}^\lambda
   \sum_kD_{m,k}^{\lambda\,*}(g')D_{m,k}^\lambda(g'')\enspace.
\end{eqnarray}
Here ${\hat f}^\lambda_{mn}$ is defined via the Fourier transformation
\begin{equation}
       f(g)=\int dE_\lambda d_\lambda
       \sum_{mn}\hat f^\lambda_{mn}D^\lambda_{mn}(g)
  \enspace,\qquad
  \hat f^\lambda_{mn}=\int_G f(g)D^{\lambda\,*}_{mn}(g^{-1})dg\enspace,
\end{equation}
and $dg$ is the invariant group (Haar) measure. $\int dE_\lambda$
stands for a Lebesque-Stieltjes integral to include discrete $(\int
dE_\lambda\to\sum_\lambda$) as well as continuous representations.
$\int dE_\lambda$ is to be taken over the complete set $\{\lambda\}$ of
class one representations. $d_\lambda$ denotes (in the compact case)
the dimension of the representation and we take
\begin{equation}
       d_\lambda\int_G D^\lambda_{mn}(g)
       D^{\lambda'\,*}_{m'n'}(g)dg=\delta(\lambda,\lambda')
  \delta_{m,m'}\delta_{n,n'}
\end{equation}
as a definition for $d_\lambda$. $\delta(\lambda,\lambda')$ can denote
a Kronecker delta, respectively, a $\delta$-function, depending whether
the quantity $\lambda$ is a discrete or continuous parameter. We have
furthermore used the group (composition) law
\begin{equation}
       D^\lambda_{mn}(g_a^{-1}g_b)
  =\sum_k D^{\lambda\,*}_{kn}(g_a) D^\lambda_{km}(g_b)\enspace.
\end{equation}
Choosing a basis $\{\vec b\,\}$ in the relevant Hilbert space fixes the
matrix elements $D_{mn}^\lambda$ through $D^\lambda_{mn}=(D^\lambda(g)
b_m ,b_n)$ of the representation $D^\lambda(g)$ of the group. In
particular the $D^\lambda_{0m}$ are called associated spherical
harmonics, and the $D^\lambda_{00}$ are the zonal harmonics. These
spherical functions are eigen-functions of the corresponding
Laplace-Beltrami operator on a, say, homogeneous space, and the entire
Hilbert space is spanned by a complete set of associated spherical
functions $D^\lambda_{0m}$ (Gelfand et al.~\cite{GEGR, GGV} and
Vilenkin \cite{VIL}). In \cite{GROab} a systematic study was presented
of path integrals of the free motion in two- and three-dimensional
spaces of constant curvature, i.e.~on homogeneous spaces.

The Smorodinsky-Winternitz potentials can also be discussed in the
context of a group path integration, provided the dynamical ``hidden''
group structure behind the potential problem is known, and the
wave-functions and the energy-levels follow from the representations
of the group \cite{INO, IKG}. However, things are not as simple as for
the free motion because we have not a problem of quantum motion on a
homogeneous space. Whereas in many coordinate system realizations we
can actually evaluate the path integral, there remain several ones,
where a direct path integral evaluation is {\it not possible\/} because
we obtain quartic, sextic, complicated Mathieu-like path integral
problems, where no solution is known. Also a consideration along the
lines of ``quasi-exactly solvable path integrals'' \cite{LUJA} can not
be applied in an obvious way. However, sometimes the zero-energy
Green's function can be evaluated. As we will see, the technique of
separation of variables in the path integral \cite{GROj} will give us a
convenient tool to exploit the actually symmetry structure in such a way
that we can systematically reduce the number of degrees of freedom by
using the appropriate path integral identities.

We will therefore provide the path integral formulations of all
Smorodinsky-Winternitz potentials in two- and three-dimensional
Euclidean space. Some can only be stated, others can be explicitly
solved. We find that each of the potentials can be at least solved in
one coordinate system. The identities arising from the relations
between the various formulations serve as a tool to analyse the
``hidden'' dynamical group structure, however, only in an implicit way.
The representations of the group in question where a path integral is
not explicitly soluble remain still unknown. Furthermore, the explicit
spectral expansions corresponding to various coordinate systems allow
to study the interbasis expansions of the problem. An example of
particular importance is the two-dimensional hydrogen atom \cite{POGOb,
MPSTAb}, respectively the ring-shaped oscillator \cite{LPSTA}, and the
Hartmann potential \cite{LPSTA, ZHE}.

The knowledge of soluble examples and their relation with each other,
in particular here the Smorodinsky-Winternitz potentials, is of great
value in many branches of theoretical physics. For instance, the
Hartmann system \cite{HART, KNe, KIWIa} with either an oscillator or an
Coulomb interaction term appears in the study of molecule physics
(modeling ring-shaped molecules like the benzene molecule), in the study
of the Aharonov-Bohm effect \cite{AB, KNf}, for path integral studies
c.f.\ e.g.~(Bernido and Inomata \cite{BEIN}, Cheng \cite{CHENGm},
Chetouani et al.~\cite{CGHc}, Gerry and Singh \cite{GSa}, Hoang and
Giang \cite{LVHG}, S\"okmen \cite{SOKc}, Tanikella and Inomata
\cite{TAIN}, and Wiegel \cite{WIEc}), and for a more recent discussion
of a deformed generalization \cite{KICA}. Evans \cite{EVA} also
mentioned cases in celestical mechanics (e.g.\ an artificial satellite
orbiting an oblate planet, dynamics of elliptical galaxies). The
consideration of the Smorodinsky-Winternitz potentials in spaces of
constant (positive and negative) curvature models a space-time of
closed, respectively open universes, c.f.\ the Kepler problem
and the Higgs oscillator problem (Granovsky et al.\ \cite{GZLb}, Higgs
\cite{HIGGS}, Infeld \cite{INFELD}, Leemon \cite{LEEMON}, Otchik and
Red'kov \cite{OTRE}, Pogosyan et al.\ \cite{POGOc}, Schr\"odinger
\cite{SCHRO}, and Steveson \cite{STEV}, and path integral discussions
c.f.\ Barut et al.~\cite{BIJa, BIJb}, and Grosche \cite{GROg}). Path
integral discussions of Smorodinsky-Winternitz potentials in space of
constant positive and negative curvature will be given in forthcoming
publications. Starting from a particular exactly soluble example one can
perturb it by a further (vector or scalar) potential which is not
destroying separability in a conveniently chosen coordinate system
\cite{MSVW}, one can expand about the exact solution of the unperturbed
one, hence giving a perturbation series in, say, either powers of the
coupling parameter (e.g.~\cite{MIZg}), respectively in powers of
$\hbar$. The latter then gives a ``pseudo'' semi-classical
expansion \'a la DeWitt-Morette \cite{MIZd}-\cite{DEWMc}.

\newpage
Our work can be seen as a generalization and extension of the earlier
work of Carpio-Bernido et al.~\cite{CARPa}-\cite{CBI}. They discussed
only some of the Smorodinsky-Winternitz potentials and they restricted
themselves only to a spherical polar coordinate path integral discussion
in $\edrei$, the three-dimensional Euclidean space. Their results are,
of course, included in our discussion. They also did not discuss a
classification scheme of the potentials. In particular, all our path
integral evaluations in parabolic coordinates will be new path integral
solutions, including the expansion into discrete and continuous
wave-functions (if present). The path integral solutions of the
two-dimensional Smorodinsky-Winternitz potential $V_4$ and
three-dimensional Smorodinsky-Winternitz potential $V_3$ are entirely
new (in Refs.\ \cite{CARPa}-\cite{CBI} special cases of the latter
have been discussed). We therefore give a combined presentation of
henceforth solutions which have been scattered around in the literature
and we also give {\it all\/} admissable solutions of each potential
problem.

Our paper will be organized as follows: In the next Section we will list
for completeness the coordinate systems in two- and three-dimensional
flat space which separate the free Helmholtz, respectively Schr\"odinger
equation. Also a  separation formula for path integrals will be stated.

In the third Section we discuss the Smorodinsky-Winternitz potentials in
two and three dimensions, respectively. This includes the formulation of
the path integral in the various separating coordinate systems, and if
possible via the basic path integrals, its solution in terms of the
propagator, respectively the Green functions, and its spectral
expansion. In particular, we will make frequently use of the separation
technique in path integrals as developed in \cite{GROj}, and will not
write down explicitly each step in the calculation of the examples of
the explicitly soluble Smorodinsky-Winternitz potentials. In our path
integral solutions we will state the Feynman kernels as explicitly as
possible, and their spectral expansions according to
\begin{eqnarray}
  K({\vec x\,}'',{\vec x\,}';T)
  &=&
  \int dE_\lambda \Psi_\lambda^*\lambda(\vec x')\Psi_\lambda(\vec x'')
  \,\e^{-\i E_\lambda T/\hbar}
  \\
  &=&\left\{\begin{array}{l}
  \displaystyle
  \sum_n\Psi_n({\vec x\,}')\Psi_n({\vec x\,}'')\,\e^{-\i E_n T/\hbar}
  \\
  \phantom{.}
  \\
  \displaystyle
  \int dp\Psi_p({\vec x\,}')\Psi_p({\vec x\,}'')\,\e^{-\i\hbar p^2 T/2M}
  \\
  \phantom{.}
  \\
  \displaystyle
  \sum_n\Psi_n({\vec x\,}')\Psi_n({\vec x\,}'')\,\e^{-\i E_n T/\hbar}
  +\int dp\Psi_p^*({\vec x\,}')\Psi_p({\vec x\,}'')
  \,\e^{-\i\hbar p^2 T/2M}\enspace,
\end{array}\right.
\end{eqnarray}
whether the spectrum is purely discrete, purely continuous, or discrete
and continuous. In these cases, where only the (energy-dependent) Green
function is available, the spectral expansions will be presented
similarly.

The fourth Section is devoted to a summary and a discussion of our
results.


\vglue0.6truecm\noindent
{\large\bf 2.~Coordinate Systems in Euclidean Space and the
          Path Integral.}
\vglue0.4truecm\noindent
{\bf 2.1.~Coordinate Systems in Euclidean Space.}
\vglue0.4truecm\noindent
{\it 2.1.1.~Coordinate Systems in $D$ Dimensions.}
In D-dimensional Euclidean space we have the coordinate system
corresponding to the elliptic coordinates
\begin{equation}
  \begin{array}{l}
  \displaystyle
  x_j^2=\frac{\prod_{i=1}^D(\rho_i-e_j)}
       {\prod_{i\not=j}^D(e_i-e_j)}\enspace,\qquad(j=1,\dots,D)
  \\
  \displaystyle
  e_1<\rho_1<e_2<\dots<e_D<\rho_D\enspace.
\end{array}
\end{equation}
This is the most general coordinate system possible, and all other
coordinate systems are included as degenerations. In the following we
will explicitly state the coordinate systems in two- and
three-dimensional flat space $\ezwei$ and $\edrei$ which separate the
Helmholtz, respectively the Schr\"odinger equation. We rewrite the
Hamiltonian, respectively the Laplacian $\Delta_L=g^{-1/2}\partial_a
g^{ab}g^{1/2}\partial_b$ in terms of the corresponding momentum
operators $p_a=-\i\hbar g^{-1/4}\partial_a g^{1/4}=-\i\hbar(\partial_a
+\Gamma_a/2)$ ($\Gamma_a=\partial_a\ln\sqrt{g}$, $g\det(g_{ab})$), where
we choose a specific ordering prescription called product ordering
\cite{GROa}. Here it is assumed that $g_{ab}=h_{ac}h_{cb}$,
$g^{ab}=h^{ac}h^{cb}$, the $h_{ac}$ being vielbeins, such that we can
write $H=-\hbarm\Delta_L={1\over2M}h^{ac}p_ap_bh^{cb}+\Delta V_{PF}$
with the quantum potential $\Delta V_{PF}$ given by
\begin{equation}
  \Delta V_{PF}={\hbar^2\over8M}
  \Big[g^{ab}\Gamma_a\Gamma_b+2(g^{ab}\Gamma_b)_{,b}
   +{g^{ab}}_{,ab}\Big]
  +{\hbar^2\over8M}
  \Big(2h^{ac}{h^{bc}}_{,ab}-{h^{ac}}_{,a}{h^{bc}}_{,b}
                -{h^{ac}}_{,b}{h^{bc}}_{,a}\Big)\enspace.
\label{numba}
\end{equation}
The potentials which may be added without spoiling the separability
were derived by Eisenhardt \cite{EIS} and Smorodinsky and Winternitz et
al.~\cite{FMSUW, MSVW}.

\vglue0.4truecm\noindent
{\it 2.1.2.~Coordinate Systems in Two Dimensions.}
\newline
{\it Cartesian Coordinates.\/}
We consider the usual cartesian coordinates $(x,y)=\vec x\in\bbbr^2
\equiv\ezwei$. The metric is given by $(g_{ab})=\bbbone_2$ and for the
momentum operators we have $p_x=-\i\hbar\partial_x$,
$p_y=-\i\hbar\partial_y$. Therefore:
\begin{equation}
  -\hbarm\Delta_\ezwei
  =-\hbarm\bigg({\partial^2\over\partial x^2}
               +{\partial^2\over\partial y^2}\bigg)
  ={1\over2M}{\vec p\,}^2\enspace.
\end{equation}
The most general potential separable in cartesian coordinates has the
form
\begin{equation}
  V(\vec x)=u(x)+v(y)\enspace.
\end{equation}

\noindent
{\it Polar Coordinates.\/}
Next, we consider two-dimensional polar coordinates
\begin{equation}
  \begin{array}{lll}
  x=\rho\cos\phi\enspace,
  &\qquad
  &\rho>0\enspace,
  \\
  y=\rho\sin\phi\enspace,
  &\qquad
  &0\leq\phi<2\pi\enspace.
\end{array}
\end{equation}
The metric is given by $(g_{ab})=\diag(1,\rho^2)$, and the momentum
operators have the form
\begin{equation}
  p_\rho=\hi\bigg({\partial\over\partial\rho}+{1\over2\rho}\bigg)
  \enspace,\qquad
  p_\phi=\hi{\partial\over\partial\phi}\enspace.
\label{numdf}
\end{equation}
This gives for the Hamiltonian
\begin{equation}
  -\hbarm\Delta_\ezwei
  =-\hbarm\bigg(
   {\partial^2\over\partial\rho^2}+{1\over\rho}
       {\partial\over\partial\rho}
   +{1\over\rho^2}{\partial^2\over\partial\phi^2}\bigg)
  ={1\over2M}\bigg(p_\rho^2+{1\over\rho^2}p_\phi^2\bigg)
   -{\hbar^2\over8M\rho^2}\enspace.
\end{equation}
A separable potential has the form
\begin{equation}
  V(\vec x)=u(\rho)+{1\over\rho^2}v(\phi)\enspace.
\end{equation}

\noindent
{\it Elliptic Coordinates.\/}
Third, we consider the coordinate system
\begin{equation}\begin{array}{lll}
  x=d\cosh\xi\cos\eta\enspace,
  &\qquad
  &\xi>0\enspace,
  \\
  y=d\sinh\xi\sin\eta\enspace,
  &\qquad
  &-\pi<\eta\leq\pi\enspace,
\end{array}
\end{equation}
and $d=R/2$, where $R$ is the interfocus distance.
The metric is $(g_{ab})=d^2(\sinh^2\xi+\sin^2\eta)\bbbone_2$ and we
obtain for the momentum operators
\begin{equation}
  p_\xi=\hi\bigg({\partial\over\partial\xi}
         +{\sinh\xi\cosh\xi\over\sinh^2\xi+\sin^2\eta}\bigg)\enspace,
  \qquad
  p_\eta=\hi\bigg({\partial\over\partial\eta}
         +{\sin\eta\cos\eta\over\sinh^2\xi+\sin^2\eta}\bigg)\enspace.
\label{numec}
\end{equation}
Consequently we have for the Hamiltonian
\begin{eqnarray}
  -\hbarm\Delta_\ezwei
  &=&-{\hbar^2\over2Md^2(\sinh^2\xi+\sin^2\eta)}
    \bigg({\partial^2\over\partial\xi^2}
         +{\partial^2\over\partial\eta^2}\bigg)
         \nonumber\\
  &=&{1\over2Md^2}(\sinh^2\xi+\sin^2\eta)^{-1/2}
  (p_\xi^2+p_\eta^2)(\sinh^2\xi+\sin^2\eta)^{-1/2}\enspace.
\end{eqnarray}
Replacing $x\mapsto x=d(\cosh\xi\cos\eta+1)$ yields a coordinate system
which we call in the following {\it elliptic II coordinates\/}
\cite{MPSTAb}. The elliptic coordinate system is useful in the study of
two center systems. Furthermore, an additional parameter is introduced,
the interfocus distance $R$. This distinguishes it from the other
two-dimensional coordinate systems which can be obtained from the
elliptic one by considering specific limits. In particular $R\to0$
reproduces polar coordinates, and $R\to\infty$ cartesian coordinates.
In the case of the elliptic II systems, $R\to\infty$ reproduces
parabolic coordinates (see below).

Note $\Delta V_{PF}=0$ which is a peculiarity of two dimensions with
metric $\propto\bbbone$, c.f.~(\ref{numba}). A potential separable in
these coordinates reads
\begin{equation}
  V(\vec x)={u(\cosh\xi)+v(\cos\eta)\over\sinh^2\xi+\sin^2\eta}\enspace.
\end{equation}

\noindent
{\it Parabolic Coordinates.\/}
Finally, we consider the coordinate system
\begin{equation}
  x=\bhalf(\eta^2-\xi^2)\enspace,\qquad
  y=\xi\eta\enspace,\qquad
   \xi\in\bbbr,\eta>0
\end{equation}
(alternatively $\xi>0$, $\eta\in\bbbr$ \cite{MESCH}). We have $(g_{ab})=
(\xi^2+\eta^2)\bbbone_2$, and consequently for the momentum operators
\begin{equation}
  p_\xi=\hi\bigg({\partial\over\partial\xi}
          +{\xi\over\xi^2+\eta^2}\bigg)\enspace,\qquad
  p_\eta=\hi\bigg({\partial\over\partial\eta}
          +{\eta\over\xi^2+\eta^2}\bigg)\enspace.
\label{numbb}
\end{equation}
This gives for the Hamiltonian
\begin{equation}
  -\hbarm\Delta_\ezwei
  =-{\hbar^2\over2M(\xi^2+\eta^2)}
   \bigg({\partial^2\over\partial\xi^2}
        +{\partial^2\over\partial\eta^2}\bigg)
  ={1\over2M}(\xi^2+\eta^2)^{-1/2}
             (p_\xi^2+p_\eta^2)(\xi^2+\eta^2)^{-1/2}\enspace.
\end{equation}
These coordinates have been used in the literature to discuss the
two-dimensional ``Cou\-lomb-pro\-blem'', c.f.~\cite{DKb, FMSUW, GROq,
INOa}. The transformation $(x,y)\mapsto(\xi,\eta)$ actually is the
two-dimensional realization of the Levi-Civita transformation.
A separable potential has the form
\begin{equation}
  V(\vec x)={u(\xi)+v(\eta)\over\xi^2+\eta^2}\enspace.
\end{equation}

\vskip0.6truecm\noindent
{\it 2.1.3.~Coordinate Systems in Three Dimensions.}
\par\par\noindent
{\it Cartesian Coordinates.\/}
Again, we start with the simplest case, cartesian coordinates
$(x,y,z)$ \linebreak $=\vec x\in\bbbr^3\equiv\edrei$. Then
$(g_{ab})=\bbbone_3$, and $\vec p=-\i\hbar\nabla$. This gives for the
Hamiltonian
\begin{equation}
  -\hbarm\Delta_\edrei
  =-\hbarm\nabla^2={1\over2M}{\vec p\,}^2\enspace,
\end{equation}
The most general potential separable in cartesian coordinates has the
form
\begin{equation}
  V(\vec x)=u(x)+v(y)+w(z)\enspace.
\end{equation}

\noindent
{\it Circular Polar Coordinates.\/}
Next, we consider circular cylinder coordinates which are very similar
to the two-dimensional polar coordinates
\begin{equation}\left.\begin{array}{lll}
  x=\rho\cos\phi\enspace,
  &\qquad
  &\rho>0\enspace,
  \\
  y=\rho\sin\phi\enspace,
  &\qquad
  &0\leq\phi<2\pi\enspace,
  \\
  z=z\enspace,
  &\qquad
  &z\in\bbbr\enspace.
\end{array}\qquad\qquad\right\}
\end{equation}
The metric reads $(g_{ab})=\diag(1,\rho^2,1)$, and the momentum
operators are given by (\ref{numdf}) and $p_z=-\i\hbar\partial_z$.
This gives for the Hamiltonian
\begin{equation}
  -\hbarm\Delta_\edrei
  =-\hbarm\bigg(
   {\partial^2\over\partial\rho^2}
   +{1\over\rho}{\partial\over\partial\rho}
    +{1\over\rho^2}{\partial^2\over\partial\phi^2}
    +{\partial^2\over\partial z^2}\bigg)
  ={1\over2M}\bigg(p_\rho^2+{1\over\rho^2}p_\phi^2+p_z^2\bigg)
   -{\hbar^2\over8M\rho^2}\enspace.
\end{equation}
A separable potential has the form
\begin{equation}
  V(\vec x)=u(\rho)+{1\over\rho^2}v(\phi)+w(z)\enspace.
\end{equation}

\noindent
{\it Circular Elliptic coordinates.\/}
We consider the coordinate system
\begin{equation}\left.\begin{array}{lll}
  x=d\cosh\xi\cos\eta\enspace,
  &\qquad
  &\xi>0\enspace,
  \\
  y=d\sinh\xi\sin\eta\enspace,
  &\qquad
  &-\pi<\eta\leq\pi\enspace,
  \\
  z=z\enspace,
  &\qquad
  &z\in\bbbr
\end{array}\qquad\qquad\right\}
\end{equation}
(alternatively $\xi\in\bbbr$, $0<\eta<\pi$ \cite{MESCH}).
The metric is $(g_{ab})=\diag[d^2(\sinh^2\xi+\sin^2\eta),
d^2(\sinh^2\xi+\sin^2\eta),1]$, and we obtain for the momentum operators
(\ref{numec}), and $p_z=-\i\hbar\partial_z$. Consequently for the
       Hamiltonian
\begin{eqnarray}
  -\hbarm\Delta_\edrei
  &=&-\hbarm\bigg[{1\over d^2(\sinh^2\xi+\sin^2\eta)}
    \bigg({\partial^2\over\partial\xi^2}
    +{\partial^2\over\partial\eta^2}\bigg)
    +{\partial^2\over\partial z^2}\bigg]
         \nonumber\\
  &=&{1\over2M}\bigg[{1\over d^2}(\sinh^2\xi+\sin^2\eta)^{-1/2}
  (p_\xi^2+p_\eta^2)(\sinh^2\xi+\sin^2\eta)^{-1/2}
  +p_z^2\bigg]\enspace.\qquad
\end{eqnarray}
Similarly as in the two-dimensional case, {\it circular elliptic II}
coordinates are described by the replacement $x\mapsto x=d(\cosh\xi
\cos\eta+1)$. A potential separable in elliptic coordinates reads
\begin{equation}
  V(\vec x)={u(\cosh\xi)+v(\cos\eta)\over\sinh^2\xi+\sin^2\eta}+w(z)
   \enspace.
\end{equation}

\noindent
{\it Circular Parabolic Coordinates.\/}
The last example for cylinder coordinates in three dimensions are the
parabolic cylinder coordinates
\begin{equation}\begin{array}{lll}
  x=\bhalf(\eta^2-\xi^2)\enspace,\qquad
  y=\xi\eta\enspace,
  &\qquad
  &\xi\in\bbbr,\eta>0\enspace,
  \\
  z=z\enspace,
  &\qquad
  &z\in\bbbr\enspace,
\end{array}
\end{equation}
which is the obvious generalization of the two-dimensional case.
We have $(g_{ab})=\diag(\xi^2+\eta^2,\xi^2+\eta^2,1)$, hence for the
momentum operators (\ref{numbb}) and $p_z=-\i\hbar\partial_z$. This
gives for the Hamiltonian
\begin{eqnarray}
  -\hbarm\Delta_\edrei
  &=&-\hbarm\bigg[{1\over\xi^2+\eta^2}
   \bigg({\partial^2\over\partial\xi^2}
        +{\partial^2\over\partial\eta^2}\bigg)
    +{\partial^2\over\partial z^2}\bigg]
         \nonumber\\
  &=&{1\over2M}\Big[(\xi^2+\eta^2)^{-1/2}
             (p_\xi^2+p_\eta^2)(\xi^2+\eta^2)^{-1/2}
             +p_z^2\Big]\enspace.
\end{eqnarray}
A separable potential has the form
\begin{equation}
  V(\vec x)={u(\xi)+v(\eta)\over\xi^2+\eta^2}+w(z)\enspace.
\end{equation}

\noindent
{\it Sphero-Conical Coordinates.\/}
We consider the coordinate system in terms of Jacobi functions
\cite{PAWI}
\begin{equation}\left.\begin{array}{lll}
  x=r\sn(\mu,k)\dn(\nu,k')\enspace,
    &\quad      &r>0\enspace,\quad k^2+{k'}^2=1\enspace,
  \\
  y=r\cn(\mu,k)\cn(\nu,k')\enspace,
    &\qquad     &-k\leq\mu\leq k\enspace
  \\
  z=r\dn(\mu,k)\sn(\nu,k')\enspace,
    &\qquad     &-2k'\leq\nu\leq2k'\enspace.
\end{array}\qquad\qquad\right\}
\end{equation}
The metric tensor $g_{ab}$ in these coordinates has the form
\begin{equation}
  (g_{ab})=\diag\big[1,r^2(k^2\cn^2\mu+{k'}^2\cn^2\nu),
  r^2(k^2\cn^2\mu+{k'}^2\cn^2\nu)]\enspace,
\end{equation}
the momentum operators are $p_r=-\i\hbar(\partial/\partial r+1/r)$,
and
\begin{equation}
  p_\mu=\hi\bigg({\partial\over\partial\mu}
         -{k^2\sn\mu\cn\mu\dn\mu\over
  k^2\cn^2\mu+{k'}^2\cn^2\nu}\bigg)\enspace,\quad
  p_\nu=\hi\bigg({\partial\over\partial\nu}
         -{{k'}^2\sn\nu\cn\nu\dn\nu\over
  k^2\cn^2\mu+{k'}^2\cn^2\nu}\bigg)\enspace.
\end{equation}
The Hamiltonian has the form
\begin{eqnarray}
  -\hbarm\Delta_\edrei
  &=&-\hbarm\left[{\partial^2\over\partial r^2}
                 +{2\over r}{\partial\over\partial r}
  +{1\over r^2}{1\over k^2\cn^2\mu+{k'}^2\cn^2\nu}\bigg(
     {\partial^2\over\partial\mu^2}+{\partial^2\over\partial\nu^2}\bigg)
   \right]
         \nonumber\\
  &=&{1\over2M}\left(p_r^2+{1\over r^2}
    {1\over\sqrt{k^2\cn^2\mu+{k'}^2\cn^2\nu}}
   (p_\mu^2+p_\nu^2){1\over\sqrt{k^2\cn^2\mu+{k'}^2\cn^2\nu}}\right)
  \enspace.
\end{eqnarray}
Here a separable potential must have the form
\begin{equation}
  V(\vec x)=u(r)+{1\over r^2}{v(\cn\alpha)+w(\cn\beta)\over
         k^2\cn^2\alpha+{k'}^2\cn^2\beta}
  \enspace.
\end{equation}
The algebraic representation of these coordinates has the form
\begin{equation}\left.\begin{array}{lll}
  \displaystyle
   x^2=r^2{(\rho_1-a_1)(\rho_2-a_1)\over(a_3-a_1)(a_2-a_1)}\enspace,
   &\qquad\phantom{\Bigg)}
   &r>0\enspace,
   \\
  \displaystyle
   y^2=r^2{(\rho_1-a_2)(\rho_2-a_2)\over(a_3-a_2)(a_1-a_2)}\enspace,
   &\qquad\phantom{\Bigg)}
   &a_1<\rho_1<a_2<\rho_2<a_3\enspace,
   \\
  \displaystyle
   z^2=r^2{(\rho_1-a_3)(\rho_2-a_3)\over(a_1-a_3)(a_2-a_3)}\enspace.
   &\qquad\phantom{\Bigg)}
   &\qquad
\end{array}\qquad\qquad\right\}
\end{equation}
Here we have $[P(\rho)=(\rho-a_1)(\rho-a_2)(\rho-a_3)]$
\begin{equation}
  (g_{ab})=\diag\bigg(1,{r^2\over4}{\rho_2-\rho_1\over P(\rho_1)},
     {r^2\over4}{\rho_1-\rho_2\over P(\rho_2)}\bigg)\enspace.
\end{equation}
The momentum operators are
\begin{equation}
  p_{\rho_1}=
  \hi\bigg({\partial\over\partial\rho_1}+
    \half{1\over\rho_1-\rho_2}-\viert{P'(\rho_1)\over P(\rho_1)}\bigg)
  \enspace,\qquad
  p_{\rho_2}=
  \hi\bigg({\partial\over\partial\rho_2}+
    \half{1\over\rho_2-\rho_1}-\viert{P'(\rho_2)\over P(\rho_2)}\bigg)
  \enspace.
\end{equation}
Together with the zweibeins
\begin{equation}
h^{\rho_1\rho_1}=\half\sqrt{P(\rho_1)\over\rho_2-\rho_1}\enspace,\qquad
h^{\rho_2\rho_2}=\half\sqrt{P(\rho_2)\over\rho_1-\rho_2}
\end{equation}
we obtain for the Hamiltonian
\begin{eqnarray}
   H&=&-\hbarm\Bigg({\partial^2\over\partial r^2}
     +{2\over r}{\partial\over\partial r}+{1\over r^2}\Delta_{LB}\Bigg)
         \nonumber\\
   &=&{1\over2M}\Bigg[p_r^2+{1\over r^2}\Bigg(
       \sum_{i=1}^2 h^{\rho_i\rho_i}p_{\rho_i}^2h^{\rho_i\rho_i}
      +\Delta V_{PF}(\rho_i)\Bigg)\Bigg]
     -{\hbar^2\over8Mr^2}\enspace,\qquad
                 \\
   \Delta_{LB}&=&{4\over\rho_2-\rho_1}\Bigg[
     \sqrt{P(\rho_1)}{\partial\over\partial\rho_1}
     \sqrt{P(\rho_1)}{\partial\over\partial\rho_1}
     +\sqrt{-P(\rho_2)}{\partial\over\partial\rho_2}
      \sqrt{-P(\rho_2)}{\partial\over\partial\rho_2}\Bigg]\enspace,
\end{eqnarray}
where the $\Delta V_{PF}(\rho_i)$ are determined by (\ref{numba}).
The relation between the coordinates $(\rho_1,\rho_2)$ and
$(\mu,\nu)$ is established via
\begin{equation}
  (a_1-\rho_1)=(a_1-a_2)\sn^2(\mu,k)\enspace,\qquad
  (a_2-\rho_2)=(a_2-a_3)\cn^2(\nu,k')\enspace.
\end{equation}

\noindent
{\it Spherical Coordinates.\/}
We consider the spherical coordinates
\begin{equation}\left.\begin{array}{lll}
   x=r\sin\theta\cos\phi\enspace,
   &\qquad
   &r>0\enspace,
   \\
   y=r\sin\theta\sin\phi\enspace,
   &\qquad
   &0<\theta<\pi\enspace,
   \\
   z=r\cos\theta\enspace,
   &\qquad
   &0\leq\phi<2\pi\enspace.
\end{array}\qquad\qquad\right\}
\end{equation}
These are the usual three-dimensional polar coordinates.
The metric tensor is $(g_{ab})=$ \linebreak
$\diag(1,r^2,r^2\sin^2\theta)$, and the
momentum operators have the form
\begin{equation}
  p_r=\hi\bigg(
    {\partial\over\partial r}+{1\over r}\bigg)\enspace,\qquad
  p_\theta=\hi\bigg(
    {\partial\over\partial\theta}+\half\cot\theta\bigg)\enspace,\qquad
  p_\phi=\hi{\partial\over\partial\phi}\enspace.
\end{equation}
For the Hamiltonian we obtain
\begin{eqnarray}
  -\hbarm\Delta_\edrei
  &=&-\hbarm\bigg[{\partial^2\over\partial r^2}
                 +{2\over r}{\partial\over\partial r}
    +{1\over r^2}\bigg({\partial^2\over\partial\theta^2}
                 +\cot\theta{\partial\over\partial\theta}\bigg)
    +{1\over r^2\sin^2\theta}{\partial^2\over\partial\phi^2}\bigg]
         \nonumber\\
  &=&{1\over2M}\bigg(p_r^2+{1\over r^2}p_\theta^2
          +{1\over r^2\sin^2\theta}p_\phi^2\bigg)
          -{\hbar^2\over8Mr^2}\bigg(1+{1\over\sin^2\theta}\bigg)
  \enspace.
\end{eqnarray}
A potential which is separable in spherical coordinates reads
\begin{equation}
  V(\vec x)=u(r)+{1\over r^2}v(\theta)+{1\over r^2\sin^2\theta}w(\phi)
  \enspace.
\end{equation}

\noindent
{\it Parabolic Coordinates.\/}
We consider the coordinate system
\begin{equation}\left.\begin{array}{lll}
  x=\xi\eta\cos\phi\enspace,\qquad
  y=\xi\eta\sin\phi\enspace,
  &\quad       &\xi,\eta>0\enspace,
  \\
  z=\half(\xi^2-\eta^2)\enspace,
  &\quad       &0\leq\phi<2\pi\enspace.
\end{array}\qquad\qquad\right\}
\end{equation}
This gives for the metric tensor $(g_{ab})=\diag(\xi^2+\eta^2,\xi^2+
\eta^2,\xi^2\eta^2)$, and for the momentum operators we get
\begin{equation}
  p_\xi=\hi\bigg({\partial\over\partial\xi}+{\xi\over\xi^2+\eta^2}
          +{1\over2\xi}\bigg)\enspace,\qquad
  p_\eta=\hi\bigg({\partial\over\partial\eta}+{\eta\over\xi^2+\eta^2}
          +{1\over2\eta}\bigg)\enspace,
\end{equation}
together with $p_\phi=-\i\hbar\partial_\phi$.
For the Hamiltonian we obtain
\begin{eqnarray}
  -\hbarm\Delta_\edrei
  &=&-\hbarm\Bigg[{1\over\xi^2+\eta^2}\bigg(
   {\partial^2\over\partial\xi^2}+{\partial\over\partial\xi}
   +{\partial^2\over\partial\eta^2}+{\partial\over\partial\eta}\bigg)
   +{1\over\xi^2\eta^2}{\partial^2\over\partial\phi^2}\Bigg]
         \nonumber\\
  &=&{1\over2M}\Bigg[{1\over\sqrt{\xi^2+\eta^2}}(p_\xi^2+p_\eta^2)
    {1\over\sqrt{\xi^2+\eta^2}}+{1\over\xi^2\eta^2}p_\phi^2\Bigg]
    -{\hbar^2\over8M\xi^2\eta^2}\enspace.
\end{eqnarray}
In parabolic coordinates a potential is separable if it has the form
\begin{equation}
  V(\vec x)={u(\xi)+v(\eta)\over\sqrt{x^2+y^2+z^2}}
   +{w(\phi)\over x^2+y^2}\enspace.
\end{equation}

\noindent
{\it Prolate Spheroidal Coordinates.\/}
We consider the coordinate system
\begin{equation}\left.\begin{array}{lll}
  x={R\over2}\sqrt{(\xi^2-1)(1-\eta^2)}\cos\phi
   =d\sinh\mu\sin\nu\cos\phi\enspace,
  &\qquad\vphantom{\bigg)}   &\mu>0\enspace,\quad 0<\nu<\pi\enspace,
  \\
  y={R\over2}\sqrt{(\xi^2-1)(1-\eta^2)}\sin\phi
   =d\sinh\mu\sin\nu\sin\phi\enspace,
  &\qquad\vphantom{\bigg)}   &\xi>1\enspace,\quad|\eta|<1\enspace,
  \\
  z={R\over2}\xi\eta
   =d\cosh\mu\cos\nu\enspace,
  &\qquad\vphantom{\bigg)}   &0\leq\phi<2\pi\enspace.
\end{array}\qquad\right\}
\end{equation}
For convenience we have also introduced the alternative
representation of the coordinates in terms of trigonometric and
hyperbolic functions via $\xi=\cosh\mu$, $\eta=\cos\nu$.
$R=2d$ is the interfocus distance.
This yields $(g_{ab})=d^2\diag(\sinh^2\mu+\sin^2\nu, \sinh^2\mu+
\sin^2\nu,\sinh^2\mu\sin^2\nu)$, and for the momentum operators
we obtain
\begin{equation}
  p_\mu=\hi\Bigg({\partial\over\partial\mu}
          +{\sinh\mu\cosh\mu\over\sinh^2\mu+\sin^2\nu}
          +\half\coth\mu\Bigg)\,,\
  p_\nu=\hi\Bigg({\partial\over\partial\nu}
          +{\sin\nu\cos\nu\over\sinh^2\mu+\sin^2\nu}
          +\half\cot\nu\Bigg)\,,
\end{equation}
and $p_\phi=-\i\hbar\partial_\phi$. The Hamiltonian has the form
\begin{eqnarray}
                       & &-\hbarm\Delta_\edrei
         \nonumber\\
                       & &
     =-{\hbar^2\over2Md^2}\Bigg[{1\over\sinh^2\mu+\sin^2\nu}\bigg(
  {\partial^2\over\partial\mu^2}+\coth\mu{\partial\over\partial\mu}+
  {\partial^2\over\partial\nu^2}+\cot\nu{\partial\over\partial\nu}\bigg)
  +{1\over\sinh^2\mu\sin^2\nu}{\partial^2\over\partial^2\phi}\Bigg]
         \nonumber\\
                       & &
     ={1\over2Md^2}\left[{1\over\sqrt{\sinh^2\mu+\sin^2\nu}}
   (p_\mu^2+p_\nu^2){1\over\sqrt{\sinh^2\mu+\sin^2\nu}}
   +{1\over\sinh^2\mu+\sin^2\nu}p_\phi^2\right]
         \nonumber\\   & &\qquad\qquad\qquad\qquad\qquad\qquad
                  \qquad\qquad\qquad\qquad\qquad\qquad
   -{\hbar^2\over8Md^2\sinh^2\mu\sin^2\nu}\enspace.
\end{eqnarray}
A separable potential must have the form
\begin{equation}
  V(\vec x)={u(\cosh\mu)+v(\cos\nu)\over\sinh^2\mu+\sin^2\nu}
   +{w(\phi)\over\sinh^2\mu\sin^2\nu}\enspace.
\end{equation}
Replacing $z\mapsto z=d(\cosh\mu\cos\nu+1)$ gives the {\sl prolate
spheroidal II coordinate system\/} \cite{CORO, KMW, KF, MPSTAa, WIN},
which is one of the four coordinate systems which separate the Coulomb
potential problem, c.f.\ the potential $V_3(\vec x)$ in 3.2.3. Note
that there is no mathematical difference between the prolate spheroidal
and prolate spheroidal II coordinate systems. The later has a shifted
origin such that the left focus is located at the coordinate origin. As
in the elliptic coordinate system in two dimensions the prolate (and
oblate) spheroidal coordinate system is a one parameter coordinate
system. The prolate spheroidal coordinate has the property of
separating the two-center Coulomb problem (Coulson and Robinson
\cite{CORO}, Morse \cite{MORa} and Teller \cite{TEL}).

\medskip\noindent
{\it Oblate Spheroidal Coordinates.\/}
We consider the coordinate system
\begin{equation}\left.\begin{array}{lll}
  x={\bar R\over2}\sqrt{(\bar\xi^2+1)(1-\bar\eta^2)}\cos\phi
   =\bar d\cosh\bar\mu\sin\bar\nu\cos\phi\enspace,
  &\qquad\vphantom{\bigg)}
  &\bar\mu>0\enspace,\quad 0<\bar\nu<\pi\enspace,
  \\
  y={\bar R\over2}\sqrt{(\bar\xi^2+1)(1-\bar\eta^2)}\sin\phi
   =\bar d\cosh\bar\mu\sin\bar\nu\sin\phi\enspace,
  &\qquad\vphantom{\bigg)}
  &\bar\xi>0\enspace,\quad|\bar\eta|<1\enspace,
  \\
  z={\bar R\over2}\bar\xi\bar\eta
  =\bar d\sinh\bar\mu\cos\bar\nu\enspace,
  &\qquad\vphantom{\bigg)}  &0\leq\phi<2\pi
\end{array}\qquad\right\}
\end{equation}
(alternatively $\bar\mu\in\bbbr$, $0<\bar\nu<\pi/2$ \cite{MESCH}). We
again use also the alternative representation $\bar\xi=\sinh\bar\mu$,
$\bar\eta=\cos\bar\nu$. $\bar R=2\bar d$ is the interfocus distance.
This yields $(g_{ab})=\bar d^2\diag(\cosh^2\bar\mu-\sin^2\bar\nu,\cosh^2
\bar \mu-\sin^2\bar\nu,\cosh^2\bar\mu\sin^2\bar\nu)$, and for the
momentum operators we obtain
\begin{equation}
  p_{\bar\mu}=\hi\Bigg({\partial\over\partial\bar\mu}
          +{\sinh\bar\mu\cosh\bar\mu\over\cosh^2\bar\mu-\sin^2\bar\nu}
          +\half\tanh\bar\mu\Bigg)\,,\
  p_{\bar\nu}=\hi\Bigg({\partial\over\partial\bar\nu}
          +{\sin\bar\nu\cos\bar\nu\over\cosh^2\bar\mu-\sin^2\bar\nu}
          +\half\cot\bar\nu\Bigg)\,,
\end{equation}
and $p_\phi=-\i\hbar\partial_\phi$. The Hamiltonian has the form
\begin{eqnarray}
                       & &-\hbarm\Delta_\edrei
         \nonumber\\
                       & &=-{\hbar^2\over2M\bar d^2}
  \Bigg[{1\over\cosh^2\bar\mu-\sin^2\bar\nu}\bigg(
  {\partial^2\over\partial\bar\mu^2}
  +\tanh\bar\mu{\partial\over\partial\bar\mu}+
  {\partial^2\over\partial\bar\nu^2}
  +\cot\bar\nu{\partial\over\partial\bar\nu}\bigg)
  +{1\over\cosh^2\bar\mu\sin^2\bar\nu}
  {\partial^2\over\partial^2\phi}\Bigg]
         \nonumber\\
                       & &={1\over2M\bar d^2}
   \left[{1\over\sqrt{\cosh^2\bar\mu-\sin^2\bar\nu}}
   (p_{\bar\mu}^2+p_{\bar\nu}^2)
     {1\over\sqrt{\cosh^2\bar\mu-\sin^2\bar\nu}}
   +{1\over\cosh^2\bar\mu-\sin^2\bar\nu}p_\phi^2\right]
         \nonumber\\   & &\qquad\qquad\qquad\qquad\qquad\qquad
                  \qquad\qquad\qquad\qquad\qquad\qquad
   -{\hbar^2\over8M\bar d^2\cosh^2\bar\mu\sin^2\bar\nu}\enspace.
\end{eqnarray}
Here a separable potential must have the form
\begin{equation}
  V(\vec x)={u(\sinh\bar\mu)
             +v(\cos\bar\nu)\over\cosh^2\bar\mu-\sin^2\bar\nu}
   +{w(\phi)\over\cosh^2\bar\mu\sin^2\bar\nu}\enspace.
\end{equation}

\noindent
{\it Paraboloidal Coordinates.\/}
The last two coordinate system are the most complicated ones and are
similar in some of their features, c.f.~(\ref{numbc}, \ref{numbd}) and
\cite{GROab, MF, OLE} for further information. First we consider the
coordinate system
\begin{equation}\left.\begin{array}{l}
  \displaystyle
  x^2={(\eta_1-a)(\eta_2-a)(\eta_3-a)\over a-b}\enspace,
  \quad\vphantom{\Bigg)}
  0<\eta_1<b<\eta_2<a<\eta_3\enspace,
  \\
  \displaystyle
  y^2={(\eta_1-b)(\eta_2-b)(\eta_3-b)\over b-a}\enspace,
  \vphantom{\Bigg)}
  \\
  \displaystyle
  z=\half\big(\eta_1+\eta_2+\eta_3-a-b\big)\enspace,
\end{array}\qquad\right\}
\label{numel}
\end{equation}
The metric tensor is given by $[P(\eta)=(\eta-a)(\eta-b)]$
\begin{equation}
  (g_{ab})=\viert\diag\bigg(
   {(\eta_1-\eta_2)(\eta_1-\eta_3)\over P(\eta_1)},
   {(\eta_2-\eta_1)(\eta_2-\eta_3)\over P(\eta_2)},
   {(\eta_3-\eta_1)(\eta_3-\eta_2)\over P(\eta_3)}\bigg)\enspace.
\end{equation}
The momentum operators are
\begin{equation}
\left.\begin{array}{l}
  \displaystyle\vphantom{\Bigg)}
  p_{\eta_1}=\hi\bigg({\partial\over\partial\eta_1}+
    \half{1\over\eta_1-\eta_2}+\half{1\over\eta_1-\eta_3}
  -\viert{P'(\eta_1)\over P(\eta_1)}\bigg)\enspace,\\
  \displaystyle\vphantom{\Bigg)}
  p_{\eta_2}=\hi\bigg({\partial\over\partial\eta_2}+
    \half{1\over\eta_2-\eta_1}+\half{1\over\eta_2-\eta_3}
  -\viert{P'(\eta_2)\over P(\eta_2)}\bigg)\enspace,\\
  \displaystyle\vphantom{\Bigg)}
  p_{\eta_3}=\hi\bigg({\partial\over\partial\eta_3}+
    \half{1\over\eta_3-\eta_1}+\half{1\over\eta_3-\eta_2}
  -\viert{P'(\eta_3)\over P(\eta_3)}\bigg)\enspace.
\end{array}\qquad\right\}
\end{equation}
Together with the dreibeins
\begin{equation}
\left.\begin{array}{l}
  \displaystyle\vphantom{\Bigg)}
h^{\eta_1\eta_1}=\half
   \sqrt{P(\eta_1)\over(\eta_1-\eta_2)(\eta_1-\eta_3} \enspace,\\
  \displaystyle\vphantom{\Bigg)}
h^{\eta_2\eta_2}=\half
   \sqrt{P(\eta_2)\over(\eta_2-\eta_1)(\eta_2-\eta_3)}\enspace,\\
  \displaystyle\vphantom{\Bigg)}
h^{\eta_3\eta_3}=\half
   \sqrt{P(\eta_3)\over(\eta_3-\eta_1)(\eta_3-\eta_2)}\enspace,
\end{array}\qquad\right\}
\end{equation}
we obtain for the Hamiltonian
\begin{eqnarray}
  -\hbarm\Delta_\edrei &=&
  -{2\hbar^2\over M}
  \Bigg[{\sqrt{P(\eta_1)}\over(\eta_1-\eta_2)(\eta_1-\eta_3)}
  {\partial\over\partial\eta_1}\sqrt{P(\eta_1)}
  {\partial\over\partial\eta_1}
  +{\sqrt{P(\eta_2)}\over(\eta_2-\eta_3)(\eta_2-\eta_1)}
  {\partial\over\partial\eta_2}\sqrt{P(\eta_2)}
  {\partial\over\partial\eta_2}
         \nonumber\\   & &\qquad\qquad\qquad\qquad\qquad\qquad
  +{\sqrt{P(\eta_3)}\over(\eta_3-\eta_1)(\eta_3-\eta_2)}
  {\partial\over\partial\eta_3}\sqrt{P(\eta_3)}
  {\partial\over\partial\eta_3}\Bigg]\enspace,
                  \\   &=&
  {1\over2M}\sum_{i=1}^3
            \Bigg(h^{\eta_i\eta_i}p_{\eta_i}^2h^{\eta_i\eta_i}
            +\Delta V(\eta_i)\Bigg)\enspace.
\end{eqnarray}
The $\Delta V_{PF}(\eta_i)$ are determined by (\ref{numba}).
A potential separable in paraboloidal coordinates must have the form
\begin{equation}
  V(\vec x)={(\eta_2-\eta_3)u(\eta_1)+(\eta_1-\eta_3)v(\eta_2)
    +(\eta_1-\eta_2)w(\eta_3)\over
     (\eta_1-\eta_2)(\eta_1-\eta_3)(\eta_2-\eta_3)}\enspace.
\label{numek}
\end{equation}

\noindent
{\it Ellipsoidal Coordinates.\/}
As the last coordinate system we consider
\begin{equation}\left.\begin{array}{l}
  \displaystyle
  x^2={(\rho_1-a_3)(\rho_2-a_3)(\rho_3-a_3)\over
           (a_3-a_1)(a_2-a_1)}\enspace,\quad
  0<a_1<\rho_1<a_2<\rho_2<a_3<\rho_3\enspace,
  \vphantom{\bigg)}\\   \displaystyle
  y^2={(\rho_1-a_2)(\rho_2-a_2)(\rho_3-a_2)\over
            (a_3-a_2)(a_1-a_2)}\enspace,
  \vphantom{\bigg)}\\   \displaystyle
  z^2={(\rho_1-a_1)(\rho_2-a_1)(\rho_3-a_1)\over
            (a_1-a_3)(a_2-a_3)}\enspace.
\end{array}\qquad\right\}
\end{equation}
For the metric tensor we have $[P(\rho)=(\rho-a_1)(\rho-a_2)(\rho-a_3)]$
\begin{equation}
  (g_{ab})=\viert\diag\bigg(
  {(\rho_1-\rho_2)(\rho_1-\rho_3)\over P(\rho_1)},
  {(\rho_2-\rho_3)(\rho_2-\rho_1)\over P(\rho_2)},
  {(\rho_3-\rho_1)(\rho_3-\rho_2)\over P(\rho_3)}\bigg)\enspace.
\end{equation}
The momentum operators are
\begin{equation}
\left.\begin{array}{l}
  \displaystyle\vphantom{\Bigg)}
  p_{\rho_1}=\hi\bigg({\partial\over\partial\rho_1}+
    \half{1\over\rho_1-\rho_2}+\half{1\over\rho_1-\rho_3}
  -\viert{P'(\rho_1)\over P(\rho_1)}\bigg)\enspace,\\
  \displaystyle\vphantom{\Bigg)}
  p_{\rho_2}=\hi\bigg({\partial\over\partial\rho_2}+
    \half{1\over\rho_2-\rho_1}+\half{1\over\rho_2-\rho_3}
  -\viert{P'(\rho_2)\over P(\rho_2)}\bigg)\enspace,\\
  \displaystyle\vphantom{\Bigg)}
  p_{\rho_3}=\hi\bigg({\partial\over\partial\rho_3}+
    \half{1\over\rho_3-\rho_1}+\half{1\over\rho_3-\rho_2}
  -\viert{P'(\rho_3)\over P(\rho_3)}\bigg)\enspace.
\end{array}\qquad\right\}
\end{equation}
Together with the dreibeins
\begin{equation}
\left.\begin{array}{l}
  \displaystyle\vphantom{\Bigg)}
h^{\rho_1\rho_1}=\half
   \sqrt{P(\rho_1)\over(\rho_1-\rho_2)(\rho_1-\rho_3} \enspace,\\
  \displaystyle\vphantom{\Bigg)}
h^{\rho_2\rho_2}=\half
   \sqrt{P(\rho_2)\over(\rho_2-\rho_1)(\rho_2-\rho_3)}\enspace,\\
  \displaystyle\vphantom{\Bigg)}
h^{\rho_3\rho_3}=\half
   \sqrt{P(\rho_3)\over(\rho_3-\rho_1)(\rho_3-\rho_2)}\enspace,
\end{array}\qquad\right\}
\end{equation}
we obtain for the Hamiltonian
\begin{eqnarray}
  -\hbarm\Delta_\edrei &=&
  -{2\hbar^2\over M}
  \Bigg[{\sqrt{P(\rho_1)}\over(\rho_1-\rho_2)(\rho_1-\rho_3)}
  {\partial\over\partial\rho_1}\sqrt{P(\rho_1)}
  {\partial\over\partial\rho_1}
  +{\sqrt{P(\rho_2)}\over(\rho_2-\rho_3)(\rho_2-\rho_1)}
  {\partial\over\partial\rho_2}\sqrt{P(\rho_2)}
  {\partial\over\partial\rho_2}
         \nonumber\\   & &\qquad\qquad\qquad\qquad\qquad\qquad
  +{\sqrt{P(\rho_3)}\over(\rho_3-\rho_1)(\rho_3-\rho_2)}
  {\partial\over\partial\rho_3}\sqrt{P(\rho_3)}
  {\partial\over\partial\rho_3}\Bigg]\enspace,
                  \\   &=&
  {1\over2M}\sum_{i=1}^3
            \Bigg(h^{\rho_i\rho_i}p_{\rho_i}^2h^{\rho_i\rho_i}
            +\Delta V(\rho_i)\Bigg)\enspace.
\end{eqnarray}
The $\Delta V_{PF}(\rho_i)$ are determined by (\ref{numba}).
A potential separable in ellipsoidal coordinates has the form of
(\ref{numek}) with $\eta_i$ replaced by $\rho_i$.

The ellipsoidal coordinate system is a two parameter coordinate
system. Denoting $R_1^2=a_2-a_1$, $R_2^2=a_3-a_2$ and $R_3^2=a_3-a_1$,
the $R_i$ must fulfill the constraint $R_1^2+R_2^2=R_3^2$. The $R_i$
describe the foci distance from the coordinate origin. By considering
the limits $R_{1,2}\to0$, respectively $R_{1,2}\to\infty$ we can
generate the seven coordinate systems described in table 1
(note $\bar\rho_i=\rho_i-a_2$).

\medskip
\centerline{{\bf Table 1:} The degenerations of the ellipsoidal
             coordinate system\hfill}
\begin{eqnarray}
\begin{array}{l}
  \vbox{\small\offinterlineskip
\halign{&\vrule#&$\strut\ \hfil\hbox{#}\hfill\ $\cr
\noalign{\hrule}
height2pt&\omit&&\omit&&\omit&&\omit&\cr
&prolate
      &&$x=(R/2)\sqrt{(\xi^2-1)(1-\eta^2)}\,\cos\phi$
      &&$R_2^2\to0$
      &&$\bar\rho_1\to R_1^2(\eta^2-1)$              &\cr
&spheroidal
      &&$y=(R/2)\sqrt{(\xi^2-1)(1-\eta^2)}\,\sin\phi$
      &&$R_1^2\to R^2/4$
      &&$\bar\rho_2\to R_2^2\sin^2\phi$              &\cr
&system
      &&$z=(R/2)\xi\eta$
      &&
      &&$\bar\rho_3\to R_1^2(\xi^2-1)$               &\cr
height2pt&\omit&&\omit&&\omit&&\omit&\cr
\noalign{\hrule}
height2pt&\omit&&\omit&&\omit&&\omit&\cr
&oblate     &&$x=(\bar R/2)\sqrt{(\bar\xi^2-1)(1-\bar\eta^2)}\,\cos\phi$
      &&$R_2^2\to0$
      &&$\bar\rho_1\to -R_1^2\sin^2\phi$             &\cr
&spheroidal &&$y=(\bar R/2)\sqrt{(\bar\xi^2-1)(1-\bar\eta^2)}\,\sin\phi$
      &&$R_1^2\to\bar R^2/4$
      &&$\bar\rho_2\to(\bar R^2/4)(1-\bar\eta^2)$     &\cr
&system     &&$z=(\bar R/2)\bar\xi\bar\eta$
      &&
      &&$\bar\rho_3\to(\bar R^2/4)(1+\bar\eta^2)$     &\cr
height2pt&\omit&&\omit&&\omit&&\omit&\cr
\noalign{\hrule}
height2pt&\omit&&\omit&&\omit&&\omit&\cr
&Spherical  &&$x=r\sin\theta\cos\phi$
      &&$R_1^2\to0$
      &&$\bar\rho_1\to -R_1^2\sin^2\theta$            &\cr
&system     &&$y=r\sin\theta\sin\phi$
      &&$R_2^2\to0$
      &&$\bar\rho_2\to R_2^2\sin^2\theta\sin^2\phi$   &\cr
&           &&$z=r\cos\theta$
      &&$R_2^2/R_1^2\to0$
      &&$\bar\rho_3\to r^2$                           &\cr
height2pt&\omit&&\omit&&\omit&&\omit&\cr
\noalign{\hrule}
height2pt&\omit&&\omit&&\omit&&\omit&\cr
&Sphero-conical
      &&$x=r\sn(\mu,k)\dn(\nu,k')$
      &&$R_1^2\to0$
      &&$\bar\rho_1\to -{k'}^2R_3^2\cn^2(\nu,k')$     &\cr
&system     &&$y=r\cn(\mu,k)\cn(\nu,k')$
      &&$R_2^2\to0$
      &&$\bar\rho_2\to k^2R_3^2\cn^2(\mu,k)$          &\cr
&           &&$z=r\dn(\mu,k)\sn(\nu,k')$
      &&$R_2^2/R_1^2\to k^2$
      &&$\bar\rho_3\to r^2$                           &\cr
height2pt&\omit&&\omit&&\omit&&\omit&\cr
\noalign{\hrule}
height2pt&\omit&&\omit&&\omit&&\omit&\cr
&Circular   &&$x=(R/2)\sinh\mu\sin\nu$
      &&$R_1^2\to\infty$
      &&$\bar\rho_1\to {z'}^2-R_1^2$                  &\cr
&elliptic   &&$y=(R/2)\cosh\mu\cos\nu$
      &&$R_2^2=R^2/4$
      &&$\bar\rho_2\to (R^2/4)\cos^2\nu$              &\cr
&system     &&$z=z'$
      &&    &&$\bar\rho_3\to (R^2/4)\cosh^2\mu$       &\cr
height2pt&\omit&&\omit&&\omit&&\omit&\cr
\noalign{\hrule}
height2pt&\omit&&\omit&&\omit&&\omit&\cr
&Circular   &&$x=\rho\cos\phi$
      &&$R_1^2\to\infty$
      &&$\bar\rho_1\to {z'}^2-R_1^2$                  &\cr
&polar      &&$y=\rho\sin\phi$
      &&$R_2^2\to0$
      &&$\bar\rho_2\to R^2\sin^2\phi$                 &\cr
&system     &&$z=z'$
      &&    &&$\bar\rho_3\to \rho^2$                  &\cr
height2pt&\omit&&\omit&&\omit&&\omit&\cr
\noalign{\hrule}
height2pt&\omit&&\omit&&\omit&&\omit&\cr
&Cartesian  &&$x=x'$
      &&$R_1^2\to\infty$
      &&$\bar\rho_1\to {z'}^2-R_1^2$                  &\cr
&system     &&$y=y'$
      &&$R_2^2\to\infty$
      &&$\bar\rho_2\to {x'}^2+R_1^2$                  &\cr
&     &&$z=z'$
      &&    &&$\bar\rho_3\to {y'}^2$                  &\cr
height2pt&\omit&&\omit&&\omit&&\omit&\cr
\noalign{\hrule}
}}\end{array}\nonumber\end{eqnarray}
We must obtain the remaining three coordinate systems which are not
listed in table 1. The paraboloidal system is obtained by the focus
translation $z'\to z-R_1$ from the ellipsoidal coordinate system by
$R_{1,2}^2\to\infty$ together with $R_2^2/R_1\to a-b$ which gives three
more coordinate systems: The parabolic system then follows from the
paraboloidal system in the limit $a\to b$ (the parabolic system can
also be obtained from the prolate spheroidal II system). The circular
parabolic system is obtained from the degeneration of the elliptic II
system, c.f.\ 2.1.2.

\vglue0.6truecm\noindent
{\bf 2.2.~Separation of Coordinates in the Path Integral.}
\vglue0.4truecm\noindent
We want to look for the coordinate systems which separate the relevant
partial differential equations, i.e.\ the Hamiltonian, and, more
important from our point of view, the path integral. In order to develop
such a theory we consider according to \cite{MF} the Lagrangian
$\CL={M\over2}\sum_{i=1}^Dh_i^2\dot x_i^2$ and the Laplacian
$\Delta_L$, respectively, in the following way (where only
orthogonal coordinate systems are taken into account)
\begin{eqnarray}
   \Delta_{LB}
   &=&\sum_{i=1}^D{1\over\prod_{j=1}^Dh_j(\{\xi\})}
   {\partial\over\partial\xi_i}\left(
   {\prod_{k=1}^Dh_k(\{\xi\})\over h_i^2(\{\xi\})}
     {\partial\over\partial \xi_i}\right)
         \nonumber\\
  &=:&\sum_{i=1}^D{1\over\prod_{j=1}^Dh_j(\{\xi\})}
   {\partial\over\partial\xi_i}\left(
   g_i(\xi_1,\dots,\xi_{i-1},\xi_{i+1},\dots,\xi_D)f(\xi_i)
   {\partial\over\partial\xi_i}\right)\enspace,
\end{eqnarray}
where $\{\xi\}$ denotes the set of variables $(\xi_1,\dots,\xi_D)$, and
the existence of the functions $f_i, g_i$ is necessary for the
separation \cite{KAL, MF}. We introduce the St\"ackel-determinant
\cite{KAL, MOON, MF}
\begin{equation}
  S(\{\xi\})=\det(\Phi_{ij})=\prod_{i=1}^D{h_i(\{\xi\})\over f_i(\xi_i)}
  \ ,\
  M_i(\xi_1,\dots,\xi_{i-1},\xi_{i+1},\dots,\xi_D)
  ={\partial S\over\partial\Phi_{i1}}
  ={S(\{\xi\})\over h_i^2(\{\xi\})}\enspace,
\end{equation}
and abbreviate  $\Gamma_i=f_i'/f_i$. Then
\begin{equation}
  g_i(\xi_1,\dots,\xi_{i-1},\xi_{i+1},\dots,\xi_D)
  =M_i(\xi_1,\dots,\xi_{i-1},\xi_{i+1},\dots,\xi_D)
  \prod_{\scriptstyle j=1 \atop\scriptstyle i\not=j}^D
  f_j(\xi_j)\enspace,
\end{equation}
which fixes the functions $g_i$. Introducing the (new) momentum
operators $P_{\xi_i}=\hi(\partial\xi_i+\half f_i'/f_i)$ we obtain
according to the general theory the following identity in the path
integral \cite{GROab} by means of the space-time transformation
technique (Duru \cite{DURb}, Fischer et al.~\cite{FLMa}, Refs.\
\cite{GRSb, GRSg}, Kleinert \cite{KLEm}, and Pak and S\"okmen \cite{PS})
\begin{eqnarray}       & &
     \prod_{i=1}^D\int\limits_{\xi_i(t')=\xi_i'}^{\xi_i(t'')=\xi_i''}
  h_i\CD\xi_i(t)\exp\left\{\ih\int_{t'}^{t''}\left[{M\over2}\sum_{i=1}^D
      h_i^2\dot\xi_i^2-\Delta V_{PF}(\{\xi\})\right]dt\right\}
         \nonumber\\   & &
  =\prod_{i=1}^D\int\limits_{\xi_i(t')=\xi_i'}^{\xi_i(t'')=\xi_i''}
   \sqrt{S\over M_i}\,\CD\xi_i(t)
  \exp\left\{\ih\int_{t'}^{t''}\left[{M\over2}S\sum_{i=1}^D
   {\dot\xi_i^2\over M_i}-\Delta V_{PF}(\{\xi\})\right]dt\right\}
         \nonumber\\   & &
  =(S'S'')^{\half(1-D/2)}\int_{\bbbr}{dE\over2\pi\hbar}
  \e^{-\i ET/\hbar}\int_0^\infty ds''
   \prod_{i=1}^D\int\limits_{\xi_i(0)=\xi_i'}^{\xi_i(s'')=\xi_i''}
   M_i^{-1/2}\CD\xi_i(s)
         \nonumber\\   & &\qquad\qquad\times
  \exp\left\{\ih\int_0^{s''}\left[{M\over2}\sum_{i=1}^D
   {\dot\xi_i^2\over M_i}+ES
    -{\hbar^2\over8M}\sum_{i=1}^DM_i\Big(\Gamma_i^2+2\Gamma_i'\Big)
    \right]ds\right\}\enspace.
\label{numbc}
\end{eqnarray}
Here we have adopted the following {\em lattice formulation} of the
path integral (c.f.\ DeWitt \cite{DEW}, Feynman \cite{FEY}, Gervais
and Jevicki \cite{GJ}, Refs.~\cite{GRSb, GRSg}, McLaughlin and
Schulman \cite{MCLS}, Mizrahi \cite{MIZa}, and Omote \cite{OMO} for
alternative lattice definitions)
\begin{eqnarray}
  K({\vec q\,}'',{\vec q\,}';T)
  &=&\int\limits_{\vec q(t')={\vec q\,}'}
                ^{\vec q(t'')={\vec q\,}''}\sqrt{g}\,\CD_{PF}\vec q\,(t)
  \exp\Bigg\{\ih\int_{t'}^{t''}
    \Bigg[{M\over2}h_{ac}({\vec q}\,)h_{cb}({\vec q}\,)\dot q^a\dot q^b
   -V({\vec q}\,)-\Delta V_{PF}({\vec q}\,)\Bigg]dt\Bigg\}
         \nonumber\\
  &\equiv&\lim_{N\to\infty}
  \bigg({M\over2\pi\i\epsilon\hbar}\bigg)^{ND/2}\prod_{j=1}^{N-1}
  \int d\vec q_j\sqrt{g(\vec q_j)}
         \nonumber\\   & &\qquad\times
  \exp\Bigg\{\ih\sum_{j=1}^N\Bigg[{M\over2\epsilon}
  h_{bc}(\vec q_j)h_{ac}(\vec q_{j-1})\Delta q_j^a\Delta q_j^b
  -\epsilon V(\vec q_j)-\epsilon\Delta V_{PF}(\vec q_j)\Bigg]\Bigg\}
  \enspace.
\label{numbd}
\end{eqnarray}
This path integral formulation will be used in the sequel.


\newpage\noindent
{\large\bf 3.~Path Integral Formulation of Smorodinsky-Winternitz
               Potentials.}
\vglue0.4truecm\noindent
{\bf 3.1.~Two-Dimensional Smorodinsky-Winternitz Potentials.}
\vglue0.4truecm\noindent
In this subsection we discuss the four two-dimensional
Smorodinsky-Winternitz potentials \cite{FMSUW}. They are characterized
by having three functionally independent integrals of motion, i.e.\
there are two more operators related to these integrals of motion which
commute with the Hamiltonian.

We will state if ever possible both the propagator and its spectral
expansion into the wave-functions and energy-levels. $\vec x$ denotes a
two-dimensional variable: $\vec x=(x,y)\equiv(x_1,x_2)$, and without
loss of generality it will be assumed that $\vec x\in\bbbr^2, \bbbr^3$.

In the table 2 we list the two-dimensional maximally super-integrable
Smorodinsky-Winternitz potentials together with the separating
coordinate systems \cite{FMSUW}. The cases where an explicit path
integration is possible are $\underline{\hbox{underlined}}$.

\hfuzz=25.0pt
\medskip
\centerline{{\bf Table 2:} The two-dimensional maximally
            super-integrable potentials\hfill}
\begin{eqnarray}
\begin{array}{l}
\vbox{\offinterlineskip
\hrule
\halign{&\vrule#&\strut\quad\hfil#\quad\hfill\quad\cr
height2pt&\omit&&\omit&\cr
&Potential $V(x,y)$
  &&Coordinate                                              &\cr
& &&\ System                                                &\cr
height2pt&\omit&&\omit&&\omit&\cr
\noalign{\hrule}
\noalign{\hrule}
height2pt&\omit&&\omit&&\omit&\cr
&$\displaystyle
  V_1={M\over2}\omega^2(x^2+y^2)+\hbarm
    \Bigg({k_1^2-\viert\over x^2}+{k_2^2-\viert\over y^2}\Bigg)$
  &&$\underline{\hbox{Cartesian}}$              &\cr
& &&$\underline{\hbox{Polar}}$                  &\cr
& &&Elliptic                                    &\cr
height2pt&\omit&&\omit&\cr
\noalign{\hrule}
height2pt&\omit&&\omit&\cr
&$\displaystyle
  V_2={M\over2}\omega^2(4x^2+y^2)+k_1x
     +\hbarm{k_2^2-\viert\over y^2}$
  &&$\underline{\hbox{Cartesian}}$              &\cr
& &&Parabolic                                   &\cr
height2pt&\omit&&\omit&\cr
\noalign{\hrule}
height2pt&\omit&&\omit&\cr
&$\displaystyle
  V_3=-{\alpha\over\sqrt{x^2+y^2}}+{\hbar^2\over4M}
    {1\over\sqrt{x^2+y^2}}\Bigg({k_1^2-\viert\over\sqrt{x^2+y^2}+x}
         +{k_2^2-\viert\over\sqrt{x^2+y^2}-x}\Bigg)$
  &&$\underline{\hbox{Polar}}$                  &\cr
& &&Elliptic II                                 &\cr
& &&$\underline{\hbox{Parabolic}}$              &\cr
height2pt&\omit&&\omit&\cr
\noalign{\hrule}
height2pt&\omit&&\omit&\cr
&$\displaystyle
  V_4=-{\alpha\over\rho}+\sqrt{2\over\rho}
   \bigg(\beta_1\cos{\phi\over2}+\beta_2\sin{\phi\over2}\bigg)$
  &&$\underline{\hbox{Mutually}}$               &\cr
& &&$\underline{\hbox{\ Parabolic}}$            &\cr
height2pt&\omit&&\omit&\cr}\hrule}
\end{array}\nonumber
\end{eqnarray}
\hfuzz=7.0pt

\vglue0.4truecm\noindent
{\it 3.1.1.}~We consider the potential ($k_{1,2}>0$)
\begin{equation}
  V_1(\vec x)={M\over2}\omega^2(x_1^2+x_2^2)+\hbarm
    \Bigg({k_1^2-\viert\over x_1^2}+{k_2^2-\viert\over x_2^2}\Bigg)
  \enspace.
\end{equation}
This potential is separable in three coordinate systems. We obtain the
path integral formulations, c.f.~(\ref{numbd}) for their lattice
definitions
\begin{eqnarray}       & &
  K^{(V_1)}({\vec x\,}'',{\vec x\,}';T)
         \nonumber\\   & &
  \underline{\hbox{Cartesian Coordinates:}}
         \nonumber\\   & &
  =\int\limits_{\vec x(t')=\vec x'}^{\vec x(t'')=\vec x''}\CD\vec x(t)
  \exp\left\{\ih\int_{t'}^{t''}\left[{M\over2}
     \big({\dot{\vec x}}^2-\omega^2\vec x^2\big)
     -\hbarm\Bigg({k_1^2-\viert\over x_1^2}
                   +{k_2^2-\viert\over x_2^2}\Bigg)\right]dt\right\}
                  \\   & &
  \phantom{\underline{\hbox{Polar Coordinates:}}}
         \nonumber\\   & &
  \underline{\hbox{Polar Coordinates:}}
         \nonumber\\   & &
  =\int\limits_{\rho(t')=\rho'}^{\rho(t'')=\rho''}\rho\CD\rho(t)
   \int\limits_{\phi(t')=\phi'}^{\phi(t'')=\phi''}\CD\phi(t)
         \nonumber\\   & &\qquad\times
  \exp\left\{\ih\int_{t'}^{t''}\left[{M\over2}
     \big(\dot \rho^2+\rho^2\dot\phi^2-\omega^2\rho^2\big)
     -{\hbar^2\over2M\rho^2}\Bigg({k_1^2-\viert\over\cos^2\phi}
         +{k_2^2-\viert\over\sin^2\phi}-\viert\Bigg)\right]dt\right\}
\label{numcc}     \\   & &
  \underline{\hbox{Elliptic Coordinates:}}
         \nonumber\\   & &
  =\int\limits_{\xi(t')=\xi'}^{\xi(t'')=\xi''}\CD\xi(t)
   \int\limits_{\eta(t')=\eta'}^{\eta(t'')=\eta''}\CD\eta(t)
   d^2(\sinh^2\xi+\sin^2\eta)
         \nonumber\\   & &\qquad\times
  \exp\Bigg\{\ih\int_{t'}^{t''}\Bigg[{M\over2}d^2
     \Big((\sinh^2\xi+\sin^2\eta)(\dot\xi^2+\dot\eta^2)
         -\omega^2(\cosh^2\xi\cos^2\eta+\sinh^2\xi\sin^2\eta)\Big)
         \nonumber\\   & &\qquad\qquad\qquad\qquad\qquad\qquad
     -{\hbar^2\over2Md^2}\Bigg({k_1^2-\viert\over\cosh^2\xi\cos^2\eta}
  +{k_2^2-\viert\over\sinh^2\xi\sin^2\eta}\Bigg)\Bigg]dt\Bigg\}
  \qquad          \\   & &
  =\int_{\bbbr}{dE\over2\pi\hbar}\e^{-\i ET/\hbar}\int_0^\infty ds''
  \int\limits_{\xi(0)=\xi'}^{\xi(s'')=\xi''}\CD\xi(s)
  \int\limits_{\eta(0)=\eta'}^{\eta(s'')=\eta''}\CD\eta(s)
         \nonumber\\   & &\qquad\times
  \exp\Bigg\{\ih\int_0^{s''}\Bigg[{M\over2}\Big((\dot\xi^2+\dot\eta^2)
         -\omega^2(\cosh^2\xi\sinh^2\xi+\sin^2\eta\cos^2\eta)\Big)
         +Ed^2(\sinh^2\xi+\sin^2\eta)
         \nonumber\\   & &\qquad\qquad
     -\hbarm\Bigg(
   (k_1^2-\bviert)\Bigg({1\over\cos^2\eta}-{1\over\cosh^2\xi}\Bigg)
  +(k_2^2-\bviert)\bigg({1\over\sin^2\eta}+{1\over\sinh^2\xi}\bigg)
                  \Bigg)\Bigg]ds\Bigg\}\enspace.
\end{eqnarray}
Note that in the last line a pure time-transformation \cite{KLEh} has
been performed.

We are going to discuss the path integral (\ref{numcc}) in some detail
in order to demonstrate the relevant technique. All subsequent path
integral evaluations are similarly done and our method remains the same
whether polar, parabolic, or other coordinate systems are used.

The path integral solution in cartesian coordinates is straightforward
by applying the path integral solution for the radial harmonic
oscillator for the coordinates $x_{1,2}$, respectively, c.f.\
(\ref{numcl}, \ref{numcg}). In the case of polar coordinates one first
separates the $\phi$-path integration by means of the path integral
identity of the P\"oschl-Teller potential ($0<\phi<\pi/2$) \cite{BJb,
DURb, INOWI}
\begin{eqnarray}       & &
  \int\limits_{\phi(t')=\phi'}^{\phi(t'')=\phi''}\CD\phi(t)
  \exp\left\{\ih\int_{t'}^{t''}\left[{M\over2}\dot\phi^2
        -\hbarm\bigg(
  {\alpha^2-\viert\over\sin^2\phi}+{\beta^2-\viert\over\cos^2\phi}\bigg)
  \right]dt\right\}
         \nonumber\\   & &\qquad
  =\sum_{n=0}^\infty\e^{-\i E_nT/\hbar}
  \Phi_n^{(\alpha,\beta)}(\phi')\Phi_n^{(\alpha,\beta)}(\phi'')
  \enspace,
\end{eqnarray}
and $\Phi^{(\alpha,\beta)}_n(\phi)$ denote the normalized
P\"oschl-Teller wave-functions
\begin{eqnarray}
  \Phi_n^{(\alpha,\beta)}(\phi)&=&\bigg[2(\alpha+\beta+2l+1)
  {l!\Gamma(\alpha+\beta+l+1)\over\Gamma(\alpha+l+1)\Gamma(\beta+l+1)}
  \bigg]^{1/2}
         \nonumber\\   & &\qquad\qquad\times
  (\sin\phi)^{\alpha+1/2}(\cos\phi)^{\beta+1/2}
  P_n^{(\alpha,\beta)}(\cos2\phi)\enspace.
\label{numca}
\end{eqnarray}
The $P_n^{(\alpha,\beta)}(x)$ are Jacobi polynomials (\cite{GRA},
p.1035). Note that the genuine P\"oschl-Teller potentials requires
$\alpha,\beta>\half$. Otherwise one can call it an attractive
P\"oschl-Teller-like potential. This rule of separation of variables in
the path integral consists of the  following path integral identity
\cite{GROj} ($\hat g=\prod g_i^2$, $\{g_i\}_i\equiv\vec g\equiv \vec
g(\vec z)$)
\begin{eqnarray}       & &
     \int\limits_{\vec z(t')=\vec z'}^{z(t'')=\vec z''}
  f^d(\vec z\,)\sqrt{\widehat{g(\vec z\,)}}\,\CD\vec z(t)
  \int\limits_{\vec x(t')=\vec x'}^{\vec x(t'')=\vec x''}\CD\vec x(t)
         \nonumber\\   & &\qquad\qquad\times
  \exp\Bigg\{\ih\int_{t'}^{t''}\Bigg[{M\over2}\Big(
  (\vec g\cdot\dot{\vec z}\,)^2+f^2(\vec z\,)\dot{\vec x}^2\Big)
  -\Bigg({V(\vec x\,)\over f^2(\vec z\,)}+W(\vec z\,)+\Delta W(\vec z\,)
  \Bigg)\Bigg]dt\Bigg\}
         \nonumber\\   & &
  =[f({\vec z\,}')f({\vec z\,}'')]^{-d/2}\int dE_\lambda
  \Phi_\lambda^*({\vec x\,}')\Phi_\lambda({\vec x\,}'')
   \int\limits_{\vec z(t')=\vec z''}^{\vec z(t'')=\vec z''}
  \sqrt{\widehat{g(\vec z\,)}}\,\CD\vec z(t)
         \nonumber\\   & &\qquad\qquad\times
  \exp\left\{\ih\int_{t'}^{t''}\Bigg[{M\over2}
  \big(\vec g\cdot\dot{\vec z}\,\big)^2
   -W(\vec z\,)-\Delta W(\vec z\,)-{E_\lambda\over f^2(\vec z\,)}
  \Bigg]dt\right\}\enspace,
\end{eqnarray}
where the $x$-path integration is given by
\begin{equation}
  \int\limits_{\vec x(t')=\vec x'}^{\vec x(t'')=\vec x''}\CD\vec x(t)
  \exp\Bigg\{\ih\int_{t'}^{t''}\Bigg[{M\over2}\dot{\vec x}^2
  -V(\vec x\,)\Bigg]dt\Bigg\}
  =\int dE_\lambda \Phi_\lambda^*({\vec x\,}')\Phi_\lambda({\vec x\,}'')
  \,\e^{-\i E_\lambda T/\hbar}\enspace.
\end{equation}
The remaining $\rho$-path integration (\ref{numcd}) is again of the
radial harmonic oscillator type. The radial $1/\rho^2$-dependence must
be incorporated into the path integral in terms of a functional weight
$\mu_\lambda[\rho^2]$ (\ref{numce}) in order to guarantee a proper
definition of the corresponding lattice formulation of the radial path
integral \cite{GRSb, STEc}. As it turns out, this Besselian path
integral formulation \cite{INO} with the functional weight
$\mu_\lambda[\rho^2]$ for radial path integrals can be interpreted as an
additional potential term, and vice versa \cite{FLMa, LI}. We also want
to point out that a similar reasoning must be done in the case of the
P\"oschl-Teller and modified P\"oschl-Teller potential, respectively
\cite{FLMb, GROe}. Hence we apply the radial harmonic oscillator the
path integral solution and its expansion into wave-functions according
to ($\lambda$ an arbitrary parameter \cite{DURd, GOOb, PI, STEc})
\begin{eqnarray}       & &
  \int\limits_{r(t')=r'}^{r(t'')=r''}\CD r(t)
  \exp\left[\ih\int_{t'}^{t''}\Bigg({M\over2}\dot r^2
  -\hbar^2{\lambda^2-\viert\over2Mr^2}
           -{M\over2}\omega^2r^2\Bigg)dt\right]\qquad\qquad
         \nonumber\\   & &\qquad\qquad
  =\int\limits_{r(t')=r'}^{r(t'')=r''}\mu_\lambda[r^2]\CD r(t)
  \exp\left[{\i M\over2\hbar}\int_{t'}^{t''}
  \big(\dot r^2-\omega^2r^2\bigg)dt\right]
                  \\   & &\qquad\qquad
  ={M\omega\sqrt{r'r''}\over\i\hbar\sin\omega T}
  \exp\bigg[-{M\omega\over2\i\hbar}({r'}^2+{r''}^2)\cot\omega T\bigg]
  I_\lambda\bigg({M\omega r'r''\over\i\hbar\sin\omega T}\bigg)
\label{numcl}
                  \\   & &\qquad\qquad
  ={2M\omega\over\hbar}\sqrt{r'r''}\sum_{n=0}^\infty
  \e^{-\i\omega T(2n+\lambda+1)}
  {n!\over\Gamma(n+\lambda+1)}
  \bigg({M\omega\over\hbar}r'r''\bigg)^\lambda
         \nonumber\\   & &\qquad\qquad\qquad\times
  \exp\bigg(-{M\omega\over2\hbar}({r'}^2+{r''}^2)\bigg)
  L_n^{(\lambda)}\bigg({M\omega\over\hbar}{r'}^2\bigg)
  L_n^{(\lambda)}\bigg({M\omega\over\hbar}{r''}^2\bigg)\enspace.
\label{numcg}
\end{eqnarray}
The expansion into the wave-functions has been obtained by means of the
Hille-Hardy formula (\cite{GRA}, p.1038)
\begin{equation}
  {t^{-\alpha/2}\over 1-t}
  \exp\bigg[ -{1\over2}(x+y){1+t\over 1-t}\bigg]
  I_\alpha\bigg({2\sqrt{xyt}\over 1-t}\bigg)
  =\sum_{n=0}^\infty{t^n n! \e^{-\half(x+y)}\over\Gamma(n+\alpha+1)}
  (xy)^{\alpha/2}L_n(x) L_n(y)\enspace.
\end{equation}
The $L_n(x)$ are Laguerre polynomials (\cite{GRA}, p.1037).
We therefore obtain
\eject\noindent
\begin{eqnarray}       & &\!\!\!\!\!\!\!\!\!\!\!\!\!\!\!
  K^{(V_1)}({\vec x\,}'',{\vec x\,}';T)
         \nonumber\\   & &\!\!\!\!\!\!\!\!\!\!\!\!\!\!\!
  \hbox{{\it Cartesian Coordinates} \cite{DURd, GOOb, PI, STEc}:}
         \nonumber\\   & &\!\!\!\!\!\!\!\!\!\!\!\!\!\!\!
  =\bigg({M\omega\over\i\hbar\sin\omega T}\bigg)^2
  \prod_{i=1}^2\sqrt{x_i'x_i''}\,
 \exp\bigg[-{M\omega\over2\i\hbar}({x_i'}^2+{x_i''}^2)\cot\omega T\bigg]
  I_{\pm k_i}\bigg({M\omega x_i'x_i''\over\i\hbar\sin\omega T}\bigg)
                  \\   & &\!\!\!\!\!\!\!\!\!\!\!\!\!\!\!
 =\sum_{n_1,n_2=0}^\infty \e^{-\i E_NT/\hbar}
  \Psi_{n_1,n_2}(x_1',x_2')\Psi_{n_1,n_2}(x_1'',x_2'')
                  \\   & &\!\!\!\!\!\!\!\!\!\!\!\!\!\!\!
  \hbox{\it Polar Coordinates:}
         \nonumber\\   & &\!\!\!\!\!\!\!\!\!\!\!\!\!\!\!
  ={1\over\sqrt{\rho'\rho''}}\sum_{n=0}^\infty
  \Phi_n^{(\pm k_2,\pm k_1)}(\phi')\Phi_n^{(\pm k_2,\pm k_1)}(\phi'')
  \!\!\int\limits_{\rho(t')=\rho'}^{\rho(t'')=\rho''}\!\!\CD\rho(t)
  \exp\Bigg\{\ih\int_{t'}^{t''}
   \Bigg[{M\over2}(\dot\rho^2-\omega^2\rho^2)
         -\hbar^2{\lambda^2-\viert\over2M\rho^2}\Bigg]dt\Bigg\}
         \nonumber\\   & &\!\!\!\!\!\!\!\!\!\!\!\!\!\!\!
\label{numcd}     \\   & &\!\!\!\!\!\!\!\!\!\!\!\!\!\!\!
  =(\rho'\rho'')^{-1/2}\sum_{n=0}^\infty
  \Phi_n^{(\pm k_2,\pm k_1)}(\phi')\Phi_n^{(\pm k_2,\pm k_1)}(\phi'')
  \int\limits_{\rho(t')=\rho'}^{\rho(t'')=\rho''}
   \mu_\lambda[\rho^2]\CD\rho(t)
  \exp\Bigg[{\i M\over2\hbar}\int_{t'}^{t''}
      \big(\dot\rho^2-\omega^2\rho^2\big)dt\Bigg]\qquad
\label{numce}     \\   & &\!\!\!\!\!\!\!\!\!\!\!\!\!\!\!
 ={M\omega\over\i\hbar\sin\omega T}\sum_{n=0}^\infty
  \Phi_n^{(\pm k_2,\pm k_1)}(\phi')\Phi_n^{(\pm k_2,\pm k_1)}(\phi'')
 \exp\bigg[-{M\omega\over2\i\hbar}\big({\rho'}^2+{\rho''}^2\big)
    \cot\omega T\bigg]
  I_\lambda\bigg({M\omega\rho'\rho''\over\i\hbar\sin\omega T}\bigg)
                  \\   & &\!\!\!\!\!\!\!\!\!\!\!\!\!\!\!
 =\sum_{m,n=0}^\infty \e^{-\i E_NT/\hbar}
  \Psi_{m,n}(\phi',\rho')\Psi_{m,n}(\phi'',\rho'')\enspace,
\end{eqnarray}
where $\lambda=2n\pm k_1\pm k_2+1$. The wave-functions and the
energy-spectra are given by
\begin{eqnarray}       & &\!\!\!\!\!\!\!\!\!\!\!\!\!\!\!
  \hbox{\it Cartesian Coordinates:}
         \nonumber\\   & &\!\!\!\!\!\!\!\!\!\!\!\!\!\!\!
  \Psi_{n_1,n_2}(x_1,x_2)={2M\omega\over\hbar}\prod_{i=1}^2
  \bigg({M\omega\over\hbar}\bigg)^{\pm k_i/2}
  \sqrt{n_i!\over\Gamma(n_i\pm k_i+1)}\,
  x_i^{1/2\pm k_i}\exp\bigg(-{M\omega\over2\hbar}x_i^2\bigg)
  L_{n_i}^{(\pm k_i)}\bigg({M\omega\over\hbar}x_i^2\bigg)\enspace,
         \nonumber\\   & &\!\!\!\!\!\!\!\!\!\!\!\!\!\!\!
                  \\   & &\!\!\!\!\!\!\!\!\!\!\!\!\!\!\!
  E_N=\hbar\omega(2N \pm k_1\pm k_2+2)\enspace,\qquad N=n_1+n_2
  \enspace,       \\   & &\!\!\!\!\!\!\!\!\!\!\!\!\!\!\!
  \hbox{\it Polar Coordinates:}
         \nonumber\\   & &\!\!\!\!\!\!\!\!\!\!\!\!\!\!\!
  \Psi_{m,n}(\rho,\phi)
  =\Phi_n^{(\pm k_2,\pm k_1)}(\phi)\sqrt{2M\omega\over\hbar}
  \bigg({M\omega\over\hbar}\rho^2\bigg)^{\lambda/2}
  \sqrt{m!\over\Gamma(m+\lambda+1)}
  \exp\bigg(-{M\omega\over2\hbar}\rho^2\bigg)
  L_{m}^{(\lambda)}\bigg({M\omega\over\hbar}\rho^2\bigg)\enspace,
         \nonumber\\   & &\!\!\!\!\!\!\!\!\!\!\!\!\!\!\!
                  \\   & &\!\!\!\!\!\!\!\!\!\!\!\!\!\!\!
  E_N=\hbar\omega(2N\pm k_1\pm k_2+2)\enspace,\qquad N=m+n\enspace,
\end{eqnarray}
and $\Phi^{(\alpha,\beta)}_n(\phi)$ of (\ref{numca}) with $\alpha=\pm
k_2$, $\beta=\pm k_1$.
Let us remark that the wave-functions have been normalized in the
domain $[0,\infty)$. The positive sign at the $k_i$ has to be taken
whenever $k_i\geq\half$, i.e.\ the radial potential term is repulsive
at the origin, and the motion takes only place in the domain $x_i>0$.
If $0<|k_i|<\half$, i.e.\ the radial potential term is attractive at
the origin, both the positive and the negative sign must be taken into
account in the solution. This is indicated by the notion $\pm k_i$ in
the formul\ae. It has also the consequence that for each $k_i$ the
motion can take place on the entire real line. In the present case this
means that in cartesian coordinates we must distinguish four cases: i)
$x_1,x_2>0$, ii) $x_1>0$, $x_2\in\bbbr$, iii) $x_1\in\bbbr$, $x_2>0$
and iv) $(x_1,x_2)\in\bbbr^2$. In polar coordinates the same feature is
recovered by the observation that the P\"oschl-Teller barriers are
absent for $|k_i|<\half$. We will keep this notion in the sequel for all
following Smorodinsky-Winternitz potentials.

\vglue0.4truecm\noindent
{\it 3.1.2.}~We consider the potential ($k_{1,2}>0$)
\begin{equation}
  V_2(\vec x)={M\over2}\omega^2(4x^2+y^2)+k_1x
     +\hbarm{k_2^2-\viert\over y^2}\enspace,
\end{equation}
which is separable in two coordinate systems. For $k_1=0$ this potential
is also called Holt-potential \cite{HOLT}. We have
\begin{eqnarray}       & &\!\!\!\!\!\!\!\!
  K^{(V_2)}({\vec x\,}'',{\vec x\,}';T)
         \nonumber\\   & &\!\!\!\!\!\!\!\!
  \underline{\hbox{Cartesian Coordinates:}}
         \nonumber\\   & &\!\!\!\!\!\!\!\!
  =\int\limits_{\vec x(t')=\vec x'}^{\vec x(t'')=\vec x''}\CD\vec x(t)
  \exp\left\{\ih\int_{t'}^{t''}\left[{M\over2}
     \Big({\dot{\vec x}}^2-\omega^2(4x^2+y^2)\Big)-k_1x
     -\hbarm{k_2^2-\viert\over y^2}\right]dt\right\}
                  \\   & &\!\!\!\!\!\!\!\!
  \underline{\hbox{Parabolic Coordinates:}}
         \nonumber\\   & &\!\!\!\!\!\!\!\!
  =\int\limits_{\eta(t')=\eta'}^{\eta(t'')=\eta''}\CD\eta(t)
   \int\limits_{\xi(t')=\xi'}^{\xi(t'')=\xi''}\CD\xi(t)
   (\xi^2+\eta^2)
         \nonumber\\   & &\!\!\!\!\!\!\!\!\ \times
  \exp\Bigg\{\ih\int_{t'}^{t''}\Bigg[{M\over2}\Big(
     (\xi^2+\eta^2)(\dot\xi^2+\dot\eta^2)
     -\omega^2\big((\xi^2-\eta^2)^2-\xi^2\eta^2\big)\Big)
     -{k_1\over2}(\xi^2-\eta^2)-\hbar^2
      {k_2^2-\viert\over2M\eta^2\xi^2}\Bigg]dt\Bigg\}
         \nonumber\\   & &\!\!\!\!\!\!\!\!
                  \\   & &\!\!\!\!\!\!\!\!
  =\int_{\bbbr}{dE\over2\pi\hbar}\e^{-\i ET/\hbar}\int_0^\infty ds''
  \int\limits_{\eta(0)=\eta'}^{\eta(s'')=\eta''}\CD\eta(s)
  \int\limits_{\xi(0)=\xi'}^{\xi(s'')=\xi''}\CD\xi(s)
         \nonumber\\   & &\!\!\!\!\!\!\!\!\ \times
  \exp\Bigg\{\ih\int_0^{s''}\Bigg[{M\over2}\Big(
     (\dot\xi^2+\dot\eta^2)-\omega^2(\xi^6+\eta^6)\Big)
         \nonumber\\   & &\!\!\!\!\!\!\!\!\qquad\qquad\qquad\qquad
     -{k_1\over2}(\xi^4-\eta^4)+E(\xi^2+\eta^2)
     -\hbarm\Bigg({k_2^2-\viert\over\eta^2}
      +{k_2^2-\viert\over\xi^2}\Bigg)\Bigg]ds\Bigg\}\enspace.
\end{eqnarray}
In the $x$-path integration one makes use of the shifted harmonic
oscillator path integral solution \cite{FEY, FH}, and in the $y$-path
integration one makes use of the radial harmonic oscillator path
integral. The expansion into the wave-functions in $x$ is obtained by
means of the Mehler formula (\cite{EMOTa}, p.272)
\begin{equation}
  \e^{-(x^2+y^2)/2}
  \sum_{n=0}^\infty{1\over n!}\bigg({z\over2}\bigg)^nH_n(x)H_n(y)
  ={1\over\sqrt{1-z^2}}
  \exp\bigg[{4xyz-(x^2+y^2)(1+z^2)\over2(1-z^2)}\Bigg]\enspace.
\end{equation}
The $H_n(x)$ are Hermite polynomials (\cite{GRA}, p.1033). We have for
the harmonic oscillator path integral the kernel and its expansion
into wave-functions according to (we omit the Maslow indices)
\begin{eqnarray}     & &\!\!\!\!\!\!\!\!\!\!\!\!\!\!\!\!\!\!\!\!\!\!\!\!
  \int\limits_{x(t')=x'}^{x(t'')=x''}\CD x(t)
   \exp\left[{\i M\over2\hbar}
   \int_{t'}^{t''}(\dot x^2-\omega^2x^2)dt\right]
       \nonumber\\   & &\!\!\!\!\!\!\!\!\!\!\!\!\!\!\!\!\!\!\!\!\!\!\!\!
  =\bigg({M\omega\over2\pi\i\hbar\sin\omega T}\bigg)^{1/2}
   \exp\bigg\{-{M\omega\over2\i\hbar}
    \bigg[({x'}^2+{x''}^2)\cot\omega T
   -2{x'x''\over\sin\omega T}\bigg]\bigg\}
                \\   & &\!\!\!\!\!\!\!\!\!\!\!\!\!\!\!\!\!\!\!\!\!\!\!\!
  =\sum_{n=0}^\infty{\e^{-\i\omega T(n+1/2)}\over2^nn!}
   \bigg({M\omega\over\pi\hbar}\bigg)^{1/2}
   H_n\left(\sqrt{M\omega\over\hbar}\,x'\right)
   H_n\left(\sqrt{M\omega\over\hbar}\,x''\right)
   \exp\bigg(-{M\omega\over2\hbar}({x'}^2+{x''}^2)\bigg)\enspace.
\end{eqnarray}
The $\xi^4,\eta^4,\xi^6,\eta^6$-terms make the path integral in
parabolic coordinates intractable. For $k_1=0$ the anharmonic sextic
potential problems in $\xi$ and $\eta$ can only be treated for a zero
separation constant, c.f.\ the appendix. The result in {\it cartesian
coordinates} has the form ($\tilde x=x+k_1/4M\omega^2$)
\eject\noindent
\begin{eqnarray}
  K^{(V_2)}({\vec x\,}'',{\vec x\,}';T)
   &=&\bigg({M\omega\over2\pi\i\hbar\sin\omega T}\bigg)^{1/2}
   \exp\bigg\{{\i M\omega\over2\hbar\sin\omega T}
   \Big[({\tilde x}^{\prime\,2}+{\tilde x}^{\prime\prime\,2})
    \cos\omega T-2\tilde x'\tilde x''\Big]\bigg\}
         \nonumber\\   & &\qquad\times
  {M\omega\over2\i\hbar\sin\omega T}
  \exp\bigg[-{M\omega\over2\i\hbar}({y'}^2+{y''}^2)\cos\omega T\bigg]
  I_{k_2}\bigg({M\omega y'y''\over\i\hbar\sin\omega T}\bigg)
  \qquad          \\
   &=&\sum_{n_1,n_2=0}^\infty\e^{-\i E_{n_1,n_2}T/\hbar}
  \Psi_{n_1,n_2}(x'',y'')\Psi_{n_1,n_2}(x',y')\enspace,
\end{eqnarray}
with the wave-functions and the energy-spectrum given by
\begin{eqnarray}
  \Psi_{n_1,n_2}(x,y)
  &=&\sqrt{2M\omega\over\hbar}
  \bigg({M\omega\over\hbar}\bigg)^{\pm k_2/2}
  \sqrt{n_2!\over\Gamma(n_2\pm k_2+1)}
  y^{1/2\pm k_2}\exp\bigg(-{M\omega\over2\hbar}y^2\bigg)
  L_{n_2}^{(\pm k_2)}\bigg({M\omega\over\hbar}y^2\bigg)
         \nonumber\\   & &        \qquad\times
  \sqrt{\sqrt{2M\omega\over\pi\hbar}{1\over2^{n_1}n_1!}}
  \exp\bigg(-{M\omega\over\hbar}{\tilde x}^2\bigg)
  H_{n_1}\left(\sqrt{2M\omega\over\hbar}\,\tilde x\right)\enspace,
                  \\
  E_{n_1,n_2}&=&\hbar\omega(n_1+2n_2\pm k_2+\hbox{${3\over2}$})
            +{k_1^2\over8M\omega^2}\enspace.
\end{eqnarray}
Note the cases whether $k_2\geq\half$, respectively $0<k_2<\half$.

\vglue0.4truecm\noindent
{\it 3.1.3.}~We consider the potential ($k_{1,2}>0$)
\begin{equation}
  V_3(\vec x)=-{\alpha\over\sqrt{x^2+y^2}}+{\hbar^2\over4M}
    {1\over\sqrt{x^2+y^2}}\Bigg({k_1^2-\viert\over\sqrt{x^2+y^2}+x}
         +{k_2^2-\viert\over\sqrt{x^2+y^2}-x}\Bigg)\enspace.
\end{equation}
This potential is separable in polar, parabolic and elliptic II
coordinates. A path integral discussion for the pure Coulomb case
is due to Inomata \cite{INOa} and Duru and Kleinert \cite{DKb}. We have
\begin{eqnarray}       & &\!\!\!\!\!\!
  K^{(V_3)}({\vec x\,}'',{\vec x\,}';T)
         \nonumber\\   & &\!\!\!\!\!\!
  \underline{\hbox{Polar Coordinates:}}
         \nonumber\\   & &\!\!\!\!\!\!
  =\int\limits_{\rho(t')=\rho'}^{\rho(t'')=\rho''}\rho\CD\rho(t)
   \int\limits_{\phi(t')=\phi'}^{\phi(t'')=\phi''}\CD\phi(t)
         \nonumber\\   & &\!\!\!\!\!\!\qquad\times
   \exp\Bigg\{\ih\int_{t'}^{t''}\Bigg[{M\over2}
     (\dot\rho^2+\rho^2\dot\phi^2)+{\alpha\over\rho}
     -{\hbar^2\over8M\rho^2}\Bigg({k_1^2-\viert\over\cos^2{\phi\over2}}
     +{k_2^2-\viert\over\sin^2{\phi\over2}}-1\Bigg)\Bigg]dt\Bigg\}
                  \\   & &\!\!\!\!\!\!
  \hbox{$\underline{\hbox{Elliptic II Coordinates}}$ \cite{MPSTAb}:}
         \nonumber\\   & &\!\!\!\!\!\!
   =\int\limits_{\xi(t')=\xi'}^{\xi(t'')=\xi''}\CD\xi(t)
   \int\limits_{\eta(t')=\eta'}^{\eta(t'')=\eta''}\CD\eta(t)
   d^2(\sinh^2\xi+\sin^2\eta)
         \nonumber\\   & &\!\!\!\!\!\!\qquad\times
  \exp\Bigg\{\ih\int_{t'}^{t''}\Bigg[{M\over2}
     (\sinh^2\xi^2+\sin^2\eta)(\dot\xi^2+\dot\eta^2)
     +{\alpha\over d(\cosh\xi+\cos\eta)}
         \nonumber\\   & &\!\!\!\!\!\!
  \qquad\qquad\qquad\qquad
    -{\hbar^2\over4Md(\cosh^2\xi-\cos^2\eta)}
  \Bigg({(k_1^2+k_2^2-\half)+(k_2^2-k_1^2)\cos\eta\over\sin^2\eta}
         \nonumber\\   & &
  \qquad\qquad\qquad\qquad\qquad\qquad\qquad\qquad
       +{(k_1^2+k_2^2-\half)+(k_2^2-k_1^2)\cosh\xi\over\sinh^2\xi}
  \Bigg)\Bigg]dt\Bigg\}
                  \\   & &\!\!\!\!\!\!
  \hbox{$\underline{\hbox{Parabolic Coordinates}}$ \cite{CIMC}:}
         \nonumber\\   & &\!\!\!\!\!\!
   =\int\limits_{\eta(t')=\eta'}^{\eta(t'')=\eta''}\CD\eta(t)
   \int\limits_{\xi(t')=\xi'}^{\xi(t'')=\xi''}\CD\xi(t)
   (\xi^2+\eta^2)
         \nonumber\\   & &\!\!\!\!\!\!\qquad\times
  \exp\Bigg\{\ih\int_{t'}^{t''}\Bigg[{M\over2}
     (\xi^2+\eta^2)(\dot\xi^2+\dot\eta^2)+{2\alpha\over\xi^2+\eta^2}
     -{\hbar^2\over2M(\xi^2+\eta^2)}\Bigg({k_1^2-\viert\over\xi^2}
     +{k_2^2-\viert\over\eta^2}\Bigg)\Bigg]dt\Bigg\}\enspace.
         \nonumber\\   & &
\end{eqnarray}
If the domain of the angular variable is $\phi\in(0,2\pi)$ there are
two separated domains (half-planes) in the case of $k_1>\half$. A
rotation of the axis about $\pi/2$ interchanges the roles of $k_1$ and
$k_2$. In order to evaluate the two path integrals in polar and
parabolic coordinates a space-time transformation must be performed.
The result in {\it polar coordinates\/} has the form
[$\kappa=\alpha\sqrt{-M/2E}/\hbar$, $\lambda=m+(1\pm k_1\pm k_2)/2$,
only the radial Green function can be explicitly evaluated]
\begin{eqnarray}       & &\!\!\!\!\!\!
  G^{(V_3)}({\vec x\,}'',{\vec x\,}';E)
  =\ih \int_0^\infty  dT\,\e^{\i TE/\hbar}
  K^{(V_3)}({\vec x\,}'',{\vec x\,}';T)
         \nonumber\\   & &\!\!\!\!\!\!
  =\half(r'r'')^{-1/2}\sum_{l=0}^\infty
   \Phi_n^{(\pm k_2,\pm k_1)}\bigg({\phi'\over2}\bigg)
   \Phi_n^{(\pm k_2,\pm k_1)}\bigg({\phi''\over2}\bigg)
         \nonumber\\   & &\!\!\!\!\!\!\qquad\qquad\times
   \ih \int_0^\infty  dT\,\e^{\i TE/\hbar}
   \int\limits_{\rho(t')=\rho'}^{\rho(t'')=\rho''}\CD\rho(t)
   \exp\Bigg[\ih\int_{t'}^{t''}
   \Bigg({M\over2}\dot\rho^2+{\alpha\over\rho}
     -\hbar^2{\lambda^2-\viert\over2M\rho^2}\Bigg)dt\Bigg]
         \nonumber\\   & &\!\!\!\!\!\!
  =\half(\rho'\rho'')^{-1/2}\sum_{n=0}^\infty
   \Phi_n^{(\pm k_2,\pm k_1)}\bigg({\phi'\over2}\bigg)
   \Phi_n^{(\pm k_2,\pm k_1)}\bigg({\phi''\over2}\bigg)
         \nonumber\\   & &\!\!\!\!\!\!\qquad\qquad\times
   {1\over\hbar}\sqrt{-{M\over2E}}\,
   {\Gamma(\half+\lambda-\kappa)\over\Gamma(2\lambda+1)}
     W_{\kappa,\lambda}\bigg(\sqrt{-8ME}\,{\rho_>\over\hbar}\bigg)
     M_{\kappa,\lambda}\bigg(\sqrt{-8ME}\,{\rho_<\over\hbar}\bigg)
                  \\   & &\!\!\!\!\!\!
  =\sum_{n=0}^\infty \left\{\sum_{m=0}^\infty
   {\Psi_{n,m}^*(\rho',\phi')\Psi_{n,m}(\rho'',\phi'')\over E_{m}-E}
   +\int_{\bbbr} dp
   {\Psi_{p,m}^*(\rho',\phi')\Psi_{p,m}^{\alpha)}(\rho'',\phi'')
  \over\hbar^2p^2/2M-E}\right\}\enspace.\qquad
\end{eqnarray}
In the $\phi$-path integration the path integral solution of the
P\"oschl-Teller potential has been used. In the radial path integration
the path integral solution for the radial Coulomb potential was used
(Chetouani and Hammann \cite{CHc}, Duru and Kleinert \cite{DKa, DKb},
and Refs.~\cite{GROm, STEb}), and the Green function is expanded
according to (Kratzer potential, $\kappa=\alpha\sqrt{-M/2E}/\hbar$)
\begin{eqnarray}       & &
  \ih \int_0^\infty  dT\,\e^{\i TE/\hbar}
  \int\limits_{x(t')=x'}^{x(t'')=x''}\CD x(t)\exp\left[\ih
  \int_{t'}^{t''}\Bigg({M\over2}\dot x^2+{\alpha\over|x|}
           -\hbarm{\lambda^2-\viert\over x^2}\Bigg)dt\right]
         \nonumber\\   & &\qquad
  ={1\over\hbar}\sqrt{-{M\over2E}}
   {\Gamma(\half+\lambda-\kappa)\over\Gamma(2\lambda+1)}
     W_{\kappa,\lambda}\bigg(\sqrt{-8ME}\,{x_>\over\hbar}\bigg)
     M_{\kappa,\lambda}\bigg(\sqrt{-8ME}\,{x_<\over\hbar}\bigg)
                  \\   & &\qquad
  =\sum_{n=0}^\infty {\Psi_n(x')\Psi_n(x'')\over E_n-E}
   +\int_{\bbbr} dp {\Psi_p^*(x')\Psi_p(x'')\over E_p-E}\enspace.
\label{numck}
\end{eqnarray}
$x_{<,>}$ denotes the smaller/larger of $x',x''$. The $W_{\kappa,\lambda
}(x)$ and $M_{\kappa,\lambda}(x)$ are Whittaker functions (\cite{GRA},
p.1059). The bound state wave-functions and the energy-spectrum of the
Kratzer potential are given by ($a=\hbar^2/M\alpha$)
\begin{eqnarray}
  \Psi_n(x)&=&{1\over n+\lambda+\half}
 \bigg[{n!\over a\Gamma(n+2\lambda+1)}\bigg]^{1/2}
         \nonumber\\   & &\qquad\times
  \bigg({2x\over a(n+\lambda+\half)}\bigg)^{\lambda+1/2}
  \exp\bigg[-{x\over a(n+1)}\bigg]
  L_n^{(2\lambda)}\bigg({2x\over a(n+1)}\bigg)\enspace,\qquad
           \\
  E_n&=&-{M\alpha^2\over2\hbar^2(n+\lambda+\half)^2}\enspace,
\end{eqnarray}
and for the continuous spectrum with $E_p=p^2\hbar^2/2M$ one has
\begin{equation}
  \Psi_p(x)={\Gamma(\half+\lambda-\i/ap)\over
              \sqrt{2\pi}\Gamma(2\lambda+1)}
  \exp\bigg({\pi\over2ap}\bigg)M_{\i/ap,\lambda}(-2\i px)\enspace.
\end{equation}
Therefore the {\it polar coordinate\/} wave-functions and the
energy-spectrum of our potential problem are given by
($a=M/\alpha\hbar^2$):
\begin{eqnarray}
  \Psi_{n,m}(\rho,\phi)&=&
  2^{-1/2}\Phi_n^{(\pm k_2,\pm k_1)}\bigg({\phi\over2}\bigg)
 \bigg[{2m!\over a^2(m+\lambda+\half)^3
        \Gamma(m+2\lambda+1)}\bigg]^{1/2}
         \nonumber\\   & & \qquad\times
  \bigg({2\rho\over a(m+\lambda+\half)}\bigg)^\lambda
  \exp\bigg(-{\rho\over a(m+\lambda+\half)}\bigg)
  L_{m}^{(2\lambda)}
  \bigg({2\rho\over a(m+\lambda+\half)}\bigg)\enspace,\qquad
                  \\
  E_n&=&-{M\alpha^2\over2\hbar^2(m+\lambda+\half)^2}\enspace,
                  \\
  \Psi_{n,p}(\rho,\phi)&=&
  2^{-1/2}\Phi_n^{(\pm k_2,\pm k_1)}\bigg({\phi\over2}\bigg)
  {\Gamma(\half+\lambda-\i/ap)\over
  \sqrt{2\pi\rho}\,\Gamma(2\lambda+1)}\exp\bigg({\pi\over2ap}\bigg)
            M_{\i/ap,\lambda}(-2\i p\rho)\enspace.
\end{eqnarray}
The $\Phi^{(\pm k_2,\pm k_1)}_n(\phi/2)$ are the P\"oschl-Teller
wave-functions (\ref{numca}).

In {\it parabolic coordinates\/} we obtain by performing a time
transformation and applying the path integral solution of the radial
harmonic oscillator (\ref{numcl}) in the $\xi$ and $\eta$ variable,
respectively ($\omega=\sqrt{-2E/M}\,$)
\begin{eqnarray}       & &\!\!\!\!\!\!
  \ih \int_0^\infty  dT\,\e^{\i TE/\hbar}
  K^{(V_3)}({\vec x\,}'',{\vec x\,}';T)
         \nonumber\\   & &\!\!\!\!\!\!
   =\int_0^\infty ds''\,e^{2\i\alpha s''/\hbar}
  \int\limits_{\eta(0)=\eta'}^{\eta(s'')=\eta''}\CD\eta(s)
  \int\limits_{\xi(0)=\xi'}^{\xi(s'')=\xi''}\CD\xi(s)
         \nonumber\\   & &\!\!\!\!\!\!\qquad\times
  \exp\Bigg\{\ih\int_0^{s''}\Bigg[{M\over2}
     (\dot\xi^2+\dot\eta^2)+E(\xi^2+\eta^2)
     -\hbarm\Bigg({k_1^2-\viert\over\xi^2}
     +{k_2^2-\viert\over\eta^2}\Bigg)\Bigg]dt\Bigg\}\enspace.
         \nonumber\\   & &\!\!\!\!\!\!
  =\bigg({M\over\i\hbar}\bigg)^2\sqrt{\xi'\xi''\eta'\eta''}
   \int_0^\infty{\omega^2ds''\over\sin^2\omega s''}
   \e^{2\i\alpha s''/\hbar}
         \nonumber\\   & &\!\!\!\!\!\!\qquad\times
 \exp\bigg[-{M\omega\over2\i\hbar}
     \big({\xi'}^2+{\xi''}^2+{\eta'}^2+{\eta''}^2\big)
 \cot\omega s''\bigg]
 I_{k_1}\bigg({M\omega\xi'\xi''\over\i\hbar\sin\omega s''}\bigg)
 I_{k_2}\bigg({M\omega\eta'\eta''\over\i\hbar\sin\omega s''}\bigg)
 \qquad  \nonumber\\   & &\!\!\!\!\!\!
  =\sum_{n_1,n_2=0}^\infty
  {\Psi_{n_1,n_2}^*(\xi',\eta')\Psi_{n_1,n_2}(\xi'',\eta'')
   \over E_{n_1,n_2}-E}+\int_0^\infty dp\int_{\bbbr}d\zeta
  {\Psi_{p,\zeta}^*(\xi',\eta')\Psi_{p,\zeta}(\xi'',\eta'')
   \over E_p-E}\enspace,
\end{eqnarray}
where are ($N=n_1+n_2+\half(\pm k_1\pm k_2)+1$, $a=\hbar^2/M\alpha$)
\begin{eqnarray}       & &\!\!\!\!\!\!\!\!
  \Psi_{n_1,n_2}(\xi,\eta)=\bigg[{2\over a^2N^3}\cdot
  {n_1!n_2!\over\Gamma(n_1\pm k_1+1)\Gamma(n_2\pm k_2+1)}\bigg]^{1/2}
         \nonumber\\   & &\!\!\!\!\!\!\!\!\qquad\qquad\qquad\times
  \bigg({\xi\over aN}\bigg)^{\pm k_1/2}
  \bigg({\eta\over aN}\bigg)^{\pm k_2/2}
  \exp\bigg(-{\xi^2+\eta^2\over2aN}\bigg)
  L_{n_1}^{(\pm k_1)}\bigg({\xi^2\over aN}\bigg)
  L_{n_2}^{(\pm k_2)}\bigg({\eta^2\over aN}\bigg)\enspace,\qquad
                  \\   & & \!\!\!\!\!\!\!\!
     E_{n_1,n_2}=-{M\alpha^2\over \hbar^2 N^2}\enspace,
                  \\   & &\!\!\!\!\!\!\!\!
         \nonumber\\   & &\!\!\!\!\!\!\!\!
  \Psi_{p,\zeta}(\xi,\eta)
  ={\Gamma[{1\pm k_2\over2}+\i(1/a+\zeta)/2p]
   \Gamma[{1\pm k_1\over2}+\i(1/a-\zeta)/2p]\over2\pi\sqrt{\xi\eta}\,
   \Gamma(1\pm k_2)\Gamma(1\pm k_1)}\e^{\pi/2ap}
         \nonumber\\   & &\!\!\!\!\!\!\!\!\qquad\qquad\qquad\times
  M_{-\i(1/a+\zeta)/2p,\pm k_1/2} (-\i p\xi^2)
  M_{-\i(1/a-\zeta)/2p,\pm k_2/2} (-\i p\eta^2)\enspace,
                  \\   & &\!\!\!\!\!\!\!\!
  E_p=\hbarm p^2\enspace.
\end{eqnarray}
In order to extract the discrete spectrum one  applies (c.f.~\cite{CHe,
GROm, GROab}) the Mehler formula \cite{EMOTa}. For the continuous
spectrum one applies the following dispersion relation
(c.f.~\cite{BUC}, p.158, \cite{EMOTb}, p.414, \cite{GRA}, p.884)
\begin{eqnarray}     & &{1\over\sin\alpha}\exp\Big[-(x+y)\cot\alpha\Big]
  I_{2\mu}\bigg({2\sqrt{xy}\over\sin\alpha}\bigg)
         \nonumber\\   & &
  ={1\over2\pi\sqrt{xy}}\int_{\bbbr}
  {\Gamma(\half+\mu+\i p)\Gamma(\half+\mu-\i p)\over\Gamma^2(1+2\mu)}
  \e^{-2\alpha p+\pi p}M_{+\i p,\mu}(-2\i x)M_{-\i p,\mu}(+2\i y)dp
  \enspace.
\label{numef}
\end{eqnarray}
and performs the variable substitution $(\nu_1,\nu_2)\mapsto(p_\xi,
p_\eta)\mapsto[(1/a+\zeta)/2p,(1/a-\zeta)/2p]$. $\zeta$ is the
parabolic separation constant. In elliptic II coordinates $K^{(V_3)}$
is not explicitly soluble and we obtain only the path integral identity
\begin{eqnarray}       & &\!\!\!\!\!\!\!\!
  K^{(V_3)}({\vec x\,}'',{\vec x\,}';T)
         \nonumber\\   & &\!\!\!\!\!\!\!\!
  =\int_{\bbbr}{dE\over2\pi\hbar}\e^{-\i ET/\hbar}\int_0^\infty ds''
  \int\limits_{\xi(0)=\xi'}^{\xi(s'')=\xi''}\CD\xi(s)
   \int\limits_{\eta(0)=\eta'}^{\eta(s'')=\eta''}\CD\eta(s)
         \nonumber\\   & &\!\!\!\!\!\!\!\!\times
  \exp\Bigg\{\ih\int_0^{s''}\Bigg[{M\over2}(\dot\xi^2+\dot\eta^2)
     +Ed^2(\cosh^2\xi-\cos^2\eta)+\alpha d(\cosh\xi+\cos\eta)
  \qquad \nonumber\\   & &\!\!\!\!\!\!\!\!\qquad\qquad\qquad\qquad
     -\hbarm\Bigg({(k_1^2-\viert)(1-\cos\eta)+(k_2^2-\viert)(1+\cos\eta)
        \over\sin^2\eta}
         \nonumber\\   & &\!\!\!\!\!\!\!\!
  \qquad\qquad\qquad\qquad\qquad\qquad
       +{(k_1^2-\viert)(1-\cosh\xi)+(k_2^2-\viert)(1+\cosh\xi)
        \over\sinh^2\xi}\Bigg)\Bigg]ds\Bigg\}\enspace.\qquad
\end{eqnarray}
No further evaluation is possible.

\vglue0.4truecm\noindent
{\it 3.1.4.\/}~We consider the potential ($\beta_{1,2}\in\bbbr$)
\begin{equation}
  V_4(\vec x)=-{\alpha\over\rho}+\sqrt{2\over\rho}
   \bigg(\beta_1\cos{\phi\over2}+\beta_2\sin{\phi\over2}\bigg)\enspace.
\end{equation}
This potential is only separable in {\it mutually orthogonal parabolic\/
} coordinates and again a time transformation must be performed. We
obtain [$\omega=\sqrt{-2E/M}\,$, only the Green function can be
explicitly evaluated, $(\tilde\xi,\tilde\eta)=(\xi-\beta_1/E,
\eta-\beta_2/E)$]
\begin{eqnarray}       & &\!\!\!\!\!\!\!
  \ih \int_0^\infty  dT\,\e^{\i TE/\hbar}
  \int\limits_{\vec x(t')=\vec x'}^{\vec x(t'')=\vec x''}\CD\vec x(t)
         \nonumber\\   & &\!\!\!\!\!\!\! \qquad\times
  \exp\Bigg\{\ih\int_{t'}^{t''}
  \Bigg[{M\over2}(\dot\rho^2+\rho^2\dot\phi^2) +{\alpha\over\rho}
    +{\alpha\over\rho}+\sqrt{2\over\rho}\bigg(\beta_1\cos{\phi\over2}
                          +\beta_2\sin{\phi\over2}\bigg)\Bigg]dt\Bigg\}
         \nonumber\\   & &\!\!\!\!\!\!\!
  =\ih \int_0^\infty  dT\,\e^{\i TE/\hbar}
   \int\limits_{\eta(t')=\eta'}^{\eta(t'')=\eta''}\CD\eta(t)
   \int\limits_{\xi(t')=\xi'}^{\xi(t'')=\xi''}\CD\xi(t)
   (\xi^2+\eta^2)
         \nonumber\\   & &\!\!\!\!\!\!\! \qquad\times
  \exp\Bigg\{\ih\int_{t'}^{t''}\Bigg[{M\over2}
     (\xi^2+\eta^2)(\dot\xi^2+\dot\eta^2)
    +2{\alpha-(\beta_1\xi+\beta_2\eta)\over\xi^2+\eta^2}\Bigg]dt\Bigg\}
                  \\   & &\!\!\!\!\!\!\!
  =\int_0^\infty ds''\,\e^{2\i\alpha s''/\hbar}
  \int\limits_{\eta(0)=\eta'}^{\eta(s'')=\eta''}\CD\eta(s)
  \int\limits_{\xi(0)=\xi'}^{\xi(s'')=\xi''}\CD\xi(s)
         \nonumber\\   & &\!\!\!\!\!\!\! \qquad\times
  \exp\Bigg\{\ih\int_0^{s''}\Bigg[{M\over2}(\dot\xi^2+\dot\eta^2)
               +E(\xi^2+\eta^2)-2(\beta_1\xi+\beta_2\eta)\Bigg]dt\Bigg\}
                  \\   & &\!\!\!\!\!\!\!
  =\int_0^\infty ds''{M\omega\over\pi\i\hbar\sin\omega s''}
  \cosh\bigg({M\omega(\tilde\xi^\prime\tilde\xi^{\prime\prime}
   +\tilde\eta^\prime\tilde\eta^{\prime\prime})
    \over\i\hbar\sin\omega s''}\bigg)
         \nonumber\\   & &\!\!\!\!\!\!\! \qquad\times
  \exp\bigg[{2\i\alpha s''\over\hbar}-\ih{\beta_1^2+\beta_2^2\over E}s''
           +{\i M\omega\over2\hbar}
    ({\tilde\xi}^{\prime\,2}+{\tilde\xi}^{\prime\prime\,2}
    +{\tilde\eta}^{\prime\,2}+{\tilde\eta}^{\prime\prime\,2})
    \cot\omega s''\bigg]
\label{numdi}     \\   & &\!\!\!\!\!\!\!
  =\sum_{n_1,n_2=0}^\infty
   {\Psi_{n_1,n_2}(\xi',\eta')\Psi_{n_1,n_2}(\xi'',\eta'')
    \over E_{n_1,n_2}-E}
  +\sum_{e,o}\int_{\bbbr} d\zeta\int_{\bbbr}dp
   {\Psi_{p,\zeta}^{(e,o)\,*}(\xi',\eta')
   \Psi_{p,\zeta}^{(e,o)}(\xi'',\eta'')\over\hbar^2p^2/2M-E}\enspace,
  \qquad
\end{eqnarray}
and $\sum_{e,o}$ denotes the summation over even and odd states,
respectively; the bound state energy-levels are determined by
($N=n_1+n_2+1$, $n_1,n_2\in\bbbn_0$, $\omega_N=\sqrt{-2E_N/M}\,$)
\begin{equation}
  \omega_N^3-{2\alpha\over N\hbar}\omega_N^2
  -2{\beta_1^2+\beta_2^2\over MN\hbar}=0\enspace.
\end{equation}
For $\alpha>0$ the discriminante of this cubic equation shows that there
is {\it one\/} real root $\omega_N$ for each $N$ such that the bound
state energy-levels $E_N$ are given by
\begin{eqnarray}
  E_N&=&E_{n_1,n_2}=
  -{M\over2}\omega_N^2\enspace,\qquad
  \omega_N=y_N+{2\alpha\over3\hbar N}\enspace,\qquad
  y_N=u_1+u_2\enspace,
  \\
  u_{1,2}&=&
  \sqrt[\scriptstyle3\,]{\bigg({2\alpha\over3\hbar N}\bigg)^3
+{\beta_1^2+\beta_2^2\over MN\hbar}
\mp\sqrt{\bigg({\beta_1^2+\beta_2^2\over MN\hbar}\bigg)^2
+2{\beta_1^2+\beta_2^2\over MN\hbar}\bigg({2\alpha\over3\hbar N}\bigg)^3
  }}\enspace.
\end{eqnarray}
The corresponding bound-state wave-functions have the form
\begin{eqnarray}
  \Psi_{n_1,n_2}(\xi,\eta)&=&
  \sqrt{{M\over\pi\hbar}\cdot{4\over n_1!n_2!2^{n_1+n_2}}}
  \left(\lim_{E\to E_N}
  {-({M\over2}\omega_N^2+E)M\omega_N^4/N\over
    \omega^3-{2\alpha\over N\hbar}\omega^2
  -2{\beta_1^2+\beta_2^2\over MN\hbar}  }\right)^{1/2}
         \nonumber\\   & &\qquad \times
  \,\exp\bigg[-{M\omega_N\over2\hbar}\big(\tilde\xi^2+\tilde\eta^2\big)
        \bigg]
  H_{n_1}\left(\sqrt{M\omega_N\over\hbar}\,\tilde\xi\right)
  H_{n_2}\left(\sqrt{M\omega_N\over\hbar}\,\tilde\eta\right)\enspace.
  \qquad
\end{eqnarray}
$N=n_1+n_2$ must be chosen in such a way that $N=n_1+n_2$ is an even
number \cite{ZZ}. The continuous functions $\Psi_{p,\zeta}^{(e,o)}(\xi,
\eta)$ are given by [$\tilde a=\hbar^2/M\big(\alpha-m(\beta_1^2+\beta_2^
2)/\hbar^2p^2\big)$]
\begin{eqnarray}
  \Psi_{p,\zeta}^{(e,o)}(\xi,\eta)&=&{\e^{\pi/2ap}\over\sqrt{2}\,4\pi^2}
   \left(\begin{array}{l}\displaystyle
    \Gamma[\bviert+\hbox{${\i\over2p}$}(1/\tilde a+\zeta)]
   E^{(0)}_{-\half+{\i\over p}(1/\tilde a+\zeta)}
                              (\e^{-\i\pi/4}\sqrt{2p}\,\tilde\xi)
  \\   \displaystyle
   \Gamma[\hbox{${3\over4}$}+\hbox{${\i\over2p}$}(1/\tilde a+\zeta)]
   E^{(1)}_{-\half+{\i\over p}(1/\tilde a+\zeta)}
                              (\e^{-\i\pi/4}\sqrt{2p}\,\tilde\xi)
  \end{array}\right)
         \nonumber\\   & &\qquad
   \times\left(\begin{array}{l}
   \displaystyle
   \Gamma[\bviert+\hbox{${\i\over2p}$}(1/\tilde a-\zeta)]
   E^{(0)}_{-\half+{\i\over p}(1/\tilde a-\zeta)}
                              (\e^{-\i\pi/4}\sqrt{2p}\,\tilde\eta)
  \\   \displaystyle
   \Gamma[\hbox{${3\over4}$}+\hbox{${\i\over2p}$}(1/\tilde a-\zeta)]
   E^{(1)}_{-\half+{\i\over p}(1/\tilde a-\zeta)}
                              (\e^{-\i\pi/4}\sqrt{2p}\,\tilde\eta)
  \end{array}\right)\enspace.
\end{eqnarray}
For the determination of the continuous spectrum one uses the
dispersion relation (\cite{GRA}, p.896)
\begin{eqnarray}       & &
     \int_{c-\i\infty}^{c+\i\infty}
  \big[D_\nu(x)D_{-\nu-1}(\i y)+D_\nu(-x)D_{-\nu-1}(-\i y)\big]
  {t^{-\nu-1}d\nu\over\sin(-\nu\pi)}
         \nonumber\\   & &\qquad\qquad\qquad\qquad
  =2\i\sqrt{2\pi\over1+t^2}\exp\bigg(
 {1-t^2\over1+t^2}\cdot{x^2+y^2\over4}+{\i txy\over1+t^2}\bigg)\enspace,
\end{eqnarray}
and performs the variable substitution $(\nu_1,\nu_2)\mapsto
(p_\xi,p_\eta)\mapsto[(1/\tilde a+\zeta)/2p,(1/\tilde a-\zeta)/2p]$.
$\zeta$ is the parabolic separation constant.
The $D_\nu(z)$ are parabolic cylinder functions, and the
$E_\nu^{(0)}(z)$ and $E_\nu^{(1)}(z)$ are even and odd parabolic
cylinder functions \cite{BUC}, respectively.

Note that in the evaluation of the path integral one has to take into
account that by using the linearly shifted harmonic oscillator solution
for the $\xi$- and $\eta$-variable, respectively, one actually uses a
double covering of the original $(x,y)\in\bbbr^2$-plane, i.e.\ $\vec u
\equiv(\tilde\xi,\tilde\eta)\in\bbbr^2$. Furthermore we have taken into
account that our mapping is of the ``square-root'' type which gives
rise to a sign ambiguity. ``Thus, if one considers all paths in the
complex $z=x+\i y $-plane  from $z'$ to $z''$, they will be mapped into
two different classes of paths in the $\vec u$-plane: Those which go
from $\vec u'$ to $\vec u''$ and those going from $\vec u'$ to $-\vec
u''$. In the cut complex $z$-plane for the function $|\vec u|=\sqrt{|z|}
$ these are the paths passing an even or odd number of times through
the square root from $| z|=0$ and $|z|=-\infty$. We may choose the
$\vec u'$ corresponding to the initial $z'$ to lie on the first sheet
i.e.\ in the right half $\vec u$-plane). The final $\vec u''$ can be in
the right as well as the left half-plane and all paths on the $z$-plane
go over into paths from $\vec u'$ to $\vec u''$ and those from $\vec
u'$ to $-\vec u''$ '' \cite{DKb}. Thus the two contributions arise in
(\ref{numdi}).

\vglue0.6truecm\noindent
{\bf 3.2.~Three-Dimensional Maximally Super-Integrable
          Smorodinsky-Winternitz Potentials.}
\vglue0.4truecm\noindent
We discuss in this subsection the five three-dimensional
maximally super-integrable potentials. They are characterized by
having five functionally independent integrals of motion.
In the sequel $\vec x$ denotes a three-dimensional coordinate:
$\vec x=(x,y,z)\equiv(x_1,x_2,x_3)$.

In table 3 we list the three-dimensional maximally super-integrable
Smorodinsky-Winternitz potentials together with the separating
coordinate systems (where the {\it italized\/} coordinates systems were
not mentioned in \cite{EVA}). The cases where an explicit path
integration is possible are $\underline{\hbox{underlined}}$.

\vglue0.4truecm\noindent
{\it 3.2.1.}~We consider the three-dimensional potential ($k_{1,2,3}>0$)
\begin{equation}
  V_1(\vec x)={M\over2}\omega^2\vec x^2
    +\hbarm\Bigg({k_1^2-\viert\over x^2}
    +{k_2^2-\viert\over y^2}+{k_3^2-\viert\over z^2}\Bigg)\enspace.
\end{equation}
Note furthermore that the pure three-dimensional harmonic oscillator
\cite{KLPS} has the same separating coordinate systems as $V_1(\vec x)$
and therefore the additional centrifugal terms do not spoil this
property.

We formulate the corresponding path integral in the coordinate
systems which separate the path integral for this potential
problem. We have
\begin{eqnarray}       & &\!\!\!\!\!\!\!\!
  K^{(V_1)}({\vec x\,}'',{\vec x\,}';T)
         \nonumber\\   & &\!\!\!\!\!\!\!\!
  \underline{\hbox{Cartesian Coordinates:}}
         \nonumber\\   & &\!\!\!\!\!\!\!\!
   \int\limits_{\vec x(t')=\vec x'}^{\vec x(t'')=\vec x''}\CD\vec x(t)
  \exp\left\{\ih\int_{t'}^{t''}\left[{M\over2}
   \big({\dot{\vec x}}^2-\omega^2\vec x^2\big)
 -\hbarm\sum_{i=1}^3{k_i^2-\viert\over x_i^2}\right]dt\right\}
                  \\   & &\!\!\!\!\!\!\!\!
  \underline{\hbox{Spherical Coordinates:}}
         \nonumber\\   & &\!\!\!\!\!\!\!\!
  =\int\limits_{r(t')=r'}^{r(t'')=r''}r^2\CD r(t)
   \int\limits_{\theta(t')=\theta'}^{\theta(t'')=\theta''}
  \sin\theta\CD\theta(t)
   \int\limits_{\phi(t')=\phi'}^{\phi(t'')=\phi''}\CD\phi(t)
         \nonumber\\   & &\!\!\!\!\!\!\!\!\qquad\times
  \exp\Bigg\{\ih\int_{t'}^{t''}\Bigg[{M\over2}
\Big(\dot r^2+r^2\dot\theta^2+r^2\sin^2\theta\dot\phi^2-\omega^2r^2\Big)
         \nonumber\\   & &\!\!\!\!\!\!\!\!\qquad\qquad\qquad
  -{\hbar^2\over2Mr^2}\Bigg({1\over\sin^2\theta}
      \bigg({k_1^2-\viert\over\cos^2\phi}
           +{k_2^2-\viert\over\sin^2\phi}-\viert\bigg)
           +{k_3^2-\viert\over\cos^2\theta}
           -{1\over4}\Bigg)\Bigg]dt\Bigg\}\qquad\quad
\end{eqnarray}

\eject\noindent
\centerline{{\bf Table 3:} The three-dimensional maximally
            super-integrable potentials\hfill}
\hfuzz=45pt
\begin{eqnarray}
\begin{array}{l}
\vbox{\offinterlineskip
\hrule
\halign{&\vrule#&
  \strut\quad\hfil#\quad\hfill\ \cr
height2pt&\omit&&\omit&\cr
&Potential $V(x,y,z)$
  &&Coordinate System                                       &\cr
height2pt&\omit&&\omit&&\omit&\cr
\noalign{\hrule}
\noalign{\hrule}
height2pt&\omit&&\omit&&\omit&\cr
&$\displaystyle
  V_1={M\over2}\omega^2\vec x^2
    +\hbarm\Bigg({k_1^2-\viert\over x^2}
    +{k_2^2-\viert\over y^2}+{k_3^2-\viert\over z^2}\Bigg)$
  &&$\underline{\hbox{Cartesian}}$              &\cr
& &&$\underline{\hbox{Spherical}}$              &\cr
& &&$\underline{\hbox{Circular Polar}}$         &\cr
& &&Circular Elliptic                           &\cr
& &&Conical                                     &\cr
& &&Oblate Spheroidal                           &\cr
& &&Prolate Spheroidal                          &\cr
& &&Ellipsoidal                                 &\cr
height2pt&\omit&&\omit&\cr
\noalign{\hrule}
height2pt&\omit&&\omit&\cr
&$\omega=0$,\quad $k_3^2-1/4=0$
  &&Paraboloidal                                &\cr
& &&$\underline{\hbox{Parabolic}}$              &\cr
&$\omega=0$, \quad $k_1^2-1/4=0$
  &&$\underline{\hbox{Circular Parabolic}}$     &\cr
height2pt&\omit&&\omit&\cr
\noalign{\hrule}
height2pt&\omit&&\omit&\cr
&$\displaystyle
  V_2={M\over2}\omega^2(x^2+y^2+4z^2)
     +\hbarm\Bigg({k_1^2-\viert\over x^2}+{k_2^2-\viert\over y^2}\Bigg)$
  &&$\underline{\hbox{Cartesian}}$              &\cr
& &&Parabolic                                   &\cr
& &&$\underline{\hbox{\it Circular Polar}}$     &\cr
& &&Circular Elliptic                           &\cr
height2pt&\omit&&\omit&\cr
\noalign{\hrule}
height2pt&\omit&&\omit&\cr
&$\displaystyle
  V_3=-{\alpha\over\sqrt{x^2+y^2+z^2}}
    +\hbarm\Bigg({k_1^2-\viert\over x^2}
    +{k_2^2-\viert\over y^2}\Bigg)$
  &&Conical                                     &\cr
& &&$\underline{\hbox{Spherical}}$              &\cr
& &&$\underline{\hbox{Parabolic}}$              &\cr
& &&{\it Prolate Spheroidal II}                 &\cr
height2pt&\omit&&\omit&\cr
\noalign{\hrule}
height2pt&\omit&&\omit&\cr
&$\displaystyle
  V_4=\hbarm\bigg(
  {k_1^2x\over y^2\sqrt{x^2+y^2}}+{k_2^2-\viert\over y^2}
  +{k_3^2-\viert\over z^2}\Bigg)$
  &&$\underline{\hbox{Spherical}}$              &\cr
&$\displaystyle\phantom{V_4}
    ={\hbar^2\over2M}\Bigg[{1\over\sqrt{x^2+y^2}}\Bigg(
   {\beta_1^2-\viert\over\sqrt{x^2+y^2}+x}
   +{\beta_2^2-\viert\over\sqrt{x^2+y^2}-x}\Bigg)
   +{k_3^2-\viert\over z^2}\Bigg]$
  &&{\it Circular Elliptic II}                  &\cr
& &&$\underline{\hbox{Circular Parabolic}}$     &\cr
& $\beta_1^2=\half(k_2^2+k_1^2+\viert)$,\quad
  $\beta_2^2=\half(k_2^2-k_1^2+\viert)$
  &&$\underline{\hbox{\it Circular Polar}}$     &\cr
height2pt&\omit&&\omit&\cr
\noalign{\hrule}
height2pt&\omit&&\omit&\cr
&$\displaystyle
  V_5=\hbarm\bigg(
  {k_1^2x\over y^2\sqrt{x^2+y^2}}+{k_2^2-\viert\over y^2}\Bigg)-k_3z$
  &&$\underline{\hbox{\it Circular Polar}}$     &\cr
&$\displaystyle\phantom{V_5}
    ={\hbar^2\over2M\sqrt{x^2+y^2}}\Bigg(
     {\beta_1^2-\viert\over\sqrt{x^2+y^2}+x}
     +{\beta_2^2-\viert\over\sqrt{x^2+y^2}-x}\Bigg)-k_3z$
  &&{\it Circular Elliptic II}                  &\cr
& &&$\underline{\hbox{Circular Parabolic}}$     &\cr
& $\beta_1^2=\half(k_2^2+k_1^2+\viert)$,\quad
  $\beta_2^2=\half(k_2^2-k_1^2+\viert)$
  &&Parabolic                                   &\cr
height2pt&\omit&&\omit&\cr}\hrule}
\end{array}
         \nonumber
\end{eqnarray}
\hfuzz=7.0pt

\begin{eqnarray}       & &\!\!\!\!\!\!\!\!
  \underline{\hbox{Circular Polar Coordinates:}}
         \nonumber\\   & &\!\!\!\!\!\!\!\!
  =\int\limits_{z(t')=z'}^{z(t'')=z''}\CD z(t)
   \int\limits_{\rho(t')=\rho'}^{\rho(t'')=\rho''}\rho\CD\rho(t)
   \int\limits_{\phi(t')=\phi'}^{\phi(t'')=\phi''}\CD\phi(t)
         \nonumber\\   & &\!\!\!\!\!\!\!\!\qquad\times
  \exp\Bigg\{\ih\int_{t'}^{t''}\Bigg[{M\over2}
     \big(\dot\rho^2+\rho^2\dot\phi^2-\omega^2\rho^2\big)
     -\hbarm\Bigg({k_1^2-\viert\over\rho^2\cos^2\phi}
         +{k_2^2-\viert\over\rho^2\sin^2\phi}
         +{k_3^2-\viert\over z^2}-{1\over4\rho^2}\Bigg)\Bigg]dt\Bigg\}
         \nonumber\\   & &
                  \\   & &\!\!\!\!\!\!\!\!
  \underline{\hbox{Circular Elliptic Coordinates:}}
         \nonumber\\   & &\!\!\!\!\!\!\!\!
  =\int\limits_{z(t')=z'}^{z(t'')=z''}\CD z(t)
   \int\limits_{\xi(t')=\xi'}^{\xi(t'')=\xi''}\CD\xi(t)
   \int\limits_{\eta(t')=\eta'}^{\eta(t'')=\eta''}\CD\eta(t)
   d^2(\sinh^2\xi+\sin^2\eta)
         \nonumber\\   & &\!\!\!\!\!\!\!\!\qquad\times
  \exp\Bigg\{\ih\int_{t'}^{t''}\Bigg[{M\over2}
     \Big(d^2(\sinh^2\xi+\sin^2\eta)
         (\dot\xi^2+\dot\eta^2)+\dot z^2\Big)
         \nonumber\\   & &\!\!\!\!\!\!\!\!\qquad\qquad\qquad
  -{M\over2}\omega^2\Big
      (d^2(\cosh^2\xi\cos^2\eta+\sinh^2\xi\sin^2\eta)+z^2\Big)
         \nonumber\\   & &\!\!\!\!\!\!\!\!\qquad\qquad\qquad
  -\hbarm\Bigg({k_1^2-\viert\over d^2\cosh^2\xi\cos^2\eta}
         +{k_2^2-\viert\over d^2\sinh^2\xi\sin^2\eta}
         +{k_3^2-\viert\over z^2}\Bigg)\Bigg]dt\Bigg\}
                  \\   & &\!\!\!\!\!\!\!\!
  \underline{\hbox{Conical Coordinates:}}
         \nonumber\\   & &\!\!\!\!\!\!\!\!
  =\int\limits_{r(t')=r'}^{r(t'')=r''}r^2\CD r(t)
  \int\limits_{\alpha(t')=\alpha'}^{\alpha(t'')=\alpha''}\CD\alpha(t)
  \int\limits_{\beta(t')=\beta'}^{\beta(t'')=\beta''}\CD\beta(t)
  (k^2\cn^2\alpha+{k'}^2\cn^2\beta)
         \nonumber\\   & &\!\!\!\!\!\!\!\!\qquad\times
  \exp\Bigg\{\ih\int^{t''}_{t'}\Bigg[{M\over2}\Big(\dot r^2+
   r^2(k^2\cn^2\alpha+{k'}^2\cn^2\beta)
   (\dot\alpha^2+\dot\beta^2)-\omega^2r^2\Big)
         \nonumber\\   & &\!\!\!\!\!\!\!\!\qquad\qquad\qquad
     -{\hbar^2\over2Mr^2}\Bigg({k_1^2-\viert\over\sn^2\alpha\dn^2\beta}
         +{k_2^2-\viert\over\cn^2\alpha\cn^2\beta}
         +{k_3^2-\viert\over\dn^2\alpha\sn^2\beta}\Bigg)\Bigg]dt\Bigg\}
                  \\   & &\!\!\!\!\!\!\!\!
  \hbox{$\underline{\hbox{Oblate Spheroidal Coordinates}}$
         ($\lambda_1=2n\pm k_1\pm k_2+1)$:}
         \nonumber\\   & &\!\!\!\!\!\!\!\!
  =\int\limits_{\bar\mu(t')=\bar\mu'}^{\bar\mu(t'')=\bar\mu''}
   \CD\bar\mu(t)
   \int\limits_{\bar\nu(t')=\bar\nu'}^{\bar\nu(t'')=\bar\nu''}
   \CD\bar\nu(t)
   \bar d^3(\cosh^2\bar\mu-\sin^2\bar\nu)\cosh\bar\mu\sin\bar\nu
  \int\limits_{\phi(t')=\phi'}^{\phi(t'')=\phi''}\CD \phi(t)
         \nonumber\\   & &\!\!\!\!\!\!\!\!\qquad\times
  \exp\Bigg\{\ih\int_{t'}^{t''}\Bigg[{M\over2}\bar d^2
  \Big((\cosh^2\bar\mu-\sin^2\bar\nu)(\dot{\bar\mu}^2+\dot{\bar\nu}^2)
        +\cosh^2\bar\mu\sin^2\bar\nu\dot\phi^2
         \nonumber\\   & &\!\!\!\!\!\!\!\!\qquad\qquad\qquad
 -\omega^2(\cosh^2\bar\mu\sin^2\bar\nu+\sinh^2\bar\mu\cos^2\bar\nu)\Big)
         \nonumber\\   & &\!\!\!\!\!\!\!\!\qquad\qquad\qquad
   -{\hbar^2\over2M\bar d^2}\Bigg({1\over\cosh^2\bar\mu\sin^2\bar\nu}
     \bigg({k_1^2-\viert\over\cos^2\phi}+{k_2^2-\viert\over\sin^2\phi}
    -\viert\bigg)
    +{k_3^2-\viert\over\sinh^2\bar\mu\cos^2\bar\nu}\Bigg)\Bigg]dt\Bigg\}
                  \\   & &\!\!\!\!\!\!\!\!
  =(\bar d^2\cosh\bar\mu'\cosh\bar\mu''\sin\bar\nu'\sin\bar\nu'')^{-1/2}
   \sum_{n=0}^\infty
  \Phi_n^{(\pm k_2,\pm k_1)}(\phi')\Phi_n^{(\pm k_2,\pm k_1)}(\phi'')
         \nonumber\\   & &\!\!\!\!\!\!\!\!\qquad\times
  \int_{\bbbr}{dE\over2\pi\hbar}\e^{-\i ET/\hbar}\int_0^\infty ds''
 \int\limits_{\bar\mu(0)=\bar\mu'}^{\bar\mu(s'')=\bar\mu''}\CD\bar\mu(s)
 \int\limits_{\bar\nu(0)=\bar\nu'}^{\bar\nu(s'')=\bar\nu''}\CD\bar\nu(s)
         \nonumber\\   & &\!\!\!\!\!\!\!\!\qquad\times
  \exp\Bigg\{\ih\int_0^{s''}
  \Bigg[{M\over2}\Big((\dot{\bar\mu}^2+\dot{\bar\nu}^2)
 -\omega^2(\cosh^2\bar\mu\sinh^2\bar\mu+\sin^2\bar\nu\cos^2\bar\nu)\Big)
      +E\bar d^2(\cosh^2\bar\mu-\sin^2\bar\nu)
         \nonumber\\   & &\!\!\!\!\!\!\!\!\qquad\qquad\qquad
   -\hbarm\Bigg((\lambda_1^2-\bviert)\bigg({1\over\cos^2\bar\nu}
              -{1\over\sinh^2\bar\mu}\bigg)
  +(k_3^2-\bviert)\bigg({1\over\sin^2\bar\nu}
              -{1\over\cosh^2\bar\mu}\bigg)
  \Bigg)\Bigg]ds\Bigg\}\qquad\quad
                  \\   & &\!\!\!\!\!\!\!\!
  \underline{\hbox{Prolate Spheroidal Coordinates:}}
         \nonumber\\   & &\!\!\!\!\!\!\!\!
  =\int\limits_{\mu(t')=\mu'}^{\mu(t'')=\mu''}\CD\mu(t)
   \int\limits_{\nu(t')=\nu'}^{\nu(t'')=\nu''}\CD\nu(t)
   d^3(\sinh^2\mu+\sin^2\nu)\sinh\mu\sin\nu
  \int\limits_{\phi(t')=\phi'}^{\phi(t'')=\phi''}\CD \phi(t)
         \nonumber\\   & &\!\!\!\!\!\!\!\!\qquad\times
  \exp\Bigg\{\ih\int_{t'}^{t''}\Bigg[{M\over2}d^2
     \Big((\sinh^2\mu+\sin^2\nu)(\dot\mu^2+\dot\nu^2)
        +\sinh^2\mu\sin^2\nu\dot\phi^2
         \nonumber\\   & &\!\!\!\!\!\!\!\!\qquad\qquad\qquad
      -\omega^2(\sinh^2\mu\sin^2\nu+\cosh^2\mu\cos^2\nu)\Big)
         \nonumber\\   & &\!\!\!\!\!\!\!\!\qquad\qquad\qquad
   -{\hbar^2\over2Md^2}\Bigg({1\over\sinh^2\mu\sin^2\nu}
    \bigg({k_1^2-\viert\over\cos^2\phi}+{k_2^2-\viert\over\sin^2\phi}
    -\viert\bigg)
    +{k_3^2-\viert\over\cosh^2\mu\cos^2\nu}\Bigg)\Bigg]dt\Bigg\}
                  \\   & &\!\!\!\!\!\!\!\!
  \underline{\hbox{Ellipsoidal Coordinates:}}
         \nonumber\\   & &\!\!\!\!\!\!\!\!
  =\int\limits_{\rho_1(t')=\rho_1'}^{\rho_1(t'')=\rho_1''}\CD\rho_1(t)
   \int\limits_{\rho_2(t')=\rho_2'}^{\rho_2(t'')=\rho_2''}\CD\rho_2(t)
   \int\limits_{\rho_3(t')=\rho_3'}^{\rho_3(t'')=\rho_3''}\CD \rho_3(t)
   {(\rho_2-\rho_1)(\rho_3-\rho_2)(\rho_3-\rho_1)\over
             8\sqrt{-P(\rho_1)P(\rho_2)P(\rho_3)}}
         \nonumber\\   & &\!\!\!\!\!\!\!\!\qquad\times
  \exp\Bigg\{\ih\sum_{i=1}^3\int_{t'}^{t''}
        \Bigg[{M\over2} g_{\rho_i\rho_i}\dot\rho_i^2
   -{1\over\prod_{i\not=j}(\rho_i-\rho_j)}
       \bigg({M\omega^2\over2}P(\rho_i)+\hbarm A(\rho_i)\bigg)
   -\Delta V_{PF}(\rho_i)\Bigg]dt\Bigg\}\enspace.
         \nonumber\\   & &\!\!\!\!\!\!\!\!
\end{eqnarray}
In this example we have explicitly written down the separated $(\bar\xi,
\bar\eta)$-path integrations in the case of the oblate spheroidal
coordinates. The case of the prolate spheroidal coordinates is, of
course, similar. For the separation in the ellipsoidal coordinates
$A(\rho)$ denotes ($a_{ik}=a_i-a_k$)
\begin{equation}
  A(\rho)=a_{31}a_{21}{k_1^2-\viert\over\rho-a_1}
         +a_{12}a_{32}{k_2^2-\viert\over\rho-a_2}
         +a_{13}a_{23}{k_3^2-\viert\over\rho-a_3}\enspace.
\end{equation}
The separation of variables then is performed by means of (\ref{numbc}).
 In the case of the conical coordinates the separation of variables in
the path integral becomes obvious if we consider for fixed $R$ the
identity
\begin{eqnarray}       & &
  \int\limits_{\alpha(t')=\alpha'}^{\alpha(t'')=\alpha''}\CD\alpha(t)
  \int\limits_{\beta(t')=\beta'}^{\beta(t'')=\beta''}\CD\beta(t)
  (k^2\cn^2\alpha+{k'}^2\cn^2\beta)
         \nonumber\\   & & \quad\times
  \exp\Bigg\{\ih\int^{t''}_{t'}\Bigg[{M\over2}
   R^2(k^2\cn^2\alpha+{k'}^2\cn^2\beta)(\dot\alpha^2+\dot\beta^2)
         \nonumber\\   & & \qquad\qquad\qquad\qquad\qquad
     -{\hbar^2\over2MR^2}\Bigg({k_1^2-\viert\over\sn^2\alpha\dn^2\beta}
         +{k_2^2-\viert\over\cn^2\alpha\cn^2\beta}
         +{k_3^2-\viert\over\dn^2\alpha\sn^2\beta}\Bigg)\Bigg]dt\Bigg\}
         \nonumber\\   & &
  =\int_{\bbbr}{dE\over2\pi\hbar}\e^{-\i ET/\hbar}\int_0^\infty ds''
  \int\limits_{\alpha(0)=\alpha'}^{\alpha(s'')=\alpha''}\CD\alpha(s)
  \int\limits_{\beta(0)=\beta'}^{\beta(s'')=\beta''}\CD\beta(s)
         \nonumber\\   & & \quad\times
  \exp\Bigg\{\ih\int_0^{s''}\Bigg[{M\over2}R^2(\dot\alpha^2+\dot\beta^2)
     -{\hbar^2\over2MR^2}\Bigg(
 (k_1^2-\bviert)\bigg({1\over\sn^2\alpha}-{k^2\over\dn^2\beta}\bigg)
         \nonumber\\   & & \qquad\qquad\qquad\qquad\qquad
 +(k_2^2-\bviert)\bigg({k^2\over\cn^2\beta}+{{k'}^2\over\cn^2\alpha}
 \bigg)
 +(k_3^2-\bviert)\bigg({1\over\sn^2\beta}-{{k'}^2\over\dn^2\alpha}\bigg)
 \Bigg)\Bigg]ds\Bigg\}\enspace.\quad\qquad
\end{eqnarray}

We find that the path integral for this potential can be explicitly
evaluated in the cartesian, circular polar and spherical coordinate
system \cite{CARPa}. The others cannot be explicitly evaluated, but are
however connected through the above path integral identities. We obtain
(c.f.\  for path integral discussions for special cases of $k_{1,2,3}$
e.g.\ Carpio-Bernido and Bernido \cite{CARPa, CBB})
\begin{eqnarray}       & &\!\!\!\!\!\!\!\!\!\!\!\!\!
  K^{(V_1)}({\vec x\,}'',{\vec x\,}';T)
         \nonumber\\   & &\!\!\!\!\!\!\!\!\!\!\!\!\!
  \hbox{\it Cartesian Coordinates:}
         \nonumber\\   & &\!\!\!\!\!\!\!\!\!\!\!\!\!
  =\bigg({M\omega\over\i\hbar\sin\omega T}\bigg)^3
  \prod_{i=1}^3\sqrt{x_i'x_i''}\,
 \exp\bigg[-{M\omega\over2\i\hbar}({x_i'}^2+{x_i''}^2)\cot\omega T\bigg]
  I_{k_i}\bigg({M\omega x_i'x_i''\over\i\hbar\sin\omega T}\bigg)
                  \\   & &\!\!\!\!\!\!\!\!\!\!\!\!\!
 =\sum_{n_1,n_2,n_3=0}^\infty \e^{-\i E_NT/\hbar}
  \Psi_{n_1,n_2,n_3}(x_1'',x_2'',x_3'')
  \Psi_{n_1,n_2,n_3}(x_1',x_2',x_3')
                  \\   & &\!\!\!\!\!\!\!\!\!\!\!\!\!
  \hbox{{\it Circular Polar Coordinates}
        ($\lambda_1=2n\pm k_1\pm k_2+1$):}
         \nonumber\\   & &\!\!\!\!\!\!\!\!\!\!\!\!\!
 =\bigg({M\omega\over\i\hbar\sin\omega T}\bigg)^2\sqrt{z'z''}\,
 \exp\bigg[-{M\omega\over2\i\hbar}({z'}^2+{z''}^2)\cot\omega T\bigg]
  I_{\pm k_3}\bigg({M\omega z'z''\over\i\hbar\sin\omega T}\bigg)
         \nonumber\\   & &\!\!\!\!\!\!\!\!\!\!\!\!\!\qquad\times
  \sum_{n=0}^\infty
  \Phi_n^{(\pm k_2,\pm k_1)}(\phi')\Phi_n^{(\pm k_2,\pm k_1)}(\phi'')
  \exp\bigg[-{M\omega\over2\i\hbar}
  ({\rho'}^2+{\rho''}^2)\cot\omega T\bigg]
  I_{\lambda_1}\bigg({M\omega\rho'\rho''\over\i\hbar\sin\omega T}\bigg)
                  \\   & &\!\!\!\!\!\!\!\!\!\!\!\!\!
  =\sum_{n,m,n_z=0}^\infty \e^{-\i E_NT/\hbar}
  \Psi_{n,m,n_z}(\phi',\rho',z')\Psi_{n,m,n_z}(\phi'',\rho'',z'')
                  \\   & &\!\!\!\!\!\!\!\!\!\!\!\!\!
  \hbox{{\it Spherical Coordinates}
        ($\lambda_2=2m+\lambda_1\pm k_3+1$):}
         \nonumber\\   & &\!\!\!\!\!\!\!\!\!\!\!\!\!
  =(r'r''\sin\theta'\sin\theta'')^{-1/2}\sum_{n=0}^\infty
  \Phi_n^{(\pm k_2,\pm k_1)}(\phi'')\Phi_n^{(\pm k_2,\pm k_1)}(\phi')
  \sum_{m=0}^\infty\Phi^{(\lambda_1,\pm k_3)}_m(\theta'')
                   \Phi^{(\lambda_1,\pm k_3)}_m(\theta')
         \nonumber\\   & &\!\!\!\!\!\!\!\!\!\!\!\!\!\qquad\times
 {M\omega\over\i\hbar\sin\omega T}
 \exp\bigg[-{M\omega\over2\i\hbar}({r'}^2+{r''}^2)\cot\omega T\bigg]
 I_{\lambda_2}\bigg({Mr'r''\over\i\hbar\sin\omega T}\bigg)
                  \\   & &\!\!\!\!\!\!\!\!\!\!\!\!\!
 =\sum_{l,n,m=0}^\infty \e^{-\i E_NT/\hbar}
  \Psi_{l,n,m}(\theta',\phi',r')\Psi_{l,n,m}(\theta'',\phi'',r'')
  \enspace,
\end{eqnarray}
with the wave-functions and the energy-spectra given by
($N$ is the principal quantum number)
\begin{eqnarray}       & &\!\!\!\!\!\!
  \hbox{\it Cartesian Coordinates:}
         \nonumber\\   & &\!\!\!\!\!\!
  \Psi_{n_1,n_2,n_3}(x_1,x_2,x_3)=
  \bigg({2M\omega\over\hbar}\bigg)^{3/2}\prod_{i=1}^3
  \bigg({M\omega\over\hbar}\bigg)^{\pm k_i/2}
         \nonumber\\   & &\!\!\!\!\!\!\qquad\qquad\times
  \sqrt{n_i!\over\Gamma(n_i\pm k_i+1)}\,
  x_i^{1/2\pm k_i}\exp\bigg(-{M\omega\over2\hbar}x_i^2\bigg)
  L_{n_i}^{(\pm k_i)}\bigg({M\omega\over\hbar}x_i^2\bigg)\enspace,
                  \\   & &\!\!\!\!\!\!
  E_N=\hbar\omega\bigg(2N+\sum_{i=1}^3(1\pm k_i)\bigg)\enspace,
      \qquad N=n_1+n_2+n_3\enspace,
                  \\   & &\!\!\!\!\!\!
  \hbox{{\it Circular Polar Coordinates}
        ($\lambda_1=2n\pm k_1\pm k_2+1$):}
         \nonumber\\   & &\!\!\!\!\!\!
  \Psi_{n,m,n_z}(\phi,\rho,z)=
  \Phi_n^{(\pm k_2,\pm k_1)}(\phi)\sqrt{2M\omega\over\hbar}
  \bigg({M\omega\over\hbar}\bigg)^{\pm k_3/2}
         \nonumber\\   & &\!\!\!\!\!\!\qquad\qquad\times
  \sqrt{n_z!\over\Gamma(n_z\pm k_3+1)}
  z^{1/2\pm k_3}\exp\bigg(-{M\omega\over2\hbar}z^2\bigg)
  L_{n_z}^{(\pm k_3)}\bigg({M\omega\over\hbar}z^2\bigg)
         \nonumber\\   & &\!\!\!\!\!\!\qquad\qquad\times
  \sqrt{2M\omega\over\hbar}
  \bigg({M\omega\over\hbar}\rho^2\bigg)^{\lambda_1/2}
  \sqrt{m!\over\Gamma(m+\lambda_1+1)}
  \exp\bigg(-{M\omega\over2\hbar}\rho^2\bigg)
  L_{m}^{(\lambda_1)}\bigg({M\omega\over\hbar}\rho^2\bigg)\enspace,
                  \\   & &\!\!\!\!\!\!
  E_N=\hbar\omega\bigg(2N+\sum_{i=1}^3(1\pm k_i)\bigg)\enspace,
      \qquad N=m+n+n_z\enspace,
                  \\   & &\!\!\!\!\!\!
  \hbox{{\it Spherical Coordinates}
         ($\lambda_2=2m+\lambda_1\pm k_3+1$):}
         \nonumber\\   & &\!\!\!\!\!\!
  \Psi_{l,n,m}(\theta,\phi,r)=(r\sin\theta)^{-1/2}
  \Phi_n^{(\pm k_2,\pm k_1)}(\phi)\Phi_m^{(\lambda_1,\pm k_3)}(\theta)
         \nonumber\\   & &\!\!\!\!\!\!\qquad\qquad\times
  \sqrt{2M\omega\over\hbar}
  \bigg({M\omega\over\hbar} r^2\bigg)^{\lambda_2/2}
  \sqrt{l!\over\Gamma(l+\lambda_2+1)}
  \exp\bigg(-{M\omega\over2\hbar} r^2\bigg)
  L_l^{(\lambda_2)}\bigg({M\omega\over\hbar} r^2\bigg)\enspace,\qquad
                  \\   & &\!\!\!\!\!\!
  E_N=\hbar\omega\bigg(2N+\sum_{i=1}^3(1\pm k_i)\bigg)\enspace,
      \qquad N=n+m+l\enspace,
\end{eqnarray}
and $\Phi^{(\pm k_2,\pm k_1)}_n(\phi)$  and $\Phi^{(\lambda_1,\pm
k_3)}_m(\theta)$ denote the wave-functions (\ref{numca}), c.f.\ Calogero
\cite{CALO} and Evans \cite{EVA, EVAb} for the Schr\"odinger approach.
Note that our solution in circular polar coordinates is new. The
$D$-dimensional generalization of this potential is also maximally
super-integrable \cite{EVAb}, and it is immediately obvious that it is
at least separable (and exactly soluble) in cartesian and spherical
coordinates (and any combination of circular spherical subsystems).

Let us furthermore consider the two following special cases of the
potential $V_1(\vec x)$ where $\omega=0$. Here we have a continuous
spectrum instead of a discrete one.

\medskip\noindent
(i) $k_3^2-\viert=0$: We have ($\lambda=2n\pm k_1\pm k_2+1$)
\begin{eqnarray}       & &\!\!\!\!\!\!
  K^{(V_1)}({\vec x\,}'',{\vec x\,}';T)
         \nonumber\\   & &\!\!\!\!\!\!
  \underline{\hbox{Paraboloidal Coordinates:}}
         \nonumber\\   & &\!\!\!\!\!\!\!\!
  =\int\limits_{\eta_1(t')=\eta_1'}^{\eta_1(t'')=\eta_1''}\CD\eta_1(t)
   \int\limits_{\eta_2(t')=\eta_2'}^{\eta_2(t'')=\eta_2''}\CD\eta_2(t)
   \int\limits_{\eta_3(t')=\eta_3'}^{\eta_3(t'')=\eta_3''}\CD \eta_3(t)
   {(\eta_2-\eta_1)(\eta_3-\eta_2)(\eta_3-\eta_1)\over
             8\sqrt{-P(\eta_1)P(\eta_2)P(\eta_3)}}
         \nonumber\\   & &\!\!\!\!\!\!\!\!\qquad\times
  \exp\Bigg\{\ih\sum_{i=1}^3\int_{t'}^{t''}
        \Bigg[{M\over2} g_{\eta_i\eta_i}\dot\eta_i^2
   -\hbarm{B(\eta_i)\over\prod_{i\not=j}(\eta_i-\eta_j)}
   -\Delta V_{PF}(\eta_i)\Bigg]dt\Bigg\}
                  \\   & &\!\!\!\!\!\!
  \underline{\hbox{Parabolic Coordinates:}}
         \nonumber\\   & &\!\!\!\!\!\!
  =\int\limits_{\eta(t')=\eta'}^{\eta(t'')=\eta''}\CD\eta(t)
   \int\limits_{\xi(t')=\xi'}^{\xi(t'')=\xi''}\CD\xi(t)
   (\xi^2+\eta^2)\xi\eta
   \int\limits_{\phi(t')=\phi'}^{\phi(t'')=\phi''}\CD\phi(t)
         \nonumber\\   & &\!\!\!\!\!\!\!\!\qquad\times
  \exp\Bigg\{\ih\int_{t'}^{t''}\Bigg[{M\over2}\Big(
     (\xi^2+\eta^2)(\dot\xi^2+\dot\eta^2)+\xi^2\eta^2\dot\phi^2\Big)
     -{\hbar^2\over2M\xi^2\eta^2}\Bigg({k_1^2-\viert\over\cos^2\phi}
      +{k_2^2-\viert\over\sin^2\phi}-\viert\Bigg)\Bigg]dt\Bigg\}
         \nonumber\\   & &\!\!\!\!\!\!
                  \\   & &\!\!\!\!\!\!\!\!
  =(\xi'\xi''\eta'\eta'')^{-1/2}\sum_{n=0}^\infty
   \Phi_n^{(\pm k_2,\pm k_1)}(\phi')\Phi_n^{(\pm k_2,\pm k_1)}(\phi'')
         \nonumber\\   & &\!\!\!\!\!\!\!\!\qquad\times
   \int_{\bbbr}{dE\over2\pi\hbar}\e^{-\i ET/\hbar}\int_0^\infty ds''
  \int\limits_{\eta(0)=\eta'}^{\eta(s'')=\eta''}\CD\eta(s)
  \int\limits_{\xi(0)=\xi'}^{\xi(s'')=\xi''}\CD\xi(s)
         \nonumber\\   & &\!\!\!\!\!\!\!\!\qquad\times
  \exp\Bigg\{\ih\int_0^{s''}\Bigg[{M\over2}\Big(
   (\dot\xi^2+\dot\eta^2)+E(\xi^2+\eta^2)
     -\hbar^2{\lambda^2-\viert\over2M}\bigg({1\over\xi^2}+{1\over\eta^2}
      \bigg)\Bigg]ds\Bigg\}\enspace,
                  \\   & &\!\!\!\!\!\!\!\!
  =\sum_{n=0}^\infty \int_{\bbbr}d\zeta\int_0^\infty dp\,
  \e^{-\i\hbar p^2T/2M}\Psi_{n,\zeta,p}^*(\phi',\xi',\eta')
  \Psi_{n,\zeta,p}(\phi'',\xi'',\eta'')\enspace.
  \end{eqnarray}
In paraboloidal coordinates $\Delta V_{PF}(\eta_i)$ are determined by
(\ref{numba}) and the $B(\eta_i)$ are given by ($i=1,2,3$, $a_{ik}=
a_i-a_k$ $a_1=b$, $a_2=a$)
\begin{equation}
  B(\eta_i)=a_{21}{k_1^2-\viert\over\eta_i-a}
           +a_{12}{k_2^2-\viert\over\eta_i-b}\enspace.
\end{equation}
The wave-functions in {\it parabolic coordinates\/} are given by
($\zeta$ is the parabolic separation constant)
\begin{eqnarray}
  \Psi_{n,\zeta,p}(\phi,\xi,\eta)
  &=&\Phi_n^{(\pm k_2,\pm k_1)}(\phi)
  {\big|\Gamma[{1+\lambda\over2}+\i\zeta/2p]\big|^2
   \over2\pi\sqrt{p}\,\Gamma^2(1+\lambda)}\e^{\pi/2ap}
         \nonumber\\   & &\qquad\qquad\times
  M_{-\i\zeta/2p,\lambda/2} (-\i p\xi^2)
  M_{-\i\zeta/2p,\lambda/2} (-\i p\eta^2)\enspace.
  \end{eqnarray}

\eject\noindent
(ii) $k_2^2-\viert=0$: We have in {\it circular parabolic coordinates}
\begin{eqnarray}       & &\!\!\!\!\!\!
  K^{(V_1)}({\vec x\,}'',{\vec x\,}';T)
         \nonumber\\   & &\!\!\!\!\!\!
  =\int\limits_{\eta(t')=\eta'}^{\eta(t'')=\eta''}\CD\eta(t)
   \int\limits_{\xi(t')=\xi'}^{\xi(t'')=\xi''}\CD\xi(t)
   (\xi^2+\eta^2)\int\limits_{z(t')=z'}^{z(t'')=z''}\CD z(t)
         \nonumber\\   & &\!\!\!\!\!\!\!\!\qquad\times
  \exp\Bigg\{\ih\int_{t'}^{t''}\Bigg[{M\over2}\Big(
     (\xi^2+\eta^2)(\dot\xi^2+\dot\eta^2)+\dot z^2\Big)
     -\hbarm\Bigg({k_1^2-\viert\over\xi^2\eta^2}
      +{k_3^2-\viert\over z^2}\Bigg)\Bigg]dt\Bigg\}
                  \\   & &\!\!\!\!\!\!\!\!
  ={M\over\i\hbar T}\exp\bigg[{\i M\over2\hbar}({z'}^2+{z''}^2)\bigg]
   I_{\pm k_3}\bigg({Mz'z''\over\i\hbar T}\bigg)
         \nonumber\\   & &\!\!\!\!\!\!\!\!\qquad\times
   \int_{\bbbr}{dE\over2\pi\hbar}\e^{-\i ET/\hbar}\int_0^\infty ds''
  \int\limits_{\eta(0)=\eta'}^{\eta(s'')=\eta''}\CD\eta(s)
  \int\limits_{\xi(0)=\xi'}^{\xi(s'')=\xi''}\CD\xi(s)
         \nonumber\\   & &\!\!\!\!\!\!\!\!\qquad\times
  \exp\Bigg\{\ih\int_0^{s''}\Bigg[{M\over2}\Big(
   (\dot\xi^2+\dot\eta^2)+E(\xi^2+\eta^2)
     -\hbar^2{k_1^2-\viert\over2M}\bigg({1\over\xi^2}+{1\over\eta^2}
      \bigg)\Bigg]ds\Bigg\}
                  \\   & &\!\!\!\!\!\!\!\!
  =\int_0^\infty dp_z\int_{\bbbr}d\zeta\int_0^\infty dp\,
  \e^{-\i ET/\hbar}\Psi_{p_z,\zeta,p}^*(z',\xi',\eta')
  \Psi_{p_z,\zeta,p}(z'',\xi'',\eta'')\enspace,
  \end{eqnarray}
with the wave-functions given by ($\zeta$ is the parabolic separation
constant)
\begin{eqnarray}
  \Psi_{n,\zeta,p}(\phi,\xi,\eta)
  &=&\sqrt{p_zz}\,J_{\pm k_3}(p_zz)
  {\big|\Gamma[{1\pm k_1\over2}+\i\zeta/2p]\big|^2
   \over2\pi\sqrt{p\xi\eta}\,\Gamma^2(1\pm k_1)}\e^{\pi/2ap}
         \nonumber\\   & &\qquad\qquad\qquad\qquad\times
  M_{-\i\zeta/2p,\pm k_1/2}(-\i p\xi^2)
  M_{-\i\zeta/2p,\pm k_1/2}(-\i p\eta^2)\enspace,
                  \\
  E&=&\hbarm(p_z^2+p^2)\enspace.
\end{eqnarray}

\vglue0.4truecm\noindent
{\it 3.2.2.}~We consider the potential ($x_3\equiv z\in\bbbr$,
$k_{1,2}>0$)
\begin{equation}
  V_2(\vec x)={M\over2}\omega^2(x_1^2+x_2^2+4x_3^2)
     +\hbarm\Bigg({k_1^2-\viert\over x_1^2}
                   +{k_2^2-\viert\over x_2^2}\Bigg)\enspace.
\end{equation}
We write down the corresponding path integral formulations
\begin{eqnarray}       & &\!\!\!\!\!\!\!\!
  K^{(V_2)}({\vec x\,}'',{\vec x\,}';T)
         \nonumber\\   & &\!\!\!\!\!\!\!\!
  \underline{\hbox{Cartesian Coordinates:}}
         \nonumber\\   & &\!\!\!\!\!\!\!\!
  =\int\limits_{\vec x(t')=\vec x'}^{\vec x(t'')=\vec x''}\CD\vec x(t)
  \exp\Bigg\{\ih\int_{t'}^{t''}\Bigg[{M\over2}
     \Big({\dot{\vec x}}^2-\omega^2(x_1^2+x_2^2+4x_3^2)\Big)
     -\hbarm\Bigg({k_1^2-\viert\over x_1^2}
     +{k_2^2-\viert\over x_2^2}\Bigg)\Bigg]dt\Bigg\}
         \nonumber\\   & &\!\!\!\!\!\!\!\!
                  \\   & &\!\!\!\!\!\!\!\!
  \hbox{$\underline{\hbox{Parabolic Coordinates}}$
        ($\lambda=2n\pm k_1\pm k_2+1$):}
         \nonumber\\   & &\!\!\!\!\!\!\!\!
  =\int\limits_{\eta(t')=\eta'}^{\eta(t'')=\eta''}\CD\eta(t)
   \int\limits_{\xi(t')=\xi'}^{\xi(t'')=\xi''}\CD\xi(t)
   (\xi^2+\eta^2)\xi\eta
   \int\limits_{\phi(t')=\phi'}^{\phi(t'')=\phi''}\CD\phi(t)
         \nonumber\\   & &\!\!\!\!\!\!\!\!\qquad\times
  \exp\Bigg\{\ih\int_{t'}^{t''}\Bigg[{M\over2}\Big(
     (\xi^2+\eta^2)(\dot\xi^2+\dot\eta^2)+\xi^2\eta^2\dot\phi^2
     -\omega^2\big(\xi^2\eta^2+(\xi^2-\eta^2)^2\big)\Big)
         \nonumber\\   & &\!\!\!\!\!\!\!\!
  \qquad\qquad\qquad\qquad\qquad\qquad\qquad\qquad
     -{\hbar^2\over2M\xi^2\eta^2}\Bigg({k_1^2-\viert\over\cos^2\phi}
      +{k_2^2-\viert\over\sin^2\phi}-\viert\Bigg)\Bigg]dt\Bigg\}
                  \\   & &\!\!\!\!\!\!\!\!
  =(\xi'\xi''\eta'\eta'')^{-1/2}\sum_{n=0}^\infty
  \Phi_n^{(\pm k_2,\pm k_1)}(\phi')\Phi_n^{(\pm k_2,\pm k_1)}(\phi'')
         \nonumber\\   & &\!\!\!\!\!\!\!\!\qquad\times
   \int_{\bbbr}{dE\over2\pi\hbar}\e^{-\i ET/\hbar}\int_0^\infty ds''
  \int\limits_{\eta(0)=\eta'}^{\eta(s'')=\eta''}\CD\eta(s)
  \int\limits_{\xi(0)=\xi'}^{\xi(s'')=\xi''}\CD\xi(s)
         \nonumber\\   & &\!\!\!\!\!\!\!\!\qquad\times
  \exp\Bigg\{\ih\int_0^{s''}\Bigg[{M\over2}\Big(
   (\dot\xi^2+\dot\eta^2)-\omega^2(\xi^6+\eta^6)\Big)
     +E(\xi^2+\eta^2)
     -\hbar^2{\lambda^2-\viert\over2M}\bigg({1\over\xi^2}+{1\over\eta^2}
      \bigg)\Bigg]ds\Bigg\}
         \nonumber\\   & &
                  \\   & &\!\!\!\!\!\!\!\!
  \underline{\hbox{Circular Polar Coordinates:}}
         \nonumber\\   & &\!\!\!\!\!\!\!\!
  =\int\limits_{\rho(t')=\rho'}^{\rho(t'')=\rho''}\rho\CD\rho(t)
   \int\limits_{\phi(t')=\phi'}^{\phi(t'')=\phi''}\CD\phi(t)
   \int\limits_{z(t')=z'}^{z(t'')=z''}\CD z(t)
         \nonumber\\   & &\!\!\!\!\!\!\!\!\qquad\times
  \exp\left\{\ih\int_{t'}^{t''}\left[{M\over2}
     \Big(\dot\rho^2+\rho^2\dot\phi^2+\dot z^2
     -\omega^2(\rho^2+4z^2)\Big)
     -{\hbar^2\over2M\rho^2}\Bigg({k_1^2-\viert\over\cos^2\phi}
         +{k_2^2-\viert\over\sin^2\phi}-\viert\Bigg)\right]dt\right\}
         \nonumber\\   & &
                  \\   & &\!\!\!\!\!\!\!\!
  \underline{\hbox{Circular Elliptic Coordinates:}}
         \nonumber\\   & &\!\!\!\!\!\!\!\!
  =\int\limits_{\xi(t')=\xi'}^{\xi(t'')=\xi''}\CD\xi(t)
   \int\limits_{\eta(t')=\eta'}^{\eta(t'')=\eta''}\CD\eta(t)
   d^2(\sinh^2\xi+\sin^2\eta)
   \int\limits_{z(t')=z'}^{z(t'')=z''}\CD z(t)
         \nonumber\\   & &\!\!\!\!\!\!\!\!\qquad\times
  \exp\Bigg\{\ih\int_{t'}^{t''}\Bigg[{M\over2}d^2
     \Big((\sinh^2\xi+\sin^2\eta)(\dot\xi^2+\dot\eta^2)
         -\omega^2(\cosh^2\xi\cos^2\eta+\sinh^2\xi\sin^2\eta)+4z^2\Big)
         \nonumber\\   & &\!\!\!\!\!\!\!\!
  \qquad\qquad\qquad\qquad\qquad\qquad\qquad\qquad
     -{\hbar^2\over2Md^2}\Bigg({k_1^2-\viert\over\cosh^2\xi\cos^2\eta}
  +{k_2^2-\viert\over\sinh^2\xi\sin^2\eta}\Bigg)\Bigg]dt\Bigg\}\enspace.
  \end{eqnarray}
and $\Phi_n^{(\alpha,\beta)}$ denote the wave-functions (\ref{numca}).
$K^{(V_2)}(T)$ is explicitly soluble in cartesian and circular polar
coordinates by means of the harmonic oscillator in $z$, and the radial
harmonic oscillator in $x_1,x_2$, respectively (note the similarity to
the Holt potential)
\begin{eqnarray}       & &\!\!\!\!\!\!\!\!\!\!\!\!\!\!
  K^{(V_2)}({\vec x\,}'',{\vec x\,}';T)
         \nonumber\\   & &\!\!\!\!\!\!\!\!\!\!\!\!\!\!
  \hbox{\it Cartesian Coordinates:}
         \nonumber\\   & &\!\!\!\!\!\!\!\!\!\!\!\!\!\!
  =\sqrt{M\omega\over\i\hbar\sin2\omega T}
  \exp\bigg\{-{M\omega\over\i\hbar\sin2\omega T}
     \Big[({z'}^2+{z''}^2)\cos2\omega T-2z'z''\Big]\bigg\}
         \nonumber\\   & &\!\!\!\!\!\!\!\!\!\!\!\!\!\!\qquad\times
  \bigg({M\omega\over\i\hbar\sin\omega T}\bigg)^2
  \prod_{i=1}^2\sqrt{x_i'x_i''}\,
 \exp\bigg[-{M\omega\over2\i\hbar}({x_i'}^2+{x_i''}^2)\cot\omega T\bigg]
  I_{k_i}\bigg({M\omega x_i'x_i''\over\i\hbar\sin\omega T}\bigg)
                  \\   & &\!\!\!\!\!\!\!\!\!\!\!\!\!\!
  =\sum_{n_z,n_1,n_2=0}^\infty \e^{-\i E_NT/\hbar}
  \Psi_{n_z,n_1,n_2}(z',x_1',x_2')\Psi_{n_z,n_1,n_2}(z'',x_1'',x_2'')
                  \\   & &\!\!\!\!\!\!\!\!\!\!\!\!\!\!
  \hbox{{\it Circular Polar Coordinates}
                  ($\lambda=2n\pm k_1\pm k_2+1$):}
         \nonumber\\   & &\!\!\!\!\!\!\!\!\!\!\!\!\!\!
  =\sqrt{M\omega\over\i\hbar\sin2\omega T}
  \exp\bigg\{-{M\omega\over\i\hbar\sin2\omega T}
     \Big[({z'}^2+{z''}^2)\cos2\omega T-2z'z''\Big]\bigg\}
         \nonumber\\   & &\!\!\!\!\!\!\!\!\!\!\!\!\!\!\qquad\times
  {M\omega\over\i\hbar\sin\omega T}\sum_{n=0}^\infty
  \Phi_n^{(\pm k_2,\pm k_1)}(\phi')\Phi_n^{(\pm k_2,\pm k_1)}(\phi'')
  \exp\bigg[-{M\omega\over2\i\hbar}\big({\rho'}^2+{\rho''}^2\big)
    \cot\omega T\bigg]
  I_\lambda\bigg({M\omega\rho'\rho''\over\i\hbar\sin\omega T}\bigg)
         \nonumber\\   & &\!\!\!\!\!\!\!\!\!\!\!\!\!\!
  \quad           \\   & &\!\!\!\!\!\!\!\!\!\!\!\!\!\!
  =\sum_{n_z,n,m=0}^\infty \e^{-\i E_NT/\hbar}
  \Psi_{n_z,n,m}(z',\phi',\rho')\Psi_{n_z,n,m}(z'',\phi'',\rho'')
  \enspace,
  \end{eqnarray}
where the wave-functions and the energy-levels are given by
($N$ is the principal quantum number)
\begin{eqnarray}       & &\!\!\!\!\!\!\!\!
  \hbox{\it Cartesian Coordinates:}
         \nonumber\\   & &\!\!\!\!\!\!\!\!
  \Psi_{n_z,n_1,n_2}(x_1,x_2,z)={2M\omega\over\hbar}\prod_{i=1}^2
  \bigg({M\omega\over\hbar}\bigg)^{\pm k_i/2}
  \sqrt{n_i!\over\Gamma(n_i\pm k_i+1)}\,
         \nonumber\\   & &\!\!\!\!\!\!\!\!\qquad\qquad\times
  x_i^{1/2\pm k_i}\exp\bigg(-{M\omega\over2\hbar}x_i^2\bigg)
  L_{n_i}^{(\pm k_i)}\bigg({M\omega\over\hbar}x_i^2\bigg)
         \nonumber\\   & &\!\!\!\!\!\!\!\!\qquad\qquad\times
  \sqrt{\sqrt{2M\omega\over\pi\hbar}{1\over2^{n_z}n_z!}}
  \exp\bigg(-{M\omega\over\hbar}z^2\bigg)
  H_{n_z}\left(\sqrt{2M\omega\over\hbar}\,z\right)\enspace,
                  \\   & &\!\!\!\!\!\!\!\!
  E_N=\hbar\omega(N\pm k_1\pm k_2+\hbox{${5\over2}$})\enspace,
  \qquad N=2(n_1+n_2)+n_z\enspace,
                  \\   & &\!\!\!\!\!\!\!\!
  \hbox{{\it Circular Polar Coordinates}
         ($\lambda=2n\pm k_1\pm k_2+1$):}
         \nonumber\\   & &\!\!\!\!\!\!\!\!
  \Psi_{n_z,n,m}(z,\phi,\rho)=\Phi_n^{(\pm k_2,\pm k_1)}(\phi)
  \sqrt{\sqrt{2M\omega\over\pi\hbar}{1\over2^{n_z}n_z!}}
  \exp\bigg(-{M\omega\over\hbar}z^2\bigg)
  H_{n_z}\left(\sqrt{2M\omega\over\hbar}\,z\right)\qquad\qquad
         \nonumber\\   & &\!\!\!\!\!\!\!\!\qquad\qquad\times
  \sqrt{2M\omega\over\hbar}
  \bigg({M\omega\over\hbar}\rho^2\bigg)^{\lambda/2}
  \sqrt{m!\over\Gamma(m+\lambda+1)}
  \exp\bigg(-{M\omega\over2\hbar}\rho^2\bigg)
  L_m^{(\lambda)}\bigg({M\omega\over\hbar}\rho^2\bigg)\enspace,
                  \\   & &\!\!\!\!\!\!\!\!
  E_N=\hbar\omega(N\pm k_1\pm k_2+\hbox{${5\over2}$})
  \enspace,\qquad N=2(n+m)+n_z\enspace,
  \end{eqnarray}
and the $\Phi_n^{(\pm k_2,\pm k_1)}(\phi)$ denote the wave-functions
(\ref{numca}).

\vglue0.4truecm\noindent
{\it 3.2.3.}~We consider the three-dimensional potential ($k_{1,2}>0$)
\begin{equation}
  V_3(\vec x)=-{\alpha\over\sqrt{x^2+y^2+z^2}}
    +\hbarm\Bigg({k_1^2-\viert\over x^2}
    +{k_2^2-\viert\over y^2}\Bigg)\enspace.
\end{equation}
This potential can be interpreted as a maximally super-integrable
generalization of the Coulomb potential. Note furthermore
that the pure Coulomb potential has the same separating coordinate
systems as $V_3(\vec x)$ and therefore the additional centrifugal terms
do not spoil this property. The pure Coulomb path integral was discussed
by many authors, e.g.\ Castrigiano and St\"ark \cite{CAST}, Chetouani
and Hammann \cite{CHc, CHe}, Duru and Grosche \cite{GROm}, Kleinert
\cite{DKa, DKb}, Inomata \cite{INOb}, Kleinert \cite{KLEh}, Pak and
S\"okmen \cite{PAKSa}, Steiner \cite{STEb}, and Storchak \cite{STORCHc}.
We formulate the path integrals in which this potential is separable
\begin{eqnarray}       & &
  K^{(V_3)}({\vec x\,}'',{\vec x\,}';T)
         \nonumber\\   & &
  =\int\limits_{\vec x(t')=\vec x'}^{\vec x(t'')=\vec x''}\CD\vec x(t)
  \exp\left\{\ih\int_{t'}^{t''}\left[{M\over2}
   {\dot{\vec x}}^2+{\alpha\over r}
 -\hbarm\sum_{i=1}^2{k_i^2-\viert\over x_i^2}\right]dt\right\}
                  \\   & &
  \underline{\hbox{Conical Coordinates:}}
         \nonumber\\   & &
  =\int\limits_{r(t')=r'}^{r(t'')=r''}r^2\CD r(t)
  \int\limits_{\alpha(t')=\alpha'}^{\alpha(t'')=\alpha''}\CD\alpha(t)
  \int\limits_{\beta(t')=\beta'}^{\beta(t'')=\beta''}\CD\beta(t)
  (k^2\cn^2\alpha+{k'}^2\cn^2\beta)
         \nonumber\\   & & \qquad\times
  \exp\Bigg\{\ih\int^{t''}_{t'}\Bigg[{M\over2}\Big(\dot r^2+
   r^2(k^2\cn^2\alpha+{k'}^2\cn^2\beta)
   (\dot\alpha^2+\dot\beta^2)\Big)+{\alpha\over r}
         \nonumber\\   & & \qquad\qquad\qquad\qquad\qquad\qquad\qquad
     -\hbarm\Bigg({k_1^2-\viert\over\sn^2\alpha\dn^2\beta}
         +{k_2^2-\viert\over\cn^2\alpha\cn^2\beta}\Bigg)\Bigg]dt\Bigg\}
                  \\   & &
  \phantom{\underline{\hbox{Spherical Coordinates:}}}
         \nonumber\\   & &
  \phantom{\underline{\hbox{Spherical Coordinates:}}}
         \nonumber\\   & &
  \underline{\hbox{Spherical Coordinates:}}
         \nonumber\\   & &
  =\int\limits_{r(t')=r'}^{r(t'')=r''}r^2\CD r(t)
   \int\limits_{\theta(t')=\theta'}^{\theta(t'')=\theta''}
  \sin\theta\CD\theta(t)
   \int\limits_{\phi(t')=\phi'}^{\phi(t'')=\phi''}\CD\phi(t)
         \nonumber\\   & & \qquad\times
  \exp\Bigg\{\ih\int_{t'}^{t''}\Bigg[{M\over2}
  \big(\dot r^2+r^2\dot\theta^2+r^2\sin^2\theta\dot\phi^2\big)
         \nonumber\\   & & \qquad\qquad\qquad\qquad\qquad\qquad\qquad
  +{\alpha\over r}-{\hbar^2\over2Mr^2\sin^2\theta}
   \Bigg({k_1^2-\viert\over\cos^2\phi}
        +{k_2^2-\viert\over\sin^2\phi}-{1\over4}\Bigg)\Bigg]dt\Bigg\}
                  \\   & &
  \underline{\hbox{Parabolic Coordinates:}}
         \nonumber\\   & &
  =\int\limits_{\eta(t')=\eta'}^{\eta(t'')=\eta''}\CD\eta(t)
   \int\limits_{\xi(t')=\xi'}^{\xi(t'')=\xi''}\CD\xi(t)
   (\xi^2+\eta^2)\xi\eta
   \int\limits_{\phi(t')=\phi'}^{\phi(t'')=\phi''}\CD\phi(t)
         \nonumber\\   & & \qquad\times
  \exp\Bigg\{\ih\int_{t'}^{t''}\Bigg[{M\over2}\Big(
     (\xi^2+\eta^2)(\dot\xi^2+\dot\eta^2)+\xi^2\eta^2\dot\phi^2\Big)
         \nonumber\\   & & \qquad\qquad\qquad\qquad\qquad\qquad\qquad
     +{2\alpha\over\xi^2+\eta^2}
     -{\hbar^2\over2M\xi^2\eta^2}\Bigg({k_1^2-\viert\over\cos^2\phi}
     +{k_2^2-\viert\over\sin^2\phi}-\viert\Bigg)\Bigg]dt\Bigg\}\qquad
                  \\   & &
  \hbox{$\underline{\hbox{Prolate Spheroidal II Coordinates}}$
        ($\lambda_1=2n\pm k_1\pm k_2+1$):}
         \nonumber\\   & &
  =\int\limits_{\mu(t')=\mu'}^{\mu(t'')=\mu''}\CD\mu(t)
   \int\limits_{\nu(t')=\nu'}^{\nu(t'')=\nu''}\CD\nu(t)
   d^3(\sinh^2\mu+\sin^2\nu)\sin\nu\sinh\mu
   \int\limits_{\phi(t')=\phi'}^{\phi(t'')=\phi''}\CD\phi(t)
         \nonumber\\   & & \qquad\times
  \exp\Bigg\{\ih\int_{t'}^{t''}\Bigg[{M\over2}d^2\Big(
  (\sinh^2\mu+\sin^2\nu)(\dot\mu^2+\dot\nu^2)
   +\sinh^2\mu\sin^2\nu\dot\phi^2\Big)
         \nonumber\\   & & \qquad\qquad\qquad\qquad\quad
  +{\alpha\over d(\cosh\mu+\cos\nu)}
  -{\hbar^2\over2Md^2\sinh^2\mu\sin^2\nu}
   \Bigg({k_1^2-\viert\over\cos^2\phi}
        +{k_2^2-\viert\over\sin^2\phi}-{1\over4}\Bigg)\Bigg]dt\Bigg\}
         \nonumber\\   & &
                  \\   & &
  =(d^2\sinh\mu'\sinh\mu''\sin\nu'\sin\nu'')^{-1/2}\sum_{n=0}^\infty
  \Phi_n^{(\pm k_2,\pm k_1)}(\phi'')\Phi_n^{(\pm k_2,\pm k_1)}(\phi')
   \int_{\bbbr}{dE\over2\pi\hbar}\e^{-\i ET/\hbar}\int_0^\infty ds''
         \nonumber\\   & & \qquad\times
   \int\limits_{\mu(0)=\mu'}^{\mu(s'')=\mu''}\CD\mu(s)
   \int\limits_{\nu(0)=\nu'}^{\nu(s'')=\nu''}\CD\nu(s)
   \exp\Bigg\{\ih\int_0^{s''}\Bigg[{M\over2}(\dot\mu^2+\dot\nu^2)
   +Ed^2(\cosh^2\mu-\cos^2\nu)\qquad\quad
         \nonumber\\   & & \qquad\qquad\qquad\qquad\quad
  +\alpha d(\cosh\mu-\cos\nu)
  -\hbarm\Bigg({\lambda_1^2-\viert\over\sinh^2\mu}
       +{\lambda_1^2-\viert\over\sin^2\nu}\Bigg)\Bigg]ds\Bigg\}\enspace.
\end{eqnarray}
We now obtain ($\lambda_1=2n\pm k_1\pm k_2+1$, $\lambda_2=m+\lambda_1+
\half$, $\kappa=\alpha\sqrt{-M/2E}/\hbar$, only the radial Green
function can be explicitly evaluated)
\begin{eqnarray}       & &\!\!\!\!\!\!\!\!\!\!\!\!\!\!\!\!
  \ih\int_0^\infty dT\,\e^{\i TE/\hbar}
  K^{(V_3)}({\vec x\,}'',{\vec x\,}';T)
         \nonumber\\   & &\!\!\!\!\!\!\!\!\!\!\!\!\!\!\!\!
  \hbox{\it Spherical Coordinates:}
         \nonumber\\   & &\!\!\!\!\!\!\!\!\!\!\!\!\!\!\!\!
  =\sum_{n=0}^\infty
  \Phi^{(\pm k_2,\pm k_1)}_n(\phi'')\Phi^{(\pm k_2,\pm k_1)}_n(\phi')
  \sum_{m=0}^\infty(m+\lambda_1+\bhalf){\Gamma(m+\lambda_1+1)\over m!}
  P_{\lambda_1+m}^{-\lambda_1}(\cos\theta'')
  P_{\lambda_1+m}^{-\lambda_1}(\cos\theta')
  \qquad \nonumber\\   & &\!\!\!\!\!\!\!\!\!\!\!\!\!\!\!\! \qquad\times
  {1\over r'r''}{1\over\hbar}\sqrt{-{M\over2E}}
  {\Gamma(\half+\lambda_2-\kappa)\over\Gamma(2\lambda_2+1)}
   W_{\kappa,\lambda_2}\bigg(\sqrt{-8ME}\,{r_>\over\hbar}\bigg)
   M_{\kappa,\lambda_2}\bigg(\sqrt{-8ME}\,{r_<\over\hbar}\bigg)
                  \\   & &\!\!\!\!\!\!\!\!\!\!\!\!\!\!\!\!
  =\sum_{n,m=0}^\infty\left\{\sum_{l=0}^\infty
   {\Psi_{n,m,l}^*(\theta',\phi',r')
    \Psi_{n,m,l}(\theta'',\phi'',r'')\over E_N-E}
   +\int_{\bbbr} dp{\Psi_{n,m,p}^*(\theta',\phi',r')
    \Psi_{n,m,p}^{\alpha)}(\theta'',\phi'',r'')
              \over\hbar^2p^2/2M-E}\right\}
                  \\   & &\!\!\!\!\!\!\!\!\!\!\!\!\!\!\!\!
  \hbox{{\it Parabolic Coordinates} ($\omega=\sqrt{-2E/M}\,)$:}
         \nonumber\\   & &\!\!\!\!\!\!\!\!\!\!\!\!\!\!\!\!
  =\bigg({M\over\i\hbar}\bigg)^2\sum_{n=0}^\infty
  \Phi^{(\pm k_2,\pm k_1)}_n(\phi'')\Phi^{(\pm k_2,\pm k_1)}_n(\phi')
   \int_0^\infty{\omega^2ds''\over\sin^2\omega s''}
   \e^{2\i\alpha s''/\hbar}
         \nonumber\\   & &\!\!\!\!\!\!\!\!\!\!\!\!\!\!\!\! \qquad\times
 I_{\lambda_1}\bigg({M\omega\eta'\eta''\over\i\hbar\sin\omega s''}\bigg)
 I_{\lambda_1}\bigg({M\omega\xi'\xi''\over\i\hbar\sin\omega s''}\bigg)
 \exp\bigg[-{M\omega\over2\i\hbar}
  \big({\xi'}^2+{\xi''}^2+{\eta'}^2+{\eta''}^2\big)
  \cot\omega s''\bigg]\qquad
                  \\   & &\!\!\!\!\!\!\!\!\!\!\!\!\!\!\!\!
  =\sum_{n=0}^\infty\Bigg[\sum_{n_1,n_2=0}^\infty
   {\Psi_{n,n_1,n_2}(\phi',\xi',\eta')
   \Psi_{n,n_1,n_2}(\phi'',\xi'',\eta'')\over E_N-E}
         \nonumber\\   & &\!\!\!\!\!\!\!\!\!\!\!\!\!\!\!\!
  \qquad\qquad\qquad\qquad\qquad\qquad
  +\int_0^\infty dp\int_{\bbbr} d\zeta
  {\Psi_{n,p,\zeta}^*(\phi',\xi',\eta')
  \Psi_{n,p,\zeta}(\phi'',\xi'',\eta'')\over\hbar^2p^2/2M-E}
                 \Bigg]\enspace.
\end{eqnarray}
The discrete state wave-functions in the {\it spherical coordinates\/}
are
\begin{eqnarray}
  \Psi_{n,m,l}(\theta,\phi,r)
  &=&\Phi^{(\pm k_2,\pm k_1)}_n(\phi)
  \sqrt{(m+\lambda_1+\bhalf){\Gamma(m+\lambda_1+1)\over m!}}
  P_{\lambda_1+m}^{-\lambda_1}(\cos\theta)
         \nonumber\\   & & \qquad\times
  {2\over(n+\lambda_1+\half)^2}
  \bigg[{2l!\over a^3(l+\lambda_2+\half)
   \Gamma(l+2\lambda_2+1)}\bigg]^{1/2}
  \bigg({2r\over a(l+\lambda_2+\half)}\bigg)^{\lambda_2}
         \nonumber\\   & & \qquad\times
  \exp\bigg(-{r\over a(l+\lambda_2+\half)}\bigg)
  L_l^{(2\lambda_2)}\bigg({2r\over a(l+\lambda_2+\half)}\bigg)\enspace,
                  \\
  E_N&=&-{M\alpha^2\over\hbar^2 N^2}\enspace,\qquad
    N=l+m+2n\pm k_1\pm k_2+1\enspace.
\end{eqnarray}
The continuous wave-functions are
\begin{eqnarray}
  \Psi_{n,m,p}(\theta,\phi,r)
  &=&\Phi^{(\pm k_2,\pm k_1)}_n(\phi)
  \sqrt{(m+\lambda_1+\bhalf){\Gamma(m+\lambda_1+1)\over m!}}
  P_{\lambda_1+m}^{-\lambda_1}(\cos\theta)
         \nonumber\\   & & \qquad\times
  {\Gamma(\half+\lambda_2-\i/ap)\over
  \sqrt{2\pi}\,r\Gamma(2\lambda_2+1)}\exp\bigg({\pi\over2ap}\bigg)
  M_{\i/ap,\lambda_2}(-2\i pr)\enspace.
\end{eqnarray}
The discrete state wave-functions in {\it parabolic coordinates\/}
are
\begin{eqnarray}
  \Psi_{n,n_1,n_2}(\phi,\xi,\eta)
  &=&\Phi^{(\pm k_2,\pm k_1)}_n(\phi)\bigg[{2\over a^2N^3}\cdot
  {2n_1!n_2!\over\Gamma(n_1+\lambda_1+1)\Gamma(n_2+\lambda_1+1)}
  \bigg]^{1/2}
         \nonumber\\   & & \qquad\times
  \bigg({\xi\eta\over(aN)^2}\bigg)^{\lambda_1}
  \exp\bigg(-{\xi^2+\eta^2\over2aN}\bigg)
  L_{n_1}^{(\lambda_1)}\bigg({\xi^2\over aN}\bigg)
  L_{n_2}^{(\lambda_1)}\bigg({\eta^2\over aN}\bigg)\enspace,
\label{numci}     \\
  E_N&=&-{M\alpha^2\over\hbar^2N^2}\enspace,\qquad
    N=n_1+n_2+2n\pm k_1\pm k_2+1\enspace.
\end{eqnarray}
The continuous wave-functions are
\begin{eqnarray}
  \Psi_{n,p,\zeta}(\phi,\xi,\eta)&=&\Phi^{(\pm k_2,\pm k_1)}_n(\phi)
   {\Gamma[{1+\lambda_1\over2}+{\i\over2p}(1/a+\zeta)]
   \Gamma[{1+\lambda_1\over2}+{\i\over2p}(1/a-\zeta)]
   \over2\pi\sqrt{p}\,\xi\eta\Gamma(1+\lambda_1)\Gamma(1+\lambda_1)}
  \e^{\pi/2ap}
         \nonumber\\   & & \qquad\times
  M_{-\i(1/a+\zeta)/2p,\lambda_1/2} (-\i p\xi^2)
  M_{-\i(1/a-\zeta)/2p,\lambda_1/2} (-\i p\eta^2)\enspace.
\label{numcj}
\end{eqnarray}
For the expansion into the wave-functions one uses (\ref{numef}) and
performs the variable substitution $(p_\xi,p_\eta)\mapsto[(1/a+\zeta)/
2p,(1/a-\zeta)/2p]$, $a=\hbar^2/M\alpha$.
$\zeta$ is the parabolic separation constant.

\vglue0.4truecm\noindent
{\it 3.2.4.}~We consider the potential ($k_{1,2,3}>0$)
\begin{equation}
  V_4(\vec x)=\hbarm\bigg(
  {k_1^2x\over y^2\sqrt{x^2+y^2}}+{k_2^2-\viert\over y^2}
  +{k_3^2-\viert\over z^2}\Bigg)\enspace.
\end{equation}
Note that for $k_3^2-\viert=0$ we obtain a trivial extension
of the $\alpha=0$ case of the two-dimensional maximally super-integrable
potential $V_3(\vec x)$. We formulate the path integral in spherical,
circular polar, circular parabolic and circular elliptic II coordinates,
respectively
\begin{eqnarray}       & &
  K^{(V_4)}({\vec x\,}'',{\vec x\,}';T)
         \nonumber\\   & &
  =\int\limits_{\vec x(t')=\vec x'}^{\vec x(t'')=\vec x''}\CD\vec x(t)
  \exp\left\{\ih\int_{t'}^{t''}\left[{M\over2}
   {\dot{\vec x}}^2-\hbarm\bigg(
  {k_1^2x\over y^2\sqrt{x^2+y^2}}+{k_2^2-\viert\over y^2}
  +{k_3^2-\viert\over z^2}\Bigg)\right]dt\right\}\qquad
                  \\   & &
  \underline{\hbox{Spherical Coordinates:}}
         \nonumber\\   & &
  =\int\limits_{r(t')=r'}^{r(t'')=r''}r^2\CD r(t)
   \int\limits_{\theta(t')=\theta'}^{\theta(t'')=\theta''}
   \sin\theta\CD\theta(t)
   \int\limits_{\phi(t')=\phi'}^{\phi(t'')=\phi''}\CD\phi(t)
         \nonumber\\   & & \qquad\times
  \exp\Bigg\{\ih\int_{t'}^{t''}\Bigg[{M\over2}
  \big(\dot r^2+r^2\dot\theta^2+r^2\sin^2\theta\dot\phi^2\big)
         \nonumber\\   & & \qquad\qquad
  -{\hbar^2\over2Mr^2}\Bigg({1\over\sin^2\theta}\bigg(
         {k_2^2+k_1^2-\viert\over4\sin^2{\phi\over2}}
        +{k_2^2-k_1^2-\viert\over4\cos^2{\phi\over2}}-\viert\Bigg)
        +{k_3^2-\viert\over\cos^2\theta}-{1\over4}\Bigg)\Bigg]dt\Bigg\}
                  \\   & &
  \underline{\hbox{Circular Polar Coordinates:}}
         \nonumber\\   & &
  =\int\limits_{\rho(t')=\rho'}^{\rho(t'')=\rho''}\rho\CD\rho(t)
   \int\limits_{\phi(t')=\phi'}^{\phi(t'')=\phi''}\CD\phi(t)
   \int\limits_{z(t')=z'}^{z(t'')=z''}\CD z(t)
         \nonumber\\   & &\qquad\times
  \exp\Bigg(\ih\int_{t'}^{t''}\Bigg\{{M\over2}
     \big(\dot \rho^2+\rho^2\dot\phi^2+\dot z^2\big)
         \nonumber\\   & &\qquad\qquad\qquad\qquad
     -\hbarm\Bigg[{1\over\rho^2}
   \Bigg({k_2^2-k_1^2-\viert\over4\cos^2{\phi\over2}}
        +{k_2^2+k_1^2-\viert\over4\sin^2{\phi\over2}}-\viert\Bigg)
   +{k_3^2-\viert\over z^2}\Bigg]\Bigg\}dt\Bigg)
                  \\   & &
  \underline{\hbox{Circular Elliptic II Coordinates:}}
         \nonumber\\   & &
   =\int\limits_{\xi(t')=\xi'}^{\xi(t'')=\xi''}\CD\xi(t)
   \int\limits_{\eta(t')=\eta'}^{\eta(t'')=\eta''}\CD\eta(t)
   d^2(\sinh^2\xi+\sin^2\eta)
   \int\limits_{z(t')=z'}^{z(t'')=z''}\CD z(t)
         \nonumber\\   & &\qquad\times
  \exp\Bigg\{\ih\int_{t'}^{t''}\Bigg[{M\over2}
     \Big((\sinh^2\xi^2+\sin^2\eta)(\dot\xi^2+\dot\eta^2)+\dot z^2\Big)
     +{\alpha\over d(\cosh\xi+\cos\eta)}
         \nonumber\\   & &\qquad\qquad
     -\hbarm\Bigg({1\over d^2\sinh^2\xi\sin^2\eta}
      \bigg(k_1^2{\cosh\xi\cos\eta+1\over\cosh\xi+\cos\eta}
      +k_2^2-\half\bigg)
     +{k_3^2-\viert\over z^2}\Bigg)\Bigg]dt\Bigg\}
                  \\   & &
  \underline{\hbox{Circular Parabolic Coordinates:}}
         \nonumber\\   & &
  =\int\limits_{\eta(t')=\eta'}^{\eta(t'')=\eta''}\CD\eta(t)
   \int\limits_{\xi(t')=\xi'}^{\xi(t'')=\xi''}\CD\xi(t)
   (\xi^2+\eta^2)
   \int\limits_{z(t')=z'}^{z(t'')=z''}\CD z(t)
         \nonumber\\   & & \qquad\times
  \exp\Bigg\{\ih\int_{t'}^{t''}\Bigg[{M\over2}\Big(
     (\xi^2+\eta^2)(\dot\xi^2+\dot\eta^2)+\dot z^2\Big)
         \nonumber\\   & & \qquad\qquad\qquad\qquad\qquad\qquad
     -\hbarm\Bigg(
     {k_1^2(\eta^2-\xi^2)\over\xi^2\eta^2(\xi^2+\eta^2)}
     +{k_2^2-\viert\over\xi^2\eta^2}
     +{k_3^2-\viert\over z^2}\Bigg)\Bigg]dt\Bigg\}\enspace.
\end{eqnarray}
We obtain for the path integral solutions ($\lambda_{\pm}^2=k_2^2\pm
k_1^2$, $\lambda_1=n+(\lambda_++\lambda_-+1)/2$, $\lambda_2=
2m+\lambda_1\pm k_3+1$)
\eject\noindent
\begin{eqnarray}       & &\!\!\!\!\!\!\!\!
  K^{(V_4)}({\vec x\,}'',{\vec x\,}';T)
         \nonumber\\   & &\!\!\!\!\!\!\!\!
  \hbox{\it Spherical Coordinates:}
         \nonumber\\   & &\!\!\!\!\!\!\!\!
  =\half(r'r''\sin\theta'\sin\theta'')^{-1/2}
   \sum_{n=0}^\infty
   \Psi^{(\lambda_+,\lambda_-)}_n\bigg({\phi'\over2}\bigg)
   \Psi^{(\lambda_+,\lambda_-)}_n\bigg({\phi''\over2}\bigg)
         \nonumber\\   & &\!\!\!\!\!\!\!\! \qquad\times
  {M\over\i\hbar T}\exp\bigg[{\i M\over2\hbar T}({r'}^2+{r''}^2)\bigg]
  \sum_{m=0}^\infty \Phi^{(\lambda_1,\pm k_3)}_m(\theta')
  \Phi^{(\lambda_1,\pm k_3)}_m(\theta'')
  I_{\lambda_2}\bigg({Mr'r''\over\i\hbar T}\bigg)
                  \\   & &\!\!\!\!\!\!\!\!
  =\sum_{n,m=0}^\infty\int_0^\infty dp\,\e^{-\i E_pT/\hbar}
  \Psi_{n,m,p}^*(\theta',\phi',r')
  \Psi_{n,m,p}(\theta'',\phi'',r'')
                  \\   & &\!\!\!\!\!\!\!\!
  \hbox{\it Circular Polar Coordinates:}
         \nonumber\\   & &\!\!\!\!\!\!\!\!
  =\bigg({M\over\i\hbar T}\bigg)^2\sqrt{z'z''}
  \exp\bigg[{\i M\over2\hbar}({z'}^2+{z''}^2+{\rho'}^2+{\rho''}^2)\bigg]
  I_{\pm k_3}\bigg({Mz'z''\over\i\hbar T}\bigg)
         \nonumber\\   & &\!\!\!\!\!\!\!\! \qquad\times
  \half\sum_{n=0}^\infty
  \Phi_n^{(\lambda_+,\lambda_-)}\bigg({\phi'\over2}\bigg)
  \Phi_n^{(\lambda_+,\lambda_-)}\bigg({\phi''\over2}\bigg)
  I_{\lambda_1}\bigg({M\rho'\rho''\over\i\hbar T}\bigg)
                  \\   & &\!\!\!\!\!\!\!\!
  =\sum_{n=0}^\infty\int_0^\infty dp_z\int_0^\infty dp\,
  \e^{-\i E_{p_z,p}T/\hbar}
  \Psi_{p_z,n,p}^*(\phi',\rho',z')
  \Psi_{p_z,n,p}(\phi'',\rho'',z'')
                  \\   & &\!\!\!\!\!\!\!\!
  \hbox{\it Circular Parabolic Coordinates:}
         \nonumber\\   & &\!\!\!\!\!\!\!\!
  ={M\over\i\hbar T}\sqrt{z'z''}
   \exp\bigg[{\i M\over2\hbar T}({z'}^2+{z''}^2)\bigg]
  I_{\pm k_3}\bigg({Mz'z''\over\i\hbar T}\bigg)
         \nonumber\\   & &\!\!\!\!\!\!\!\! \qquad\times
  \int_{\bbbr} d\zeta\int_0^\infty{dp\over p}
  {\big|\Gamma\big({1+\lambda_+\over2}+{\i\zeta\over2p})\big|^2
   \big|\Gamma\big({1+\lambda_-\over2}-{\i\zeta\over2p})\big|^2
   \e^{\pi/ap}\over4\pi^2\sqrt{\xi'\xi''\eta'\eta''}\,
   \Gamma^2(1+\lambda_-)\Gamma^2(1+\lambda_+)}\e^{-\i\hbar p^2T/2M}
  \qquad\qquad\qquad\qquad
         \nonumber\\   & &\!\!\!\!\!\!\!\! \qquad\times
  M_{-\i\zeta/2p,\lambda_+/2}(-\i p{\xi''}^2)
  M_{\i\zeta/2p,\lambda_+/2}(\i p{\xi'}^2)
  M_{\i\zeta/2p,\lambda_-/2}(-\i p{\eta''}^2)
  M_{-\i\zeta/2p,\lambda_-/2}(\i p{\eta'}^2)\qquad
                  \\   & &\!\!\!\!\!\!\!\!
  =\int_0^\infty dp_z\int_{\bbbr} d\zeta\int_0^\infty dp\,
  \e^{-\i E_{p_z,p}T/\hbar}\Psi_{p_z,p,\zeta}^*(z',\xi',\eta')
  \Psi_{p_z,p,\zeta}(z'',\xi'',\eta'')\enspace,
\end{eqnarray}
($\zeta$ is the parabolic separation constant)
where the wave-functions and the energy-spectra are given by
\begin{eqnarray}       & &\!\!\!\!\!\!\!\!
  \hbox{\it Spherical Coordinates:}
         \nonumber\\   & &\!\!\!\!\!\!\!\!
  \Psi_{n,m,p}(\theta,\phi,r)=(2r\sin\theta)^{-1/2}
   \Phi_n^{(\lambda_+,\lambda_-)}\bigg({\phi\over2}\bigg)
  \Phi_m^{(\lambda_1,\pm k_3)}(\theta)\sqrt{p}\,J_{\lambda_2}(pr)
  \enspace,       \\   & &\!\!\!\!\!\!\!\!
  E_p=\hbarm p^2\enspace,
                  \\   & &\!\!\!\!\!\!\!\!
  \hbox{\it Circular Polar Coordinates:}
         \nonumber\\   & &\!\!\!\!\!\!\!\!
  \Psi_{n,p,p_z}(\phi,\rho,z)
  =\sqrt{p_zzp}\,J_{\pm k_3}(p_zz)J_{\lambda_1}(p\rho)
  2^{-1/2}\Phi_n^{(\lambda_+,\lambda_-)}\bigg({\phi\over2}\bigg)
  \enspace,       \\   & &\!\!\!\!\!\!\!\!
  E_{p_z,p}=\hbarm(p_z^2+p^2)\enspace,
                  \\   & &\!\!\!\!\!\!\!\!
  \hbox{\it Circular Parabolic Coordinates:}
         \nonumber\\   & &\!\!\!\!\!\!\!\!
  \Psi_{p_z,p,\zeta}(z,\xi,\eta)=\sqrt{p_zz}\,J_{\pm k_3}(p_zz)
   {\Gamma\big({1+\lambda_+\over2}+{\i\zeta\over2p})
   \Gamma\big({1+\lambda_-\over2}-{\i\zeta\over2p})
   \e^{\pi/2ap}\over2\pi \sqrt{p\xi \eta }\,
   \Gamma(1+\lambda_-)\Gamma(1+\lambda_+)}
   \qquad\qquad\qquad\qquad
         \nonumber\\   & &\!\!\!\!\!\!\!\!
  \qquad\qquad\qquad\qquad\qquad\qquad\times
  M_{-\i\zeta/2p,\lambda_+/2}(-\i p \xi^2)
  M_{-\i\zeta/2p,\lambda_-/2}(-\i p \eta^2)\enspace,
                  \\   & &\!\!\!\!\!\!\!\!
  E_{p_z,p}=\hbarm(p_z^2+p^2)\enspace.
\end{eqnarray}
In the path integral evaluation subsequently \cite{GROj} the path
integral solution for the P\"oschl-Teller potential and the radial
harmonic oscillator have been used. The spectral expansion in the $z$-
and $\rho$-dependent kernels is done by means of (analytical
continuation \cite{PI} of) the Weber formula
\begin{equation}
  \int_0^\infty r\e^{\i\alpha r^2}I_\nu(-\i ar)I_\nu(-\i br)dr
  ={\i\over2\alpha}\e^{(a^2+b^2)/4\alpha i}
          I_\nu\bigg({ab\over2\alpha\i}\bigg)
\end{equation}
and we obtain the path integral identity (we use the functional
weight formulation)
\begin{eqnarray}
  \int\limits_{r(t')=r'}^{r(t'')=r''}\mu_\lambda[r^2]\CD r(t)
  \exp\left({\i M\over2\hbar}\int_{t'}^{t''}\dot r^2dt\right)
                       &=&
  \sqrt{r'r''}{M\over\i\hbar T}
  \exp\bigg[{\i M\over2\hbar T}({r'}^2+{r''}^2)\bigg]
  I_\lambda\bigg({Mr'r''\over\i\hbar T}\bigg)
  \qquad          \\   &=&
  \sqrt{r'r''}\,\int_0^\infty pdp\,\e^{-\i\hbar p^2T/2M}
    J_\lambda(pr')J_\lambda(pr'')\enspace.
\label{numch}
\end{eqnarray}
Applying (\ref{numch}) gives in an obvious way the spectral expansion
of the $z$- and $\rho$-dependent kernel. For the extraction of the
continuous states in parabolic one applies (\ref{numef}) by introducing
a Coulomb coupling $\alpha$ and sets $a=\infty$, i.e.\ infinite Bohr
radius, in the in the final formul\ae.

In circular elliptic II coordinates the $z$-path integration is
separated in the same way as for the circular parabolic case, and for
the remaining path integrations we obtain the identity
\begin{eqnarray}       & &
  K^{(V_4)}({\vec x\,}'',{\vec x\,}';T)
         \nonumber\\   & &
  ={M\over\i\hbar T}\sqrt{z'z''}
   \exp\bigg[{\i M\over2\hbar T}({z'}^2+{z''}^2)\bigg]
  I_{\pm k_3}\bigg({Mz'z''\over\i\hbar T}\bigg)
         \nonumber\\   & & \qquad\times
  \int_{\bbbr}{dE\over2\pi\hbar}\e^{-\i ET/\hbar}\int_0^\infty ds''
  \int\limits_{\xi(0)=\xi'}^{\xi(s'')=\xi''}\CD\xi(s)
   \int\limits_{\eta(0)=\eta'}^{\eta(s'')=\eta''}\CD\eta(s)
         \nonumber\\   & &\qquad\times
  \exp\Bigg\{\ih\int_0^{s''}\Bigg[{M\over2}(\dot\xi^2+\dot\eta^2)
     +Ed^2(\cosh^2\xi-\cos^2\eta)
         \nonumber\\   & &\qquad\qquad\qquad\qquad\qquad
     -\hbarm\Bigg({(k_2^2-\half)+k_1^2\cos\eta\over\sin^2\eta}
       +{(k_2^2-\half)+k_1^2\cosh\xi\over\sinh^2\xi}\Bigg)
  \Bigg]ds\Bigg\}\enspace.\qquad
\end{eqnarray}
For the spherical coordinates the expansion into radial wave-functions,
and for the circular parabolic coordinates the expansion of the
$z$-dependent kernel are performed by means of the Hille-Hardy formula
(\ref{numcg}).

\vglue0.4truecm\noindent
{\it 3.2.5.}~We consider the potential ($k_{1,2}>0$)
\begin{equation}
  V_5(\vec x)=\hbarm\bigg(
  {k_1^2x\over y^2\sqrt{x^2+y^2}}+{k_2^2-\viert\over y^2}\Bigg)-k_3z
  \enspace.
\end{equation}
Note that for $k_3=0$ we obtain a trivial extension of the $\alpha=0$
two-dimensional maximally super-integrable potential $V_3(\vec x)$.
We obtain the path integral formulations
\begin{eqnarray}       & &\!\!\!\!\!\!\!\!\!
  K^{(V_5)}({\vec x\,}'',{\vec x\,}';T)
         \nonumber\\   & &\!\!\!\!\!\!\!\!\!
  \underline{\hbox{Circular Polar Coordinates:}}
         \nonumber\\   & &\!\!\!\!\!\!\!\!\!\!
  =\int\limits_{\rho(t')=\rho'}^{\rho(t'')=\rho''}\rho\CD\rho(t)
   \int\limits_{\phi(t')=\phi'}^{\phi(t'')=\phi''}\CD\phi(t)
   \int\limits_{z(t')=z'}^{z(t'')=z''}\CD z(t)
         \nonumber\\   & &\!\!\!\!\!\!\!\!\!\!\qquad\times
  \exp\Bigg\{\ih\int_{t'}^{t''}\Bigg[{M\over2}
  \big(\dot\rho^2+\rho^2\dot\phi^2+\dot z^2)-k_3z
  -{\hbar^2\over2M\rho^2}
   \Bigg({k_2^2-k_1^2-\viert\over4\cos^2{\phi\over2}}
        +{k_2^2+k_1^2-\viert\over4\sin^2{\phi\over2}}-\viert\Bigg)
  \Bigg]dt\Bigg\}
         \nonumber\\   & &\!\!\!\!\!\!\!\!\!\!
                  \\   & &\!\!\!\!\!\!\!\!\!\!
  \phantom{\underline{\hbox{Circular Elliptic II Coordinates:}}}
         \nonumber\\   & &\!\!\!\!\!\!\!\!\!\!
  \underline{\hbox{Circular Elliptic II Coordinates:}}
         \nonumber\\   & &\!\!\!\!\!\!\!\!\!\!
   =\int\limits_{\xi(t')=\xi'}^{\xi(t'')=\xi''}\CD\xi(t)
   \int\limits_{\eta(t')=\eta'}^{\eta(t'')=\eta''}\CD\eta(t)
   d^2(\sinh^2\xi+\sin^2\eta)
   \int\limits_{z(t')=z'}^{z(t'')=z''}\CD z(t)
         \nonumber\\   & &\!\!\!\!\!\!\!\!\!\!\qquad\times
  \exp\Bigg\{\ih\int_{t'}^{t''}\Bigg[{M\over2}
     \Big((\sinh^2\xi^2+\sin^2\eta)(\dot\xi^2+\dot\eta^2)+\dot z^2\Big)
     -k_3z
         \nonumber\\   & &\!\!\!\!\!\!\!\!\!\!
  \qquad\qquad\qquad\qquad\qquad\qquad
     -\hbarm{1\over d^2\sinh^2\xi\sin^2\eta}
      \Bigg(k_1^2{\cosh\xi\cos\eta+1\over\cosh\xi+\cos\eta}
      +k_2^2-\bhalf\Bigg)\Bigg]dt\Bigg\}
                  \\   & &\!\!\!\!\!\!\!\!\!\!
  \underline{\hbox{Circular Parabolic Coordinates:}}
         \nonumber\\   & &\!\!\!\!\!\!\!\!\!\!
  =\int\limits_{\eta(t')=\eta'}^{\eta(t'')=\eta''}\CD\eta(t)
   \int\limits_{\xi(t')=\xi'}^{\xi(t'')=\xi''}\CD\xi(t)
   (\xi^2+\eta^2)\int\limits_{z(t')=z'}^{z(t'')=z''}\CD z(t)
         \nonumber\\   & &\!\!\!\!\!\!\!\!\!\!\qquad\times
  \exp\Bigg\{\ih\int_{t'}^{t''}\Bigg[{M\over2}\Big(
     (\xi^2+\eta^2)(\dot\xi^2+\dot\eta^2)+\dot z^2\Big)
     -\hbarm\Bigg(
     k_1^2{\eta^2-\xi^2\over\xi^2\eta^2(\xi^2+\eta^2)}
     +{k_2^2-\viert\over\xi^2\eta^2}\Bigg)-k_3z\Bigg]dt\Bigg\}
         \nonumber\\   & &
                  \\   & &\!\!\!\!\!\!\!\!\!\!
  \hbox{$\underline{\hbox{Parabolic Coordinates}}$
         ($\lambda^2=k_1^2+k_2^2$):}
         \nonumber\\   & &\!\!\!\!\!\!\!\!\!\!
  =\int\limits_{\eta(t')=\eta'}^{\eta(t'')=\eta''}\CD\eta(t)
   \int\limits_{\xi(t')=\xi'}^{\xi(t'')=\xi''}\CD\xi(t)
   (\xi^2+\eta^2)\xi\eta
   \int\limits_{\phi(t')=\phi'}^{\phi(t'')=\phi''}\CD\phi(t)
         \nonumber\\   & &\!\!\!\!\!\!\!\!\!\!\qquad\times
  \exp\Bigg\{\ih\int_{t'}^{t''}\Bigg[{M\over2}\Big(
     (\xi^2+\eta^2)(\dot\xi^2+\dot\eta^2)+\xi^2\eta^2\dot\phi^2\Big)
         \nonumber\\   & &\!\!\!\!\!\!\!\!\!\!
  \qquad\qquad\qquad\qquad\qquad\qquad
     -{\hbar^2\over2M\xi^2\eta^2}\Bigg(
       {k_1^2+k_2^2-\viert\over\sin^2\phi}-k_1^2-\viert\Bigg)
       -{k_3\over2}(\xi^2-\eta^2)\Bigg]dt\Bigg\}
                  \\   & &\!\!\!\!\!\!\!\!\!\!
  =(\xi'\xi''\eta'\eta'')^{-1/2}\sum_{n=0}^\infty
   (n+\lambda+\bhalf){\Gamma(n+2\lambda+1)\over n!}
 P_{\lambda+n}^{-\lambda}(\cos\phi')P_{\lambda+n}^{-\lambda}(\cos\phi'')
         \nonumber\\   & &\!\!\!\!\!\!\!\!\!\!\qquad\times
   \int_{\bbbr}{dE\over2\pi\hbar}\e^{-\i ET/\hbar}\int_0^\infty ds''
  \int\limits_{\eta(0)=\eta'}^{\eta(s'')=\eta''}\CD\eta(s)
  \int\limits_{\xi(0)=\xi'}^{\xi(s'')=\xi''}\CD\xi(s)
         \nonumber\\   & &\!\!\!\!\!\!\!\!\!\!\qquad\times
  \exp\Bigg\{\ih\int_0^{s''}\Bigg[{M\over2}(\dot\xi^2+\dot\eta^2)
     -{k_3\over2}(\xi^4-\eta^4)+E(\xi^2+\eta^2)
         \nonumber\\   & &\!\!\!\!\!\!\!\!\!\!
  \qquad\qquad\qquad\qquad\qquad\qquad
     -\hbarm\Bigg({\lambda^2-k_1^2-\viert\over\xi^2}
      +{\lambda^2-k_1^2-\viert\over\eta^2}\Bigg)\Bigg]ds\Bigg\}\enspace.
\end{eqnarray}
The path integral in parabolic coordinates cannot be done due to the
$\xi^4,\eta^4$ dependence. Actually $V_5$ represents a Stark effect.
In the cases of the circular polar and circular parabolic coordinates
the corresponding Green function can be evaluated by means of the path
integral solution for the linear potential \cite{FH} for the $z$-path
integration, and the radial harmonic oscillator in the $(r,\phi)$, and
in the $\xi$- and $\eta$-path integration [$\lambda_{\pm}^2=k_2^2\pm
k_1^2$, $\lambda_1=n+(\lambda_++\lambda_-+1)/2$]
\begin{eqnarray}       & &\!\!\!\!\!\!\!\!\!\!
  K^{(V_5)}({\vec x\,}'',{\vec x\,}';T)
         \nonumber\\   & &\!\!\!\!\!\!\!\!\!\!
  \hbox{\it Circular Polar Coordinates:}
         \nonumber\\   & &\!\!\!\!\!\!\!\!\!\!
  =\bigg({M\over2\pi\i\hbar T}\bigg)^{1/2}
  \exp\Bigg[\ih\Bigg({M\over2T}(z''-z')^2-{k_3T\over2}(z'+z'')
      -{k_3^2T^3\over24M}\Bigg)\Bigg]
         \nonumber\\   & &\!\!\!\!\!\!\!\!\!\!\qquad\times
  {M\over2\i\hbar T}\sum_{n=0}^\infty
   \Phi_n^{(\lambda_+,\lambda_-)}\bigg({\phi'\over2}\bigg)
   \Phi_n^{(\lambda_+,\lambda_-)}\bigg({\phi''\over2}\bigg)
  \exp\bigg[-{M\over2\i\hbar T}({\rho'}^2+{\rho''}^2)\bigg]
  I_{\lambda_1}\bigg({M\rho'\rho''\over\i\hbar T}\bigg)
                  \\   & &\!\!\!\!\!\!\!\!\!\!
  =\sum_{n=0}^\infty\int_0^\infty dp\int_{\bbbr} dE_z\,
  \e^{-\i E_{E_z,n,p}T/\hbar}
  \Psi_{E_z,n,p}^*(z',\phi',\rho')\Psi_{E_z,n,p}(z'',\phi'',\rho'')
                  \\   & &\!\!\!\!\!\!\!\!\!\!
  \hbox{\it Circular Parabolic Coordinates:}
         \nonumber\\   & &\!\!\!\!\!\!\!\!\!\!
  =\bigg({M\over2\pi\i\hbar T}\bigg)^{1/2}
  \exp\Bigg[\ih\Bigg({M\over2T}(z''-z')^2-{k_3T\over2}(z'+z'')
      -{k_3^2T^3\over24M}\Bigg)\Bigg]
         \nonumber\\   & &\!\!\!\!\!\!\!\!\!\!\qquad\times
  \int_{\bbbr} d\zeta\int_0^\infty{dp\over p}
  {\big|\Gamma\big({1+\lambda_+\over2}+{\i\zeta\over2p})\big|^2
   \big|\Gamma\big({1+\lambda_-\over2}-{\i\zeta\over2p})\big|^2
   \e^{\pi/ap}\over4\pi^2\sqrt{\xi'\xi''\eta'\eta''}\,
   \Gamma^2(1+\lambda_-)\Gamma^2(1+\lambda_+)}\e^{-\i\hbar p^2T/2M}
         \nonumber\\   & &\!\!\!\!\!\!\!\!\!\!\qquad\times
  M_{-\i\zeta/2p,\lambda_+/2}(-\i p{\xi''}^2)
  M_{\i\zeta/2p,\lambda_+/2}(\i p{\xi''}^2)
  M_{\i\zeta/2p,\lambda_-/2}(-\i p{\eta''}^2)
  M_{-\i\zeta/2p,\lambda_-/2}(\i p{\eta''}^2)\qquad
                  \\   & &\!\!\!\!\!\!\!\!\!\!
  =\int_{\bbbr} d\zeta\int_0^\infty dp\int_{\bbbr} dE_z\,
  \e^{-\i E_{E_z,p}T/\hbar}\Psi_{E_z,p,\zeta}^*(z',\xi',\eta')
  \Psi_{E_z,p,\zeta}(z'',\xi'',\eta'')\enspace,
\end{eqnarray}
($\zeta$ is the parabolic separation constant)
where the wave-functions and the energy-spectra are given by
\begin{eqnarray}       & &\!\!\!\!\!\!\!\!\!\!\!\!\!\!\!\!\!\!\!\!\!\!\!
  \hbox{\it Circular Polar Coordinates:}
         \nonumber\\   & &\!\!\!\!\!\!\!\!\!\!\!\!\!\!\!\!\!\!\!\!\!\!\!
  \Psi_{E_z,n,p}(z,\phi,\rho)
  =\bigg({2M\over\hbar^2\sqrt{k}}\bigg)^{1/3}
   \Ai\left[\bigg(z-{E_z\over k}\bigg)
       \bigg({2Mk\over\hbar^2}\bigg)^{1/3}\right]
  2^{-1/2}\Phi_n^{(\lambda_+,\lambda_-)}\bigg({\phi\over2}\bigg)
   \sqrt{p\rho}\,J_{\lambda_1}(p\rho)
  \enspace,       \\   & &\!\!\!\!\!\!\!\!\!\!\!\!\!\!\!\!\!\!\!\!\!\!\!
  E_{E_z,p}=E_z+\hbarm p^2\enspace,
                  \\   & &\!\!\!\!\!\!\!\!\!\!\!\!\!\!\!\!\!\!\!\!\!\!\!
  \hbox{\it Circular Parabolic Coordinates:}
         \nonumber\\   & &\!\!\!\!\!\!\!\!\!\!\!\!\!\!\!\!\!\!\!\!\!\!\!
  \Psi_{E_z,p,\zeta}(z,\xi,\eta)
  =\bigg({2M\over\hbar^2\sqrt{k}}\bigg)^{1/3}
   \Ai\left[\bigg(z-{E_z\over k}\bigg)
  \bigg({2Mk\over\hbar^2}\bigg)^{1/3 }\right]
         \nonumber\\   & &\!\!\!\!\!\!\!\!\!\!\!\!\!\!\!\!\!\!\!\!\!\!\!
  \qquad\qquad\times
  {\Gamma\big({1+\lambda_+\over2}+{\i\zeta\over2p})
   \Gamma\big({1+\lambda_-\over2}-{\i\zeta\over2p})
   \e^{\pi/2ap}\over2\pi\sqrt{p\xi\eta}\,
   \Gamma(1+\lambda_-)\Gamma(1+\lambda_+)}
  M_{-\i\zeta/2p,\lambda_+/2}(-\i p \xi^2)
  M_{\i\zeta/2p,\lambda_-/2}(-\i p \eta^2)\enspace,
                  \\   & &\!\!\!\!\!\!\!\!\!\!\!\!\!\!\!\!\!\!\!\!\!\!\!
  E_{E_z,p}=E_z+\hbarm p^2\enspace.
\end{eqnarray}
$\zeta$ is the parabolic separation constant.

The expansion into the wave-function in the $z$ coordinates is done
according to (\cite{FH}, $\Ai(z)$ are Airy functions)
\begin{eqnarray}       & &\!\!\!\!
   \int\limits_{z(t')=z'}^{z(t'')=z''}
   \CD z(t)\exp\left[\ih\int_{t'}^{t''}
              \bigg({m\over2}\dot z^2-kz\bigg)dt\right]
         \nonumber\\   & &\!\!\!\!
  =\bigg({M\over2\pi\i\hbar T}\bigg)^{1/2}
  \exp\Bigg[\ih\Bigg({M\over2T}(z''-z')^2-{kT\over2}(z'+z'')
      -{k^2T^3\over24M}\Bigg)\Bigg]
                  \\   & &\!\!\!\!
  =\int_{\bbbr} dE\,\e^{-\i ET/\hbar}
  \bigg({2M\over\hbar^2\sqrt{k}}\bigg)^{2/3}
    \Ai\left[\bigg(z'-{E\over k}\bigg)
       \bigg({2Mk\over\hbar^2}\bigg)^{1/3}\right]
    \Ai\left[\bigg(z''-{E\over k}\bigg)
       \bigg({2Mk\over\hbar^2}\bigg)^{1/3}\right]\enspace.
\qquad\qquad
\end{eqnarray}
The expansion into the radial wave-functions for the circular polar
coordinates is performed by the Weber formula (\ref{numch}).

\vglue0.6truecm\noindent
{\bf 3.3.~Three-Dimensional Minimally Super-Integrable
          Smorodinsky-Winternitz Potentials.}
\vglue0.4truecm\noindent
We discuss in this subsection the eight three-dimensional
minimally super-integrable potentials. The are characterized by
having four functionally independent integrals of motion.
In the sequel $\vec x$ denotes a three-dimensional coordinate:
$\vec x=(x,y,z)\equiv(x_1,x_2,x_3)$.

In table 4 we list the three-dimensional minimally super-integrable
Smorodinsky-Winternitz potentials together with the separating
coordinate systems (where the {\it italized\/} coordinates systems were
not mentioned in \cite{EVA}. The cases where an explicit path
integration is possible are $\underline{\hbox{underlined}}$.

\hfuzz=10pt
\bigskip
\centerline{{\bf Table 4:} The three-dimensional minimally
                           super-integrable potentials\hfill}
\begin{eqnarray}
\begin{array}{l}
\vbox{\offinterlineskip
\hrule
\halign{&\vrule#&\strut\quad\hfil#\quad\hfill\cr
height2pt&\omit&&\omit&\cr
&Potential $V(x,y,z)$
  &&Coordinate System                                       &\cr
height2pt&\omit&&\omit&&\omit&\cr
\noalign{\hrule}
\noalign{\hrule}
height2pt&\omit&&\omit&&\omit&\cr
&$\displaystyle
  V_1=F(r)+\hbarm\Bigg({k_1^2-\viert\over x^2}
    +{k_2^2-\viert\over y^2}+{k_3^2-\viert\over z^2}\Bigg)$
  &&$\underline{\hbox{Spherical}}$              &\cr
& &&Conical                                     &\cr
height2pt&\omit&&\omit&\cr
\noalign{\hrule}
height2pt&\omit&&\omit&\cr
&$\displaystyle
  V_2={M\over2}\omega^2(x^2+y^2)
    +\hbarm\Bigg({k_1^2-\viert\over x^2}
    +{k_2^2-\viert\over y^2}\Bigg)+F(z)$
  &&$\underline{\hbox{Cartesian}}$              &\cr
& &&$\underline{\hbox{Circular Polar}}$         &\cr
& &&Circular Elliptic                           &\cr
height2pt&\omit&&\omit&\cr
\noalign{\hrule}
height2pt&\omit&&\omit&\cr
&$\displaystyle
  V_3={M\over2}\omega^2(4x^2+y^2)
     +\hbarm{k_2^2-\viert\over y^2}+F(z)$
  &&$\underline{\hbox{Cartesian}}$              &\cr
& &&Circular Parabolic                          &\cr
height2pt&\omit&&\omit&\cr
\noalign{\hrule}
height2pt&\omit&&\omit&\cr
&$\displaystyle
  V_4=-{\alpha\over\sqrt{x^2+y^2}}
     +\hbarm\bigg(
  {k_1^2x\over y^2\sqrt{x^2+y^2}}+{k_2^2-\viert\over y^2}\bigg)+F(z)$
  &&$\underline{\hbox{Circular Polar}}$         &\cr
& &&{\it Circular Elliptic II}                  &\cr
& &&$\underline{\hbox{Circular Parabolic}}$     &\cr
height2pt&\omit&&\omit&\cr
\noalign{\hrule}
height2pt&\omit&&\omit&\cr
&$\displaystyle
  V_5={M\over2}\omega^2(x^2+y^2+z^2)
  +\hbarm\Bigg({k_3^2-\viert\over z^2}
  +{F(y/x)\over x^2+y^2}\bigg)$
  &&$\underline{\hbox{Spherical}}$              &\cr
& &&$\underline{\hbox{Circular Polar}}$         &\cr
& &&Oblate Spheroidal                           &\cr
& &&Prolate Spheroidal                          &\cr
height2pt&\omit&&\omit&\cr
\noalign{\hrule}
height2pt&\omit&&\omit&\cr
&$\displaystyle
  V_6={M\over2}\omega^2(x^2+y^2+4z^2)
     +\hbarm{F(y/x)\over x^2+y^2}$
  &&$\underline{\hbox{Circular Polar}}$         &\cr
& &&Parabolic                                   &\cr
height2pt&\omit&&\omit&\cr
\noalign{\hrule}
height2pt&\omit&&\omit&\cr
&$\displaystyle
  V_7=-{\alpha\over r}
     +{\hbar^2\over2M(x^2+y^2)}\Bigg(
  {k_1^2z\over r}+F\bigg({y\over x}\bigg)\Bigg)$
  &&$\underline{\hbox{Spherical}}$              &\cr
& &&$\underline{\hbox{Parabolic}}$              &\cr
& &&{\it Prolate Spheroidal II}                 &\cr
height2pt&\omit&&\omit&\cr
\noalign{\hrule}
height2pt&\omit&&\omit&\cr
&$\displaystyle
  V_8=-{\alpha\over\rho}+\sqrt{2\over\rho}
   \bigg(\beta_1\cos{\phi\over2}+\beta_2\sin{\phi\over2}\bigg)+F(z)$
  &&$\underline{\hbox{Mutually Circular}}$      &\cr
& &&$\underline{\hbox{\ Parabolic}}$            &\cr
height2pt&\omit&&\omit&\cr}\hrule}
\end{array}
         \nonumber
\end{eqnarray}
\hfuzz=7.0pt

\vglue0.4truecm\noindent
{\it 3.3.1.}~We consider the potential ($r=|\vec x|$, $k_{1,2,3}>0$)
\begin{equation}
  V_1(\vec x)=F(r)+\hbarm\Bigg({k_1^2-\viert\over x^2}
    +{k_2^2-\viert\over y^2}+{k_3^2-\viert\over z^2}\Bigg)\enspace.
\end{equation}
\eject\noindent
We formulate the path integral in the separating coordinate
systems and obtain
\begin{eqnarray}       & &\!\!
  K^{(V_1)}({\vec x\,}'',{\vec x\,}';T)
         \nonumber\\   & &\!\!
  =\int\limits_{\vec x(t')=\vec x'}^{\vec x(t'')=\vec x''}\CD\vec x(t)
  \exp\left\{\ih\int_{t'}^{t''}\left[{M\over2}{\dot{\vec x}}^2-F(r)
    -\hbarm\Bigg({k_1^2-\viert\over x^2}
    +{k_2^2-\viert\over y^2}+{k_3^2-\viert\over z^2}\Bigg)
  \right]dt\right\}
                  \\   & &\!\!
  \underline{\hbox{Spherical Coordinates:}}
         \nonumber\\   & &\!\!
  =\int\limits_{r(t')=r'}^{r(t'')=r''}r^2\CD r(t)
   \int\limits_{\theta(t')=\theta'}^{\theta(t'')=\theta''}
  \sin\theta\CD\theta(t)
   \int\limits_{\phi(t')=\phi'}^{\phi(t'')=\phi''}\CD\phi(t)
         \nonumber\\   & &\!\! \qquad\times
  \exp\Bigg\{\ih\int_{t'}^{t''}\Bigg[{M\over2}
\big(\dot r^2+r^2\dot\theta^2+r^2\sin^2\theta\dot\phi^2\big)-F(r)
         \nonumber\\   & &\!\! \qquad\qquad\qquad\qquad\qquad
  -{\hbar^2\over2Mr^2}\Bigg({1\over\sin^2\theta}
      \bigg({k_1^2-\viert\over\cos^2\phi}
           +{k_2^2-\viert\over\sin^2\phi}-\viert\bigg)
           +{k_3^2-\viert\over\cos^2\theta}
           -{1\over4}\Bigg)\Bigg]dt\Bigg\}\qquad\qquad
                  \\   & &\!\!
  \underline{\hbox{Conical Coordinates:}}
         \nonumber\\   & &\!\!
  =\int\limits_{r(t')=r'}^{r(t'')=r''}r^2\CD r(t)
  \int\limits_{\alpha(t')=\alpha'}^{\alpha(t'')=\alpha''}\CD\alpha(t)
  \int\limits_{\beta(t')=\beta'}^{\beta(t'')=\beta''}\CD\beta(t)
  (k^2\cn^2\alpha+{k'}^2\cn^2\beta)
         \nonumber\\   & &\!\! \qquad\times
  \exp\Bigg\{\ih\int^{t''}_{t'}\Bigg[{M\over2}\Big(\dot r^2+
   r^2(k^2\cn^2\alpha+{k'}^2\cn^2\beta)
   (\dot\alpha^2+\dot\beta^2)\Big)-F(r)
         \nonumber\\   & &\!\! \qquad\qquad\qquad\qquad\qquad
     -{\hbar^2\over2Mr^2}\Bigg({k_1^2-\viert\over\sn^2\alpha\dn^2\beta}
         +{k_2^2-\viert\over\cn^2\alpha\cn^2\beta}
         +{k_3^2-\viert\over\dn^2\alpha\sn^2\beta}\Bigg)\Bigg]dt\Bigg\}
  \enspace.
\end{eqnarray}
For this potential the angular path integrations can be easily done and
one is left with a radial path integral \cite{CARPa}, i.e.\ ($\lambda_1=
2n\pm k_1\pm k_2+1$, $\lambda_2=2m+\lambda_1\pm k_3+1$)
\begin{eqnarray}       & &\!\!\!\!\!\!\!\!\!\!
  K^{(V_1)}({\vec x\,}'',{\vec x\,}';T)
         \nonumber\\   & &\!\!\!\!\!\!\!\!\!\!
  =({r'}^2{r''}^2\sin\theta'\sin\theta'')^{-1/2}\sum_{n=0}^\infty
  \Phi^{(\pm k_2,\pm k_1)}_n(\phi'')\Phi^{(\pm k_2,\pm k_1)}_n(\phi')
  \sum_{m=0}^\infty\Phi^{(\lambda_1,\pm k_3)}_m(\theta'')
                   \Phi^{(\lambda_1,\pm k_3)}_m(\theta')
         \nonumber\\   & &\!\!\!\!\!\!\!\!\!\! \qquad\times
   \int\limits_{r(t')=r''}^{r(t'')=r''}\CD r(t)
  \exp\Bigg[\ih\int_{t'}^{t''}\Bigg({M\over2}\dot r^2-F(r)
     -\hbarm{\lambda_2^2-\viert\over r^2}\Bigg)dt\Bigg]
                  \\   & &\!\!\!\!\!\!\!\!\!\!
  \equiv({r'}^2{r''}^2\sin\theta'\sin\theta'')^{-1/2}\sum_{n=0}^\infty
  \Phi^{(\pm k_2,\pm k_1)}_n(\phi'')\Phi^{(\pm k_2,\pm k_1)}_n(\phi')
  \sum_{m=0}^\infty\Phi^{(\lambda_1,\pm k_3)}_m(\theta'')
                   \Phi^{(\lambda_1,\pm k_3)}_m(\theta')
  \qquad\qquad
         \nonumber\\   & &\!\!\!\!\!\!\!\!\!\! \qquad\times
  \int\limits_{r(t')=r''}^{r(t'')=r''}\mu_{\lambda_2}[r^2]\CD r(t)
  \exp\Bigg[\ih\int_{t'}^{t''}\Bigg({M\over2}\dot r^2-F(r)
      \Bigg)dt\Bigg]
                  \\   & &\!\!\!\!\!\!\!\!\!\!
  =\sum_{m,n=0}^\infty\int dE_{\lambda_r}\,\e^{-\i E_{\lambda_r}/\hbar}
   \Phi_{\lambda_r,n,m}^*(\theta',\phi',r')
   \Phi_{\lambda_r,n,m}(\theta'',\phi'',r'')\enspace.
\end{eqnarray}
Here use has been made of the functional weight formulation in path
integrals according to \cite{GRSb, GRSg, STEc}.
The wave-functions have the form
\begin{equation}
   \Phi_{\lambda_r,n,l,}(\theta,\phi,r)=(r^2\sin\theta)^{-1/2}
  \Phi^{(\pm k_2,\pm k_1)}_n(\phi)\Phi^{(\lambda_1,\pm k_3)}_m(\theta)
    \Phi_{\lambda_r}(r)\enspace,
\end{equation}
where $\Phi_{\lambda_r}(r)$ are the radial wave-functions.
Of special interest are the cases $F(r)=m\omega^2r^2/2$ and $F(r)=-
\alpha/r$, c.f.\ 3.2.1 and 3.2.3, respectively.
\goodbreak

\eject\noindent
{\it 3.3.2.}~The next three potentials represent a simple $z$-dependence
extension of the two-dimensional potentials $V_1$--$V_3$.  We consider
the potential ($k_{1,2}>0$)
\begin{equation}
  V_2(\vec x)={M\over2}\omega^2(x^2+y^2)
    +\hbarm\Bigg({k_1^2-\viert\over x^2}
    +{k_2^2-\viert\over y^2}\Bigg)+F(z)\enspace.
\end{equation}
For the case of $F(z)=2M\omega^2z^2$ there is an additional coordinate
system, the parabolic coordinates, which separate $V_2$, c.f.\ the
three-dimensional maximally super-integrable potential $V_2$. Here now
we have
\begin{eqnarray}       & &\!\! \!\! \!\! \!\!
  K^{(V_2)}({\vec x\,}'',{\vec x\,}';T)
         \nonumber\\   & & \!\! \!\! \!\! \!\!
  \underline{\hbox{Cartesian Coordinates:}}
         \nonumber\\   & & \!\! \!\! \!\! \!\!
  =\int\limits_{\vec x(t')=\vec x'}^{\vec x(t'')=\vec x''}\CD\vec x(t)
  \exp\left\{\ih\int_{t'}^{t''}\left[{M\over2}
   \Big({\dot{\vec x}}^2-\omega^2(x^2+y^2)\Big)
 -\hbarm\sum_{i=1}^2{k_i^2-\viert\over x_i^2}-F(z)
  \right]dt\right\}
                  \\   & & \!\! \!\! \!\! \!\!
  \underline{\hbox{Circular Polar Coordinates:}}
         \nonumber\\   & & \!\! \!\! \!\! \!\!
  =\int\limits_{z(t')=z'}^{z(t'')=z''}\CD z(t)
   \int\limits_{\rho(t')=\rho'}^{\rho(t'')=\rho''}\rho\CD\rho(t)
   \int\limits_{\phi(t')=\phi'}^{\phi(t'')=\phi''}\CD\phi(t)
         \nonumber\\   & & \!\! \!\! \!\! \!\!\quad\times
  \exp\Bigg\{\ih\int_{t'}^{t''}\Bigg[{M\over2}
     \big(\dot\rho^2+\rho^2\dot\phi^2-\omega^2\rho^2\big)
     -{\hbar^2\over2M\rho^2}\Bigg({k_1^2-\viert\over\cos^2\phi}
         +{k_2^2-\viert\over\sin^2\phi}-\viert\Bigg)-F(z)\Bigg]dt\Bigg\}
  \qquad          \\   & & \!\! \!\! \!\! \!\!
  \underline{\hbox{Circular Elliptic Coordinates:}}
         \nonumber\\   & & \!\! \!\! \!\! \!\!
  =\int\limits_{z(t')=z'}^{z(t'')=z''}\CD z(t)
   \int\limits_{\xi(t')=\xi'}^{\xi(t'')=\xi''}\CD\xi(t)
   \int\limits_{\eta(t')=\eta'}^{\eta(t'')=\eta''}\CD\eta(t)
   d^2(\sinh^2\xi+\sin^2\eta)
         \nonumber\\   & & \!\! \!\! \!\! \!\!\quad\times
  \exp\Bigg\{\ih\int_{t'}^{t''}\Bigg[{M\over2}
     \Big(d^2(\sinh^2\xi+\sin^2\eta)
         (\dot\xi^2+\dot\eta^2)+\dot z^2
  -\omega^2d^2(\cosh^2\xi\cos^2\eta+\sinh^2\xi\sin^2\eta)\Big)
         \nonumber\\   & & \!\! \!\! \!\! \!\!
  \qquad\qquad\qquad\qquad\qquad\qquad
  -F(z)-{\hbar^2\over2Md^2}\Bigg({k_1^2-\viert\over\cosh^2\xi\cos^2\eta}
  +{k_2^2-\viert\over\sinh^2\xi\sin^2\eta}\Bigg)\Bigg]dt\Bigg\}\enspace.
\end{eqnarray}
Except for the $z$-path integration, we can perform in the case of
cartesian and circular polar polar the path integration
explicitly yielding
\begin{eqnarray}       & &\!\!\!\!\!\!\!\!\!\!\!\!\!\!\!\!
  K^{(V_2)}({\vec x\,}'',{\vec x\,}';T)
         \nonumber\\   & &\!\!\!\!\!\!\!\!\!\!\!\!\!\!\!\!
  \hbox{\it Cartesian Coordinates:}
         \nonumber\\   & &\!\!\!\!\!\!\!\!\!\!\!\!\!\!\!\!
  =\bigg({M\omega\over\i\hbar\sin\omega T}\bigg)^3
  \prod_{i=1}^2\sqrt{x_i'x_i''}\,
 \exp\bigg[-{M\omega\over2\i\hbar}({x_i'}^2+{x_i''}^2)\cot\omega T\bigg]
  I_{k_i}\bigg({M\omega x_i'x_i''\over\i\hbar\sin\omega T}\bigg)
  \qquad\qquad
         \nonumber\\   & &\!\!\!\!\!\!\!\!\!\!\!\!\!\!\!\!\qquad\times
  \int\limits_{z(t')=z'}^{z(t'')=z''}\CD z(t)
  \exp\Bigg[\ih\int_{t'}^{t''}\Bigg({M\over2}\dot z^2-F(z)\Bigg)dt\Bigg]
                  \\   & &\!\!\!\!\!\!\!\!\!\!\!\!\!\!\!\!
 =\sum_{n_1,n_2=0}^\infty \e^{-\i E_{n_1,n_2}T/\hbar}
  \Psi_{n_1,n_2}(x_1',x_2')\Psi_{n_1,n_2}(x_1'',x_2'')
  \int dE_{\lambda_z}\,\e^{-\i E_{\lambda_z}T/\hbar}
  \Phi_{\lambda_z}^*(z')\Phi_{\lambda_z}(z'')
                  \\   & &\!\!\!\!\!\!\!\!\!\!\!\!\!\!\!\!
  \hbox{{\it Circular Polar Coordinates}
         ($\lambda_1=2n\pm k_1\pm k_2+1)$):}
         \nonumber\\   & &\!\!\!\!\!\!\!\!\!\!\!\!\!\!\!\!
 ={M\omega\over\i\hbar\sin\omega T}\sum_{n=0}^\infty
  \Phi^{(\pm k_2,\pm k_1)}(\phi')\Phi^{(\pm k_2,\pm k_1)}(\phi'')
 \exp\bigg[-{M\omega\over2\i\hbar}\big({\rho'}^2+{\rho''}^2\big)
    \cot\omega T\bigg]
  I_{\lambda_1}\bigg({M\omega\rho'\rho''\over\i\hbar\sin\omega T}\bigg)
  \qquad \nonumber\\   & &\!\!\!\!\!\!\!\!\!\!\!\!\!\!\!\!\qquad\times
  K_F(z'',z';T)
                  \\   & &\!\!\!\!\!\!\!\!\!\!\!\!\!\!\!\!
 =\sum_{n,m=0}^\infty \e^{-\i E_{n,m}T/\hbar}
  \Psi_{n,m}(\phi',\rho')\Psi_{n,m}(\phi'',\rho'')
  \int dE_{\lambda_z}\,\e^{-\i E_{\lambda_z} T/\hbar}
  \Phi_{\lambda_z}^*(z')\Phi_{\lambda_z}(z'')
\end{eqnarray}
and in the sequel we denote the remaining $z$-path integral by
\begin{equation}
  K_F(z'',z';T)=
  \int dE_{\lambda_z} \e^{-\i E_{\lambda_z} T/\hbar}
  \Phi_{\lambda_z}^*(z')\Phi_{\lambda_z}(z'')\enspace.
\end{equation}
The wave-functions in $(x_1,x_2)$ and $(\rho,\phi)$ are the same as
in 3.1.1.

\vglue0.4truecm\noindent
{\it 3.3.3.}~We consider the potential ($k_2>0$)
\begin{equation}
  V_3(\vec x)={M\over2}\omega^2(4x^2+y^2)
     +\hbarm{k_2^2-\viert\over y^2}+F(z)\enspace.
\end{equation}
Due to the similarity of the corresponding two-dimensional case
in 3.1.2 we just give the path integral formulations which have the form
\begin{eqnarray}       & &\!\!\!\!\!\!\!\!\!\!
  K^{(V_3)}({\vec x\,}'',{\vec x\,}';T)
         \nonumber\\   & &\!\!\!\!\!\!\!\!\!\!
  \underline{\hbox{Cartesian Coordinates:}}
         \nonumber\\   & &\!\!\!\!\!\!\!\!\!\!
  =\int\limits_{\vec x(t')=\vec x'}^{\vec x(t'')=\vec x''}\CD\vec x(t)
  \exp\left\{\ih\int_{t'}^{t''}\left[{M\over2}
     \Big({\dot{\vec x}}^2-\omega^2(4x^2+y^2)\Big)-F(z)
     -\hbarm{k_2^2-\viert\over y^2}\right]dt\right\}
                  \\   & &\!\!\!\!\!\!\!\!\!\!
  =\sum_{n,l=0}^\infty\e^{-\i E_{n,l}T/\hbar}
  \Psi_{n,l}(x'',y'')\Psi_{n,l}(x',y')
  \int dE_{\lambda_z}\,\e^{-\i E_{\lambda_z}T/\hbar}
  \Phi_{\lambda_z}^*(z')\Phi_{\lambda_z}(z'')
                  \\   & &\!\!\!\!\!\!\!\!\!\!
  \underline{\hbox{Circular Parabolic Coordinates:}}
         \nonumber\\   & &\!\!\!\!\!\!\!\!\!\!
  =\int\limits_{\eta(t')=\eta'}^{\eta(t'')=\eta''}\CD\eta(t)
   \int\limits_{\xi(t')=\xi'}^{\xi(t'')=\xi''}\CD\xi(t)
   (\xi^2+\eta^2)
         \nonumber\\   & &\!\!\!\!\!\!\!\!\!\!\ \times
  \exp\Bigg\{\ih\int_{t'}^{t''}\Bigg[{M\over2}\Big(
     (\xi^2+\eta^2)(\dot\xi^2+\dot\eta^2)
     -\omega^2\big((\xi^2-\eta^2)^2-\xi^2\eta^2\big)\Big)
     -{k_1\over2}(\xi^2-\eta^2)-\hbar^2
      {k_2^2-\viert\over2M\eta^2\xi^2}\Bigg]dt\Bigg\}
         \nonumber\\   & &\!\!\!\!\!\!\!\!\!\!\ \times
  K_F(z'',z';T)\enspace,
\end{eqnarray}
\nobreak
with the same Feynman kernel and wave-functions in $(x,y)$ as in 3.1.2.

\vglue0.4truecm\noindent
{\it 3.3.4.\/}~We consider the potential ($k_{1,2}>0$)
\begin{equation}
  V_4(\vec x)=-{\alpha\over\sqrt{x^2+y^2}}
     +\hbarm\bigg(
  {k_1^2x\over y^2\sqrt{x^2+y^2}}+{k_2^2-\viert\over y^2}
  \bigg)+F(z)\enspace,
\end{equation}
We formulate the path integral in circular polar and circular parabolic
coordinates, respectively
\begin{eqnarray}       & &\!\!\!\!\!\!\!\!
  K^{(V_4)}({\vec x\,}'',{\vec x\,}';T)
         \nonumber\\   & &\!\!\!\!\!\!\!\!
  =\int\limits_{\vec x(t')=\vec x'}^{\vec x(t'')=\vec x''}\CD\vec x(t)
  \exp\Bigg\{\ih\int_{t'}^{t''}\Bigg[{M\over2}
   {\dot{\vec x}}^2+{\alpha\over\sqrt{x^2+y^2}}
  -\hbarm\Bigg({k_1^2x\over y^2\sqrt{x^2+y^2}}
   +{k_2^2-\viert\over y^2}\Bigg)-F(z)\Bigg]dt\Bigg\}
         \nonumber\\   & &
                  \\   & &\!\!\!\!\!\!\!\!
  \underline{\hbox{Circular Polar Coordinates:}}
         \nonumber\\   & &\!\!\!\!\!\!\!\!
  =\int\limits_{\rho(t')=\rho'}^{\rho(t'')=\rho''}\rho\CD\rho(t)
   \int\limits_{\phi(t')=\phi'}^{\phi(t'')=\phi''}\CD\phi(t)
         \nonumber\\   & &\!\!\!\!\!\!\!\!\qquad\times
  \exp\Bigg\{\ih\int_{t'}^{t''}\Bigg[{M\over2}
  \big(\dot\rho^2+\rho^2\dot\phi^2\big)+{\alpha\over\rho}
  -{\hbar^2\over2M\rho^2}
   \Bigg({k_2^2-k_1^2-\viert\over4\cos^2{\phi\over2}}
        +{k_2^2+k_1^2-\viert\over4\sin^2{\phi\over2}}-\viert\Bigg)
  \Bigg]dt\Bigg\}
         \nonumber\\   & &\!\!\!\!\!\!\!\!\qquad\times
  K_F(z'',z';T)
                  \\   & &\!\!\!\!\!\!\!\!
  \underline{\hbox{Circular Elliptic II Coordinates:}}
         \nonumber\\   & &\!\!\!\!\!\!\!\!
   =\int\limits_{\xi(t')=\xi'}^{\xi(t'')=\xi''}\CD\xi(t)
   \int\limits_{\eta(t')=\eta'}^{\eta(t'')=\eta''}\CD\eta(t)
   d^2(\sinh^2\xi+\sin^2\eta)\int\limits_{z(t')=z'}^{z(t'')=z''}\CD z(t)
         \nonumber\\   & &\!\!\!\!\!\!\!\!\qquad\times
  \exp\Bigg\{\ih\int_{t'}^{t''}\Bigg[{M\over2}
     \Big((\sinh^2\xi^2+\sin^2\eta)(\dot\xi^2+\dot\eta^2)+\dot z^2\Big)
     +{\alpha\over d(\cosh\xi+\cos\eta)}-F(z)
         \nonumber\\   & &\!\!\!\!\!\!\!\!
  \qquad\qquad\qquad\qquad\qquad\qquad
     -\hbarm{1\over d^2\sinh^2\xi\sin^2\eta}
      \Bigg(k_1^2{\cosh\xi\cos\eta+1\over\cosh\xi+\cos\eta}
      +k_2^2-\bhalf\Bigg)\Bigg]dt\Bigg\}\qquad
                  \\   & &\!\!\!\!\!\!\!\!
  \underline{\hbox{Circular Parabolic Coordinates:}}
         \nonumber\\   & &\!\!\!\!\!\!\!\!
  =\int\limits_{\eta(t')=\eta'}^{\eta(t'')=\eta''}\CD\eta(t)
   \int\limits_{\xi(t')=\xi'}^{\xi(t'')=\xi''}\CD\xi(t)
   (\xi^2+\eta^2)\int\limits_{z(t')=z'}^{z(t'')=z''}\CD z(t)
         \nonumber\\   & &\!\!\!\!\!\!\!\!\qquad\times
  \exp\Bigg\{\ih\int_{t'}^{t''}\Bigg[{M\over2}\Big(
     (\xi^2+\eta^2)(\dot\xi^2+\dot\eta^2)+\dot z^2\Big)
     +{2\alpha\over\xi^2+\eta^2}-F(z)
         \nonumber\\   & &\!\!\!\!\!\!\!\!
  \qquad\qquad\qquad\qquad\qquad\qquad\qquad\qquad\qquad
     -\hbarm\Bigg(k_1^2{\eta^2-\xi^2\over\xi^2\eta^2(\xi^2+\eta^2)}
     +{k_2^2-\viert\over\xi^2\eta^2}\Bigg)\Bigg]dt\Bigg\}\enspace.
\end{eqnarray}
The circular polar an circular parabolic path integral formulations can
be explicitly evaluated in terms of the corresponding Green function
($\lambda_{\pm}^2=k_2^2\pm k_1^2$, $\lambda_1=n+(\lambda_++\lambda_-
+1)/2$, $\kappa=\alpha\sqrt{-M/2E}/\hbar$)
\begin{eqnarray}       & &\!\!\!\!\!\!\!\!\!\!\!\!\!\!
  K^{(V_4)}({\vec x\,}'',{\vec x\,}';T)
         \nonumber\\   & &\!\!\!\!\!\!\!\!\!\!\!\!\!\!
  \hbox{\it Circular Polar Coordinates:}
         \nonumber\\   & &\!\!\!\!\!\!\!\!\!\!\!\!\!\!
  =K_F(z'',z';T)(\rho'\rho'')^{-1/2}\half\sum_{n=0}^\infty
   \Phi_n^{(\lambda_+,\lambda_-)}\bigg({\phi'\over2}\bigg)
   \Phi_n^{(\lambda_+,\lambda_-)}\bigg({\phi''\over2}\bigg)
         \nonumber\\   & &\!\!\!\!\!\!\!\!\!\!\!\!\!\!\qquad\times
  {1\over\hbar}\int_{\bbbr}{dE\over2\pi\hbar}\e^{-\i ET/\hbar}
  \sqrt{-{M\over2E}}
  {\Gamma(\half+\lambda_1-\kappa)\over\Gamma(1+2\lambda_1)}
   W_{\kappa,\lambda_1}\bigg(\sqrt{-8ME}\,{\rho_>\over\hbar}\bigg)
   M_{\kappa,\lambda_1}\bigg(\sqrt{-8ME}\,{\rho_<\over\hbar}\bigg)
  \qquad          \\   & &\!\!\!\!\!\!\!\!\!\!\!\!\!\!
  =\int dE_{\lambda_z}\,\e^{-\i E_{\lambda_z} T/\hbar}
  \Phi_{\lambda_z}^*(z')\Phi_{\lambda_z}(z'')
         \nonumber\\   & &\!\!\!\!\!\!\!\!\!\!\!\!\!\!\qquad\times
  \sum_{m=-0}^\infty\left\{\sum_{n=0}^\infty
   {\Phi_{\lambda_1,m}^*(\rho',\phi')
    \Phi_{\lambda_1,m}(\rho'',\phi'')\over E_n-E}
   +\int_{\bbbr} dp{\Phi_{\lambda_z,p}^*(\rho',\phi')
    \Phi_{\lambda,p}^{\alpha)}(\rho'',\phi'')
  \over\hbar^2p^2/2M-E}\right\}
                  \\   & &\!\!\!\!\!\!\!\!\!\!\!\!\!\!
  \hbox{\it Circular Parabolic Coordinates:}
         \nonumber\\   & &\!\!\!\!\!\!\!\!\!\!\!\!\!\!
  =\int dE_{\lambda_z}\,\e^{-\i E_{\lambda_z}T/\hbar}
  \Phi_{\lambda_z}^*(z')\Phi_{\lambda_z}(z'')
         \nonumber\\   & &\!\!\!\!\!\!\!\!\!\!\!\!\!\!
  \qquad\times\left\{\sum_{n_1,n_2=0}^\infty
  {\Psi_{n_1,n_2}^*(\xi',\eta')
   \Psi_{n_1,n_2}(\xi'',\eta'')\over E_{n_1,n_2}-E}
  +\int_0^\infty dp\int_{\bbbr}d\zeta {\Psi_{p,\zeta}^*(\xi',\eta')
  \Psi_{p,\zeta}(\xi'',\eta'')\over \hbar^2p^2/2M-E}\right\}\enspace,
\label{numcb}
\end{eqnarray}
with the same wave-functions in ($\phi,\rho$) and $(\xi,\eta)$ as in
3.2.3.

\vglue0.4truecm\noindent
{\it 3.3.5.}~The next three potential involve an arbitrary
$\phi$-dependence where $F(y/x)=\gamma^2>0$ describes the well-known
double ring-shaped oscillator (c.f.~e.g.\ Carpio-Bernido et al.\
\cite{CBB, CBBI}, Kibler et al.\ \cite{KICA, KLW, KIWIa}, Lutsenko et
al.\ \cite{LMPS}, and Quesne \cite{QUES}). We consider the potential
($k_3>0$)
\begin{equation}
  V_5(\vec x)={M\over2}\omega^2(x^2+y^2+z^2)
  +\hbarm\Bigg({k_3^2-\viert\over z^2}
  +{F(y/x)\over x^2+y^2}\bigg)\enspace.
\end{equation}
We obtain the following path integral formulations in the
corresponding separating coordinate systems
\begin{eqnarray}       & &
  K^{(V_5)}({\vec x\,}'',{\vec x\,}';T)
         \nonumber\\   & &
  \underline{\hbox{Spherical Coordinates:}}
         \nonumber\\   & &
  =\int\limits_{r(t')=r'}^{r(t'')=r''}r^2\CD r(t)
   \int\limits_{\theta(t')=\theta'}^{\theta(t'')=\theta''}
  \sin\theta\CD\theta(t)
   \int\limits_{\phi(t')=\phi'}^{\phi(t'')=\phi''}\CD\phi(t)
         \nonumber\\   & &\qquad\times
  \exp\Bigg\{\ih\int_{t'}^{t''}\Bigg[{M\over2}
\Big(\dot r^2+r^2\dot\theta^2+r^2\sin^2\theta\dot\phi^2-\omega^2r^2\Big)
         \nonumber\\   & &\qquad\qquad\qquad\qquad\qquad\qquad\qquad
  -{\hbar^2\over2Mr^2}\Bigg({F(\tan\phi)-\viert\over\sin^2\theta}
  +{k_3^2-\viert\over\cos^2\theta}-{1\over4}\Bigg)\Bigg]dt\Bigg\}
                  \\   & &
  \underline{\hbox{Circular Polar Coordinates:}}
         \nonumber\\   & &
  =\int\limits_{z(t')=z'}^{z(t'')=z''}\CD z(t)
   \int\limits_{\rho(t')=\rho'}^{\rho(t'')=\rho''}\rho\CD\rho(t)
   \int\limits_{\phi(t')=\phi'}^{\phi(t'')=\phi''}\CD\phi(t)
         \nonumber\\   & &\qquad\times
  \exp\Bigg\{\ih\int_{t'}^{t''}\Bigg[{M\over2}
     \Big(\dot\rho^2+\rho^2\dot\phi^2+\dot z^2-\omega^2(\rho^2+z^2)\Big)
     -\hbarm\Bigg({F(\tan\phi)-\viert\over\rho^2}
         +{k_3^2-\viert\over z^2}\Bigg)\Bigg]dt\Bigg\}
         \nonumber\\   & &
                  \\   & &
  \underline{\hbox{Oblate Spheroidal Coordinates:}}
         \nonumber\\   & &=
\int\limits_{\bar\mu(t')=\bar\mu'}^{\bar\mu(t'')=\bar\mu''}\CD\bar\mu(t)
\int\limits_{\bar\nu(t')=\bar\nu'}^{\bar\nu(t'')=\bar\nu''}\CD\bar\nu(t)
   \bar d^3(\cosh^2\bar\mu-\sin^2\bar\nu)\cosh\bar\mu\sin\bar\nu
  \int\limits_{\phi(t')=\phi'}^{\phi(t'')=\phi''}\CD \phi(t)
         \nonumber\\   & & \qquad\times
  \exp\Bigg\{\ih\int_{t'}^{t''}\Bigg[{M\over2}\bar d^2
  \Big((\cosh^2\bar\mu-\sin^2\bar\nu)(\dot{\bar\mu}^2+\dot{\bar\nu}^2)
        +\cosh^2\bar\mu\sin^2\bar\nu\dot\phi^2
         \nonumber\\   & & \qquad\qquad\qquad\qquad\qquad\qquad\qquad
  -\omega^2(\cos^2\bar\mu\sin^2\bar\nu+\sinh^2\bar\mu\cos^2\bar\nu)\Big)
         \nonumber\\   & & \qquad\qquad\qquad\qquad\qquad\qquad\qquad
 -{\hbar^2\over2M\bar d^2}
   \Bigg({F(\tan\phi)-\viert\over\cosh^2\bar\mu\sin^2\bar\nu}
   +{k_3^2-\viert\over\sinh^2\bar\mu\cos^2\bar\nu}\Bigg)\Bigg]dt\Bigg\}
                  \\   & &
  \underline{\hbox{Prolate Spheroidal Coordinates:}}
         \nonumber\\   & &
  =\int\limits_{\mu(t')=\mu'}^{\mu(t'')=\mu''}\CD\mu(t)
   \int\limits_{\nu(t')=\nu'}^{\nu(t'')=\nu''}\CD\nu(t)
   d^3(\sinh^2\mu+\sin^2\nu)\sinh\mu\sin\nu
  \int\limits_{\phi(t')=\phi'}^{\phi(t'')=\phi''}\CD \phi(t)
         \nonumber\\   & & \qquad\times
  \exp\Bigg\{\ih\int_{t'}^{t''}\Bigg[{M\over2}d^2
     \Big((\sinh^2\mu+\sin^2\nu)(\dot\mu^2+\dot\nu^2)
        +\sinh^2\mu\sin^2\nu\dot\phi^2
         \nonumber\\   & & \qquad\qquad\qquad\qquad\qquad\qquad\qquad
      -\omega^2(\sinh^2\mu\sin^2\nu+\cosh^2\mu\cos^2\nu)\Big)
         \nonumber\\   & & \qquad\qquad\qquad\qquad\qquad\qquad\qquad
-{\hbar^2\over2Md^2}\Bigg({F(\tan\phi)-\viert\over\sinh^2\mu\sin^2\nu}
 +{k_3^2-\viert\over\cosh^2\mu\cos^2\nu}\Bigg)\Bigg]dt\Bigg\}\enspace.
\end{eqnarray}
We consider the spherical and circular polar coordinate cases, the
others can be treated in a similar way as far as the separation of the
$\phi$-path integration is concerned. We denote the energy-spectrum of
the $\phi$-dependent problem by $E_{\lambda_\phi}=\hbar^2\lambda_\phi^2/
2M$. The most important case occurs when $F(\tan\phi)=constant$. Then
the quantum motion in $\phi$ are just circular waves and we have in this
case $\Phi_{\lambda_\phi}(\phi)\equiv\Phi_\nu(\phi)=\e^{\i\nu\phi}/
\sqrt{2\pi}$ $(\nu\in\bbbz)$. This is also the case for the potentials
$V_6$ and $V_7$. We have ($\lambda_2=2n+\lambda_\phi\pm k_3+1$)
\eject\noindent
\begin{eqnarray}       & &\!\!\!\!\!\!\!\!\!\!\!\!
  K^{(V_5)}({\vec x\,}'',{\vec x\,}';T)
         \nonumber\\   & &\!\!\!\!\!\!\!\!\!\!\!\!
  \hbox{{\it Spherical Coordinates} \cite{CARPa}:}
         \nonumber\\   & &\!\!\!\!\!\!\!\!\!\!\!\!
  =(r'r''\sin\theta'\sin\theta'')^{-1/2}\int dE_{\lambda_\phi}
  \Phi^*_{\lambda_\phi}(\phi')\Phi_{\lambda_\phi}(\phi'')
         \nonumber\\   & &\!\!\!\!\!\!\!\!\!\!\!\!\qquad\times
  \sum_{n=0}^\infty\Phi_n^{(\lambda_\phi,\pm k_3)}(\theta')
  \Phi_n^{(\lambda_\phi,\pm k_3)}(\theta'')
  {M\omega\over\i\hbar\sin\omega T}
  \exp\bigg[{\i M\over2\hbar}({r'}^2+{r''}^2)\cot\omega T\bigg]
  I_{\lambda_2}\bigg({M\omega r'r''\over\i\hbar\sin\omega T}\bigg)
  \qquad\quad     \\   & &\!\!\!\!\!\!\!\!\!\!\!\!
  =\int dE_{\lambda_\phi}\sum_{n,l=0}^\infty\e^{-\i E_NT/\hbar}
  \Psi^*_{\lambda_\phi,n,l}(\phi',\theta',r')
  \Psi_{\lambda_\phi,n,l}(\phi'',\theta'',r'')
                  \\   & &\!\!\!\!\!\!\!\!\!\!\!\!
  \hbox{\it Circular Polar Coordinates:}
         \nonumber\\   & &\!\!\!\!\!\!\!\!\!\!\!\!
  =\bigg({M\omega\over\i\hbar\sin\omega T}\bigg)^2\sqrt{z'z''}\,
  \exp\bigg[-{M\omega\over2\i\hbar}
     ({z'}^2+{z''}^2+{\rho'}^2+{\rho''}^2)\cot\omega T\bigg]
  I_{\pm k_3}\bigg({M\omega z'z''\over\i\hbar\sin\omega T}\bigg)
         \nonumber\\   & &\!\!\!\!\!\!\!\!\!\!\!\!\qquad\times
  \int dE_{\lambda_\phi}
  \Phi^*_{\lambda_\phi}(\phi')\Phi_{\lambda_\phi}(\phi'')
  I_{\lambda_\phi}
  \bigg({M\omega\rho'\rho''\over\i\hbar\sin\omega T}\bigg)
                  \\   & &\!\!\!\!\!\!\!\!\!\!\!\!
  =\int dE_{\lambda_\phi}\sum_{m,n_z=0}^\infty\e^{-\i E_NT/\hbar}
  \Psi^*_{\lambda_\phi,m,n_z}(\phi',\rho',z')
  \Phi_{\lambda_\phi,m,n_z}(\phi'',\rho'',z'')\enspace.
\end{eqnarray}
The wave-functions and energy-levels are given by
\begin{eqnarray}       & &
  \hbox{\it Spherical Coordinates:}
         \nonumber\\   & &
  \Phi_{\lambda_\phi,n,l}(\phi,\theta,r)
  =(r\sin\theta)^{-1/2}\Phi_{\lambda_\phi}(\phi)
  \Phi_m^{(\lambda_\phi,\pm k_3)}(\theta')
         \nonumber\\   & &\qquad\times
  \sqrt{2M\omega\over\hbar}\bigg({M\omega\over\hbar}\bigg)^{\lambda_2/2}
  \sqrt{l!\over\Gamma(l+\lambda_2+1)}\,
  r^{\lambda_2+1/2}\exp\bigg(-{M\omega\over2\hbar}r^2\bigg)
  L_l^{(\lambda_2)}\bigg({M\omega\over\hbar}r^2\bigg)
  \enspace,\qquad\qquad
                  \\   & &
  E_N=\hbar\omega(N\pm k_3+2)\enspace,\qquad
    N=2l+2n+\lambda_\phi\enspace,
                  \\   & &
  \hbox{\it Circular Polar Coordinates:}
         \nonumber\\   & &
  \Psi_{\lambda_\phi,m,n_z}(\phi,\rho,z)=\Phi_{\lambda_\phi}(\phi)
  {2M\omega\over\hbar}\bigg({M\omega\over\hbar}\bigg)^{\pm k_3/2}
         \nonumber\\   & &\qquad\times
  \sqrt{n_z!\over\Gamma(n_z\pm k_3+1)}
  z^{1/2\pm k_3}\exp\bigg(-{M\omega\over2\hbar}z^2\bigg)
  L_{n_z}^{(\pm k_3)}\bigg({M\omega\over\hbar}z^2\bigg)
  \qquad\qquad\qquad\qquad
         \nonumber\\   & &\qquad\times
  \bigg({M\omega\over\hbar}\rho^2\bigg)^{\lambda_\phi/2}
  \sqrt{m!\over\Gamma(m+\lambda_\phi+1)}
  \exp\bigg(-{M\omega\over2\hbar}\rho^2\bigg)
  L_m^{(\lambda_\phi)}\bigg({M\omega\over\hbar}\rho^2\bigg)\enspace,
                  \\   & &
  E_N=\hbar\omega(N\pm k_3+2)\enspace,\qquad
    N=2m+2n_z+\lambda_\phi\enspace.
\end{eqnarray}
For $F(\tan\phi)=constant$, $\lambda_\phi=|\nu|$, c.f.\ also $V_6$ and
$V_7$.

\vglue0.4truecm\noindent
{\it 3.3.6.}~We consider the potential
\begin{equation}
  V_6(\vec x)={M\over2}\omega^2(x^2+y^2+4z^2)
     +\hbarm{F(y/x)\over x^2+y^2}\enspace.
\end{equation}
We obtain the following path integral formulations in the
corresponding separable coordinate systems
We write down the corresponding path integral formulations
\begin{eqnarray}       & &\!\!\!\!\!\!\!\!
  K^{(V_6)}({\vec x\,}'',{\vec x\,}';T)
         \nonumber\\   & &\!\!\!\!\!\!\!\!
  \underline{\hbox{Circular Polar Coordinates:}}
         \nonumber\\   & &\!\!\!\!\!\!\!\!
  =\int\limits_{z(t')=z'}^{z(t'')=z''}\CD z(t)
   \int\limits_{\rho(t')=\rho'}^{\rho(t'')=\rho''}\rho\CD\rho(t)
   \int\limits_{\phi(t')=\phi'}^{\phi(t'')=\phi''}\CD\phi(t)
         \nonumber\\   & &\!\!\!\!\!\!\!\!\qquad\times
  \exp\Bigg\{\ih\int_{t'}^{t''}\Bigg[{M\over2}
     \Big(\dot\rho^2+\rho^2\dot\phi^2+\dot z^2
                                -\omega^2(\rho^2+4z^2)\Big)
     -\hbar^2{F(\tan\phi)-\viert\over2M\rho^2}\Bigg]dt\Bigg\}
                  \\   & &\!\!\!\!\!\!\!\!
  =(\rho'\rho'')^{-1/2}\int dE_{\lambda_\phi}
  \Phi^*_{\lambda_\phi}(\phi')\Phi_{\lambda_\phi}(\phi'')
         \nonumber\\   & &\!\!\!\!\!\!\!\!\qquad\times
  \int\limits_{z(t')=z'}^{z(t'')=z''}\CD z(t)
  \int\limits_{\rho(t')=\rho'}^{\rho(t'')=\rho''}
  \mu_{\lambda_\phi}[\rho^2]\CD\rho(t)
  \exp\Bigg[{\i M\over2\hbar}\int_{t'}^{t''}
     \Big(\dot\rho^2+\dot z^2-\omega^2(\rho^2+4z^2)\Big)dt\Bigg]
                  \\   & &\!\!\!\!\!\!\!\!
  =\int dE_{\lambda_\phi}\sum_{n_z,l=0}^\infty
   \e^{-\i E_{n_z,m}T/\hbar}
   \Psi_{n_z,\lambda_\phi,m}^*(\phi',\rho',z')
   \Psi_{n_z,\lambda_\phi,m}(\phi'',\rho'',z'')
                  \\   & &\!\!\!\!\!\!\!\!
  \underline{\hbox{Parabolic Coordinates:}}
         \nonumber\\   & &\!\!\!\!\!\!\!\!
  =(\xi'\xi''\eta'\eta'')^{-1/2}\int dE_{\lambda_\phi}
  \Phi^*_{\lambda_\phi}(\phi')\Phi_{\lambda_\phi}(\phi'')
         \nonumber\\   & &\!\!\!\!\!\!\!\!\qquad\times
   \int_{\bbbr}{dE\over2\pi\hbar}\e^{-\i ET/\hbar}\int_0^\infty ds''
  \int\limits_{\eta(0)=\eta'}^{\eta(s'')=\eta''}\CD\eta(s)
  \int\limits_{\xi(0)=\xi'}^{\xi(s'')=\xi''}\CD\xi(s)
         \nonumber\\   & &\!\!\!\!\!\!\!\!\qquad\times
  \exp\Bigg\{\ih\int_0^{s''}\Bigg[{M\over2}\Big(
   (\dot\xi^2+\dot\eta^2)-\omega^2(\xi^6+\eta^6)\Big)
     +E(\xi^2+\eta^2)
     -\hbar^2{\lambda_\phi^2-\viert\over2M}
     \bigg({1\over\xi^2}+{1\over\eta^2}\bigg)\Bigg]ds\Bigg\}\enspace.
         \nonumber\\   & &\!\!\!\!\!\!\!\!
\end{eqnarray}
Here we have used the same notation as in the previous case. The
wave-functions and the energy-spectrum in {\it circular polar
coordinates\/} are given by
\begin{eqnarray}       & &\!\!\!\!\!\!\!\!\!\!\!\!
   \Psi_{n_z,\lambda_\phi,m}(\phi,\rho,z)
  =\Phi_{\lambda_\phi}(\phi)\sqrt{2M\omega\over\hbar}
  \bigg({M\omega\over\hbar}\rho^2\bigg)^{\lambda_\phi/2}
  \sqrt{m!\over\Gamma(m+\lambda_\phi+1)}
  \exp\bigg(-{M\omega\over2\hbar}\rho^2\bigg)
  L_m^{(\lambda_\phi)}\bigg({M\omega\over\hbar}\rho^2\bigg)
         \nonumber\\   & &\!\!\!\!\!\!\!\!\!\!\!\!\qquad\times
  \sqrt{\sqrt{2M\omega\over\pi\hbar}{1\over2^{n_z}n_z!}}
  \exp\bigg(-{M\omega\over\hbar}z^2\bigg)
  H_{n_z}\left(\sqrt{2M\omega\over\hbar}\,z\right)\enspace,
                  \\   & &\!\!\!\!\!\!\!\!\!\!\!\!
  E_{n_z,m}=\hbar\omega(2m+\lambda_\phi+n_z+\hbox{${3\over2}$})\enspace.
\end{eqnarray}

\vglue0.4truecm\noindent
{\it 3.3.7.}~We consider ($k_1>0$)
\begin{equation}
  V_7(\vec x)=-{\alpha\over\sqrt{x^2+y^2+z^2}}
     +{\hbar^2\over2M(x^2+y^2)}\Bigg(
  {k_1^2z\over\sqrt{x^2+y^2+y^2}}+F\bigg({y\over x}\bigg)\Bigg)\enspace.
\end{equation}
For $F(y/x)=\gamma^2>0$ this potential is known as the ring-shaped
Hartmann potential (c.f.\ Calogero \cite{CALO}, Carpio-Bernido et al.\
\cite{CARPa, CARPb, CBBI, CBB}, Chetouani at al.\ \cite{CGHa}, Gerry
\cite{GERRY}, Granovsky et al.\ \cite{GZLa}, Grosche \cite{GROm}, Guha
and Mukherjee \cite{GUMU}, Hartmann \cite{HART}, Kibler et al.\
\cite{KICA}-\cite{KNe}, \cite{KIWIa}, Lutsenko et al.\ \cite{LPSTA},
Vaidya and Boschi-Filho \cite{VBH}, and Zhedanov \cite{ZHE}; compare
also the connection with a Coulomb plus Aharonov-Bohm solenoid, e.g.\
Chetouani et al.\ \cite{CGHc}, Kibler and Negadi \cite{KNf}, and
S\"okmen \cite{SOKc}). We write down the corresponding  path integral
formulations where only  the corresponding Green function can be
evaluated [$\lambda_\pm^2=\lambda_\phi^2\pm k_1^2$, $\lambda_1=n+
(\lambda_++\lambda_-+1)/2$, $\kappa=\alpha\sqrt{-M/2E}/\hbar$]
\begin{eqnarray}       & &\!\!\!\!\!\!\!\!
  \ih\int_0^\infty dT\,\e^{\i ET/\hbar}
  K^{(V_7)}({\vec x\,}'',{\vec x\,}';T)
         \nonumber\\   & &\!\!\!\!\!\!\!\!
  \underline{\hbox{Spherical Coordinates:}}
         \nonumber\\   & &\!\!\!\!\!\!\!\!
  =\ih\int_0^\infty dT\,\e^{\i ET/\hbar}
   \int\limits_{r(t')=r'}^{r(t'')=r''}r^2\CD r(t)
   \int\limits_{\theta(t')=\theta'}^{\theta(t'')=\theta''}
  \sin\theta\CD\theta(t)
   \int\limits_{\phi(t')=\phi'}^{\phi(t'')=\phi''}\CD\phi(t)
         \nonumber\\   & &\!\!\!\!\!\!\!\!\qquad\times
  \exp\Bigg\{\ih\int_{t'}^{t''}\Bigg[{M\over2}
  \big(\dot r^2+r^2\dot\theta^2+r^2\sin^2\theta\dot\phi^2\big)
         \nonumber\\   & &\!\!\!\!\!\!\!\!
  \qquad\qquad\qquad\qquad\qquad\qquad
  +{\alpha\over r}
  -{\hbar^2\over2Mr^2}\Bigg({k_1^2\cos\theta\over\sin^2\theta}
         +{F(\tan\phi)-\viert\over\sin^2\theta}
         -{1\over4}\Bigg)\Bigg]dt\Bigg\}
                  \\   & &\!\!\!\!\!\!\!\!
  =({r'}^2{r''}^2\sin\theta'\sin\theta'')^{-1/2}\int dE_{\lambda_\phi}
   \Phi^*_{\lambda_\phi}(\phi')\Phi_{\lambda_\phi}(\phi'')
   \half\sum_{n=0}^\infty
   \Phi_n^{(\lambda_+,\lambda_-)}\bigg({\theta'\over2}\bigg)
   \Phi_n^{(\lambda_+,\lambda_-)}\bigg({\theta''\over2}\bigg)
         \nonumber\\   & &\!\!\!\!\!\!\!\!\qquad\times
   {1\over\hbar}\sqrt{-{M\over2E}}\,
  {\Gamma(\half+\lambda_1-\kappa)\over\Gamma(2\lambda_2+1)}
   W_{\kappa,\lambda_1}\bigg(\sqrt{-8ME}\,{r_>\over\hbar}\bigg)
   M_{\kappa,\lambda_1}\bigg(\sqrt{-8ME}\,{r_<\over\hbar}\bigg)
                  \\   & &\!\!\!\!\!\!\!\!
  =\sum_{n,m=0}^\infty\left\{\sum_{l=0}^\infty
   {\Psi_{n,m,l}^*(\theta',\phi',r')
    \Psi_{n,m,l}(\theta'',\phi'',r'')\over E_N-E}
   +\int_{\bbbr} dp{\Psi_{n,m,p}^*(\theta',\phi',r')
    \Psi_{n,m,p}(\theta'',\phi'',r'')\over\hbar^2p^2/2M-E}\right\}
         \nonumber\\   & &\!\!\!\!\!\!\!\!
                  \\   & &\!\!\!\!\!\!\!\!
  \underline{\hbox{Parabolic Coordinates:}}
         \nonumber\\   & &\!\!\!\!\!\!\!\!
  =\ih\int_0^\infty dT\,\e^{\i ET/\hbar}
  \int\limits_{\eta(t')=\eta'}^{\eta(t'')=\eta''}\CD\eta(t)
   \int\limits_{\xi(t')=\xi'}^{\xi(t'')=\xi''}\CD\xi(t)
   (\xi^2+\eta^2)\xi\eta
   \int\limits_{\phi(t')=\phi'}^{\phi(t'')=\phi''}\CD\phi(t)
         \nonumber\\   & &\!\!\!\!\!\!\!\!\qquad\times
  \exp\Bigg\{\ih\int_{t'}^{t''}\Bigg[{M\over2}\Big(
     (\xi^2+\eta^2)(\dot\xi^2+\dot\eta^2)+\xi^2\eta^2\dot\phi^2\Big)
         \nonumber\\   & &\!\!\!\!\!\!\!\!
  \qquad\qquad\qquad\qquad\qquad\qquad
     +{2\alpha\over\xi^2+\eta^2}-{\hbar^2\over2M\xi^2\eta^2}
     \Bigg(k_1^2{\eta^2-\xi^2\over\eta^2+\xi^2}
     +F(\tan\phi)-\viert\Bigg)\Bigg]dt\Bigg\}\qquad
                  \\   & &\!\!\!\!\!\!\!\!
  =\sum_{n=0}^\infty\Bigg[\sum_{n_1,n_2=0}^\infty
   {\Psi_{n,n_1,n_2}(\phi',\xi',\eta')
   \Psi_{n,n_1,n_2}(\phi'',\xi'',\eta'')\over E_{n_1,n_2}-E}
         \nonumber\\   & &\!\!\!\!\!\!\!\!
  \qquad\qquad\qquad\qquad\qquad\qquad
  +\int_0^\infty dp\int_{\bbbr} d\zeta
  {\Psi_{n,p,\zeta}^*(\phi',\xi',\eta')
   \Psi_{n,p,\zeta}(\phi'',\xi'',\eta'')
  \over\hbar^2p^2/2M-E}\Bigg]\enspace.
\end{eqnarray}
The bound state wave-functions in the {\it polar coordinates \/} are
\begin{eqnarray}
  \Psi_{n,m,l}(\theta,\phi,r)
  &=&(2\sin\theta)^{-1/2}\Psi^*_{\lambda_\phi}(\phi)
   \Phi_m^{(\lambda_+,\lambda_-)}\bigg({\theta\over2}\bigg)
         \nonumber\\   & & \qquad\times
  {2\over(l+\lambda_1+\half)^2}\bigg[{l!\over a^3(l+\lambda_1+\half)
   \Gamma(l+2\lambda_1+1)}\bigg]^{1/2}
         \nonumber\\   & & \qquad\times
  \bigg({2r\over a(l+\lambda_1+\half)}\bigg)^{\lambda_1}
  \exp\bigg(-{r\over a(l+\lambda_1+\half)}\bigg)
  L_l^{(2\lambda_1)}\bigg({2r\over a(l+\lambda_1+\half)}\bigg)\enspace,
         \nonumber\\   & &
                  \\
  E_N&=&-{M\alpha^2\over\hbar^2N^2}\enspace,\qquad
    N=l+n+\bhalf(1+\lambda_-+\lambda_+)\enspace.
\end{eqnarray}
The continuous wave-functions are
\begin{eqnarray}
  \Psi_{n,l,p}(\theta,\phi,r)
  &=&=(2\sin\theta)^{-1/2}\Phi_{\lambda_\phi}(\phi)
   \Phi_n^{(\lambda_+,\lambda_-)}\bigg({\theta\over2}\bigg)
         \nonumber\\   & & \qquad\times
  {\Gamma(\half+\lambda_2-\i/ap)\over
  \sqrt{2\pi}\,r\Gamma(2\lambda_2+1)}\exp\bigg({\pi\over2ap}\bigg)
  M_{\i/ap,\lambda_2}(-2\i pr)\enspace.
\end{eqnarray}
The wave-functions in {\it parabolic coordinates\/} are the same as in
3.2.3, except for the modification that $\lambda_\pm^2=\lambda_\phi^2\pm
k_1^2$ instead of $\lambda_\pm^2=k_2\pm k_1^2$.

The potential $V_7$ is also separable in prolate spheroidal II
coordinates and we have the identity
\begin{eqnarray}       & &
  K^{(V_7)}({\vec x\,}'',{\vec x\,}';T)
         \nonumber\\   & &
  =\int\limits_{\mu(t')=\mu'}^{\mu(t'')=\mu''}\CD\mu(t)
   \int\limits_{\nu(t')=\nu'}^{\nu(t'')=\nu''}\CD\nu(t)
   d^3(\sinh^2\mu+\sin^2\nu)\sin\nu\sinh\mu
   \int\limits_{\phi(t')=\phi'}^{\phi(t'')=\phi''}\CD\phi(t)
         \nonumber\\   & & \quad\times
  \exp\Bigg\{\ih\int_{t'}^{t''}\Bigg[{M\over2}d^2\Big(
  (\sinh^2\mu+\sin^2\nu)(\dot\mu^2+\dot\nu^2)
   +\sinh^2\mu\sin^2\nu\dot\phi^2\Big)
         \nonumber\\   & & \qquad
  +{\alpha\over d(\cosh\mu+\cos\nu)}
  -{\hbar^2\over2Md^2\sinh^2\mu\sin^2\nu}
   \Bigg({k_1^2(\cosh\mu\cos\nu+1)\over\cosh\mu+\cos\nu}
        +F(\tan\phi)-{1\over4}\Bigg)\Bigg]dt\Bigg\}
         \nonumber\\   & &
                  \\   & &
  =(d^2\sinh\mu'\sinh\mu''\sin\nu'\sin\nu'')^{-1/2}
   \int dE_{\lambda_\phi}
   \Phi^*_{\lambda_\phi}(\phi')\Phi_{\lambda_\phi}(\phi'')
   \int_{\bbbr}{dE\over2\pi\hbar}\e^{-\i ET/\hbar}\int_0^\infty ds''
         \nonumber\\   & & \quad\times
   \int\limits_{\mu(0)=\mu'}^{\mu(s'')=\mu''}\CD\mu(s)
   \int\limits_{\nu(0)=\nu'}^{\nu(s'')=\nu''}\CD\nu(s)
   \exp\Bigg\{\ih\int_0^{s''}\Bigg[{M\over2}(\dot\mu^2+\dot\nu^2)
   +Ed^2(\cosh^2\mu-\cos^2\nu)
         \nonumber\\   & & \qquad
  +\alpha d(\cosh\mu-\cos\nu)
  -\hbarm\Bigg({\lambda^2-\viert+k_1^2\cosh\mu\over\sinh^2\mu}
   +{\lambda^2-\viert+k_1^2\cos\nu\over\sin^2\nu}\bigg)\Bigg]ds\Bigg\}
   \enspace.
\end{eqnarray}
Unfortunately we cannot solve this path integral.

\vglue0.4truecm\noindent
{\it 3.3.8.}~We consider the potential ($\rho=\sqrt{x^2+y^2}$\,)
\begin{equation}
  V_8(\vec x)=-{\alpha\over\rho}+\sqrt{2\over\rho}
   \bigg(\beta_1\cos{\phi\over2}+\beta_2\sin{\phi\over2}\bigg)
   +F(z)\enspace.
\end{equation}
This potential is the same potential as in 3.1.4 with an additional
$z$-dependence via $F(z)$. The corresponding path integral formulations
and  solutions in mutually orthogonal circular parabolic coordinates
are therefore a straightforward combination of the results of 3.1.4 and
$K_F(z'',z';T)$.


\vglue0.6truecm\noindent
{\large\bf 4.~Summary.}
\vglue0.4truecm\noindent
\noindent
In the present work we have made the first step in the classification of
all the presently possible path integral solutions of the
Smorodinsky-Winternitz potentials in two and three dimensions. This
approach as compared with the Schr\"odinger approach has advantages as
well as disadvantages. The advantages are that the path integral gives
a global view of the problem in question, whereas the Schr\"odinger
equation only a local one. This property enables one to incorporate
into the Feynman path integral many important features like topological
effects, perturbation theory as well as the investigation of
non-perturbative effects, a semi-classical expansion of the propagator,
and many more. It is quite likely that the Feynman path integral is
possibly the only consistent way to incorporate renormalization theory
in a unified field theory. Therefore it is always important to
investigate solvable quantum systems ``from the point of view of
fluctuating paths'' \cite{DKb}. A further advantage is that the
explicit computation of the propagator (the Feynman kernel),
respectively the (energy-dependent) Green's function, gives
simultaneously the spectral expansion into the wave-functions and the
energy spectrum.

A disadvantage is the fact that an explicit path integration can be
done only in a limited number of coordinate systems, namely in polar,
spherical, and some problems in parabolic coordinates, respectively
coordinate systems which are related to them. Path integral evaluations
in, say, elliptic, paraboloidal and other parametric coordinate systems
are not accessible. One may say that one of the very advantages of the
Feynman path integral, i.e.\ its global approach to a quantum mechanical
problem, leads in these cases to a disadvantage, because the actual
path integration would require ``addition theorems'' in terms of special
functions of parametric coordinate systems, i.e.\ their corresponding
spherical functions of the Laplace operator. Such ``addition theorems''
are not to our disposal presently. However, it seems reasonable that a
path integral identity in conical coordinates can be derived via the
path integral on the sphere $S^{(6)}$, where for the spherical
functions a representation in terms of Jacobi functions is chosen.

Explicitly solvable models always involve solutions in terms of
hypergeometric and confluent hypergeometric functions. We have seen
that in at least one coordinate system each of the discussed
Smorodinsky-Winternitz potentials can be solved in terms of these
higher transcendental functions. In the parameter dependent coordinates
the solution cannot be written in closed form. Therefore an
investigation of the Feynman path integral in such coordinate systems
is desirable and further studies along these lines will be subject to
future work. This is not idle doing, then, for instance, the spheroidal
coordinate system is only one into which the two-center problem
separates. In future investigations we will try to discuss these more
complicated problems.

The present work does not claim a full analysis of the quantum
mechanical Smorodinsky-Winternitz system. Such an analysis must include
not only a comprehensive classification but also the determination of
the interbasis coefficients of the wave-functions of the same energy.
This requires the knowledge of the dynamical symmetry group which
is responsible for the accidental degeneracy of the energy spectrum of
a Smorodinsky-Winternitz potential.

In due time we will continue our work along the lines described above,
and we will also try to include an analysis in which quantum mechanical
problems in spaces of (positive and negative) constant curvature are
taken into account. This will include a path integral approach of
parametric coordinate systems on the sphere and pseudo-sphere, the
study of the Coulomb- and the Higgs oscillator problem in particular,
and of Smorodinsky-Winternitz potentials in spaces of constant curvature
in general.

\vglue0.6truecm\noindent
{\large\bf Acknowledgement.}
\vglue0.4truecm\noindent
\noindent
G.S.Pogosyan would like to thank the members of the II.Institut f\"ur
Theoretische Physik, Hamburg University, for their kind hospitality.


\vglue0.6truecm\noindent
{\large\bf Appendix.}
\vglue0.4truecm\noindent
In this appendix we want to give a short discussion of the path integral
problem of the anharmonic sextic oscillator as encountered in parabolic
coordinates for the Holt potential. Let us consider the one-dimensional
path integral
\begin{equation}
  K(x'',x';T)=\int\limits_{x(t')=x'}^{x(t'')=x''}\CD x(t)
  \exp\left\{\ih\int_{t'}^{t''}
  \Bigg[{M\over2}\big(\dot x^2-\omega^2x^6\big)
     +{\hbar^2 k^2\over2M}x^2-{\hbar^2\beta^2\over2Mx^2}\Bigg]dt\right\}
\label{numAb}
\end{equation}
Performing a combined space-time transformation with the new coordinate
$z=x^4$, i.e.\ $x=F(z)=z^{1/4}$, and the new ``pseudo-time'' defined by
$s_j-s_{j-1}=\delta_j=16x_j^3x_{j-1}^3(t_j-t_{j-1})$ on the lattice,
respectively $s''=16\int_{t'}^tx^6(s)ds$, we obtain the transformtion
formul\ae\ (c.f.\ \cite{DURb, FLMa, GRSb, GRSg, KLEm, PS})
\begin{eqnarray}
  K(x'',x';T)
  &=&\int_{-\infty}^\infty{dE\over2\pi\i}\e^{-\i TE/\hbar} G(z'',z';E)
               \\
  G(z'',z';E)&=&\Big[F'(z'')F'(z')\Big]^{1/2}
     \ih\int_0^\infty \hat K(z'',z';s'')ds''\enspace,
\label{numAa}
\end{eqnarray}
and the kernel $\hat K(z'',z';s'')$ is given by
[$\lambda^2-\viert=(\beta^2+15/4)16$]
\begin{equation}
  \hat K(z'',z',s'')=\int\limits_{z(0)=z'}^{z(s'')=z''}\CD z(s)
  \exp\Bigg\{\ih\int_0^{s''}\Bigg[{M\over2}\dot z^2
  +{\hbar^2k^2\over32Mz}+{E\over16 z^{3/2}}
  -{\hbar^2\over2M}{\lambda^2-\viert\over z^2}\Bigg]ds\Bigg\}\enspace.
\end{equation}
We see that this path integral cannot be evaluated for $E\not=0$.
This corresponds in the case of the Holt potential (and all related
ones as well) that we can only ``solve'' it for the parabolic
separation parameter equal to zero.

If we now set $E=0$ the path integral $\hat K(s'')$ is actually
a Coulomb potential problem. For the Coulomb potential the
Feynman kernel is not known in closed form, however the corresponding
Green's function $G^{(C)}(E)$ is. Insertion in (\ref{numAa}) gives the
the zero-energy Green's function of (\ref{numAb})
\begin{eqnarray}
  G(z'',z';0)&=&\Big[F'(z'')F'(z')\Big]^{1/2}
  G^{(C)}\bigg(z'',z';-{M\omega^2\over32}\bigg)
                  \\   &=&(z'z'')^{1/8}
  {4\over\hbar\omega}{\Gamma(\half+\lambda+\hbar k^2/8M\omega)\over
   \Gamma(1+2\lambda)}
         \nonumber\\   & &\qquad\qquad\times
   W_{-\hbar k^2/8M\omega,\lambda}\bigg({M\omega\over2\hbar}z_>\bigg)
   M_{-\hbar k^2/8M\omega,\lambda}\bigg({M\omega\over2\hbar}z_<\bigg)
  \enspace.
\end{eqnarray}
Note that this formula is a generalization of the result of Steiner
\cite{STEc} where only one power potential term was taken into account.
It is obvious that in a similar way an $x^{14}$ potential term can be
considered instead of a $x^6$, leading to a (zero-energy) Green's
function of a radial harmonic oscillator. Furthermore, higher power
potentials can be also treated by choosing another power dependence
in the transformation $x=F(z)$.

Note that the zero-energy Green's function can also be evaluated in
cases, where coordinate systems in flat space separate the Laplace
equation (and not the Helmhotz, respectively the Schr\"odinger
equation), i.e.\ coordinate systems which are R-separable, as for
instance bispherical or toroidal coordinates.


\newpage\noindent
\renewcommand{\baselinestretch}{0.95}

\end{document}